%% file: fullpaper.tex
\documentclass[aps,pre,reprint,amsmath,amssymb,floats,floatfix, twocolumn,
superscriptaddress,nofootinbib,showpacs]{revtex4-1}

\pdfoutput=1

\usepackage[usenames,dvipsnames]{xcolor}
\usepackage{hyperref}
\definecolor{cobalt}{RGB}{44, 98, 120}
\hypersetup{
  colorlinks,
  citecolor=Violet,
  linkcolor=cobalt,
  urlcolor=Blue
}
\usepackage[capitalise]{cleveref}
\usepackage[normalem]{ulem}
\usepackage[caption=false]{subfig}
\usepackage{tensor}

\usepackage{soul}

\usepackage{tikz}
\usetikzlibrary{positioning}
\usetikzlibrary{decorations.pathmorphing}
\usetikzlibrary{decorations.text}
\usetikzlibrary{shapes.misc}
\usetikzlibrary{calc}
\usepackage{tikzscale}

\graphicspath{{figures/}}

\newcommand{\Cornell}{\affiliation{Cornell Center for Astrophysics
    and Planetary Science, Cornell University, Ithaca, New York 14853, USA}}

\newcommand{\ts}{\textsuperscript}
\newcommand{\SpEC}{\textsc{SpEC}}

\usepackage{microtype}

\begin{document}

\title{
  Toroidal Horizons in Binary Black Hole Mergers
}

\author{Andy Bohn}
\email[Contact email: ]{adb228@cornell.edu}
\author{Lawrence E. Kidder, Saul A. Teukolsky} \Cornell

\date{\today}

\begin{abstract}
\input{abstract.tex}
\end{abstract}


\maketitle

\newcommand{\tikzprefix}{Tikz}
\input{body.tex}

\begin{acknowledgments}
\input{acknowledgments.tex}
\end{acknowledgments}

\bibliography{References/References}

\end{document}

%% file: abstract.tex
We find the first binary black hole event horizon with a toroidal topology.
It had been predicted that generically
the event horizons of merging black holes should
briefly have a toroidal topology, but such a phase has never been
seen prior to this work.
In all previous binary black hole simulations,
in the coordinate slicing used to evolve the black holes,
the topology of the event
horizon
transitions directly from two spheres during the inspiral to a single sphere
as the black holes merge.
We present a coordinate transformation to a foliation of spacelike hypersurfaces
that ``cut a hole'' through the event horizon surface, resulting in a toroidal
event horizon.
A torus could potentially provide a mechanism for violating topological
censorship.
However, these toroidal event horizons satisfy topological censorship by
construction, because we can always trivially apply the inverse
coordinate transformation to remove the topological feature.

%% file: body.tex
\section{Introduction}

It is well established that stationary
black hole spacetimes contain an
event horizon with a spherical topology, assuming the dominant energy
condition holds~\cite{hawkingellis, Hawking1972, Chrusciel1994}.
If the black hole is allowed to be dynamical, Gannon~\cite{Gannon1976}
showed that smooth black hole event horizons
could have either a spherical or a toroidal topology.
Topological censorship
places an upper bound on the lifetime of any topological structure
such as a toroidal event horizon, where the torus must collapse
faster than it would take light to traverse
it~\cite{Friedman1993, Galloway1995, Jacobson1995}.
Otherwise an observer would be able to probe the topological structure
of the torus by passing a light ray through the hole.
Equivalently, a different foliation
of the spacetime can always be chosen such that the toroidal event horizon has
a spherical topology~\cite{Siino1998a, Siino1998b}.
Numerical simulations of the collapse of a rotating distribution of matter
showed that
event horizons can indeed initially form with a short-lived toroidal topology
that quickly transitions to a sphere~\cite{Hughes1994, Shapiro1995}.

The situation with merging black holes is more complicated.
Siino~\cite{Siino1998b} and Husa and Winicour~\cite{Husa-Winicour:1999}
predicted that the event horizon of a
generic binary black hole
system should briefly exhibit a toroidal
topology during the merger.
However, no
toroidal event horizons have been found in numerical simulations of
merging black holes,
where the topology has only been seen to transition from
two spheres during the inspiral to a single sphere after
the merger.\footnote{We are specifically discussing the topology of
slices of the event horizon on Cauchy surfaces as opposed to
the global topology of the $2+1$-dimensional event horizon hypersurface.
The topology of the event horizon has only been seen initially
as the disjoint union of two spheres~($\mathcal{S}^2 \sqcup \mathcal{S}^2$)
that transitions to a single sphere~($\mathcal{S}^2$) through
an instantaneous state called the wedge sum of two
spheres~($\mathcal{S}^2~\vee~\mathcal{S}^2$)~\cite{Hatcher2001}.
We will ignore the fine distinction between a disjoint union and a
wedge sum and just consider the union hereafter.
}
Cohen~\textit{et al.}~\cite{Cohen2012} found that the spatial cross section of
the
event horizon during merger has spherical topology, but the horizon
structure suggested that a different spacetime foliation should reveal
a torus.
Simulations of three black holes~\cite{Diener:2003} and eight black holes
in a ring~\cite{ponce:11} similarly did not exhibit a toroidal event
horizon.

\begin{figure}
 \centering
  \includegraphics[width=\columnwidth]{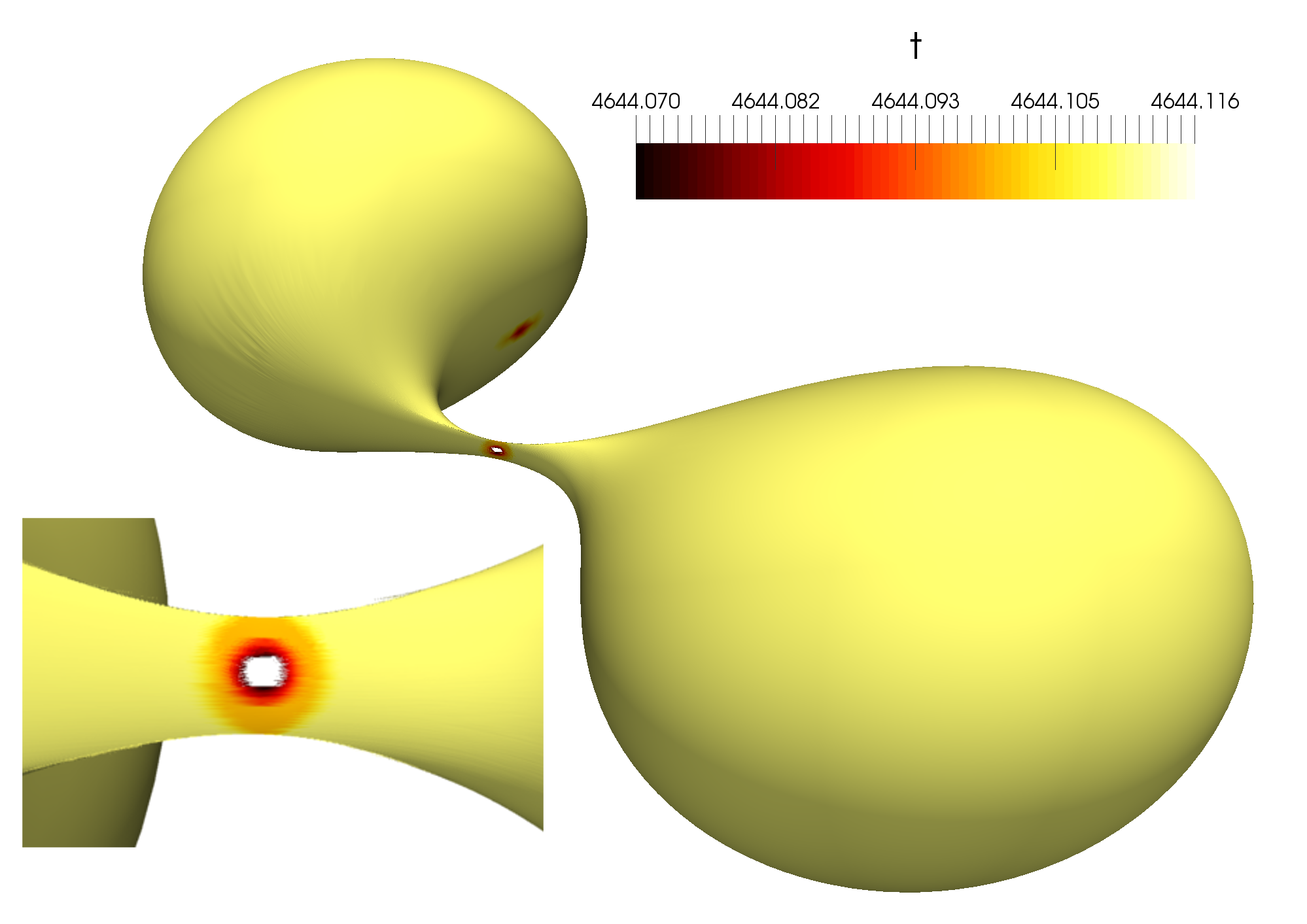}
  \caption[Toroidal event horizon found in a GW150914 consistent simulation]
  {Event horizon with a toroidal topology, shown in a different
    time slicing than the one used in the \SpEC{} simulation.
    The binary black hole simulation
    has a mass ratio of $1.25$ and spin
    parameters consistent with the
    first BBH system Advanced LIGO detected~\cite{Abbott:2016blz}.
    The inset figure in the bottom left corner shows a zoomed in
    and slightly rotated viewpoint of the hole in the event horizon.
    The horizon is colored by \SpEC{} simulation time $t$, which we will
    show in \cref{secReslicing} should have smaller values
    near the hole in this slicing.
  }
  \label{figLIGOTorus}
\end{figure}

For the results in this paper, we locate event horizons in binary black hole
(BBH) mergers by utilizing a theorem
stating that the event horizon is generated by null geodesics having no
future end point~\cite{Penrose1968, hawkingellis, Wald84}, meaning
they will never leave the EH surface in the future.
The method is based on choosing a set of outgoing null geodesics that
lie on the apparent horizon of the remnant black hole at the end of the
BBH simulation when the horizon is nearly stationary~\cite{Anninos1995}, and
integrating the geodesics backwards in time~\cite{Shapiro-Teukolsky:1980,
  Hughes1994,
  Shapiro1995, Anninos1995, Libson96, CohenPfeiffer2008, Cohen2012}.
The convention that we will follow in this paper
is to call these geodesics event horizon generators,
although they are only very good approximations to the true
generators~\cite{Cohen2012}.
Whereas generators of the horizon have no future endpoint, while tracing
the generators backwards in time, some may ``leave'' the event horizon
surface where they meet other generators of the horizon.
These meeting points are important in the study of event horizon topologies
and are called \textit{caustics} where infinitesimally neighboring generators
join together, and \textit{crossover points} where non-neighboring
generators cross paths~\cite{Shapiro1995,
Siino1998b, Husa-Winicour:1999, Lehner1999, Cohen2012}.
After they leave the event horizon surface backwards in time, generators
are known as \textit{future generators} of the horizon.

When viewing the event horizon forwards in time, future generators become
generators of the event horizon after they join at either caustics or
crossover points.
Browdy~\textit{et al.}~\cite{Browdy1995} found that the topology of the
event horizon must be spherical once future event horizon generators
cease joining the event horizon, which limits
any potential toroidal topology to times when future generators
are still joining the horizon.
Therefore it is critical to accurately
identify the time and location of caustics and crossover points.

In this paper, we find that the topology of the event horizon
for binary black hole (BBH) systems transitions from
two spheres ($2\times \mathcal{S}^2$)
to a single sphere ($\mathcal{S}^2$) in the gauge
used to merge the binary with the Spectral Einstein
Code~(\SpEC{})~\cite{SpECwebsite,
Szilagyi:2009qz, Hemberger:2012jz, SXSCatalog},
in agreement with previous results~\cite{Cohen2012}.
However, the event horizon is a $2+1$-dimensional hypersurface where
the topology of the event horizon depends on the foliation of the
spacetime~\cite{Siino1998a, Siino1998b}.
When considering how future generators join the event horizon,
the set of crossover points is known to live on a spacelike hypersurface
that becomes asymptotically null as this hypersurface approaches a set of
caustics~\cite{Shapiro1995}.
Therefore there must exist a spacelike foliation
that cuts a hole out of the spacelike surface of crossover points, resulting
in a short-lived toroidal event horizon.
We show explicitly that the event horizon topology can be
toroidal~($\mathcal{T}^2$) in a spacelike foliation of the spacetime, as
shown in \cref{figLIGOTorus},
by applying a coordinate transformation to the coordinate system
used in \SpEC{} to evolve the binary.

The organization of this paper is as follows: In \cref{secReslicing}
we present a coordinate transformation designed to find
a new spacetime foliation where the event horizon has
a toroidal topology.
We begin in \cref{secSphericalModel} by studying a toy model horizon of
a spherical wavefront in flat spacetime, where there are no crossovers.
In \cref{secHOMerger}, we analyze a head-on BBH merger and find a future
generator structure similar
to the spherical wavefront model that prohibits the possibility of a toroidal
event horizon in any spacelike foliation of the spacetime.
However, in \cref{secEllipsoidalModel} we show a toy model horizon of an
ellipsoidal\footnote{Here ``ellipsoidal'' refers to an
oblate ellipsoid that is not a coordinate sphere.}
wavefront in flat spacetime where the caustic and crossover
distribution allows for a torodial reslicing.
Utilizing what we learn with the ellipsoidal model, we are able to
directly reslice an equal mass inspiral EH into a short-lived
torus in \cref{secEMI}.
Finally, in \cref{secBabies},
we show that a similar coordinate transformation of the EH
can produce a ``baby'' event horizon that appears briefly during
BBH mergers, before all three surfaces connect.

\section{Reslicing the event horizon}
\label{secReslicing}

The binary black hole event horizons we simulated for this work
do not show a toroidal topology
using the \SpEC{} time coordinate.
However, the event horizon is a $2+1$-dimensional hypersurface,
and the simulation time coordinate describes only one possible
spacelike foliation of the hypersurface.
The generalized harmonic time slicing of our binary black hole
simulations~\cite{Lindblom2006}
may not be conducive to producing toroidal event
horizons~\cite{Cohen2012, CohenPfeiffer2008}.
We specify in this section
a coordinate transformation from the coordinate system
of the BBH evolution to a new coordinate system
to explore the possibility of another time slicing yielding a toroidal event
horizon.

In the companion~\cite{BohnMethods2016} to this paper, we
introduce a complete replacement for the previous
event horizon finding code in \SpEC{}~\cite{CohenPfeiffer2008, Cohen2012}.
The overall method is the same as before, where we evolve a set of event horizon
generators backwards in time to trace out the horizon surface.
At each time,
we connect the generators together to form a polygon approximating a
smooth surface with the topology
of a sphere that may be self-intersecting.
This surface does not approximate the event horizon only, but
the union of the true event horizon and
the locus of the future generators~\cite{ThorneSuggestion}.
The new event horizon finder is fully adaptive and so can resolve
fine-scale features of the event horizon.
This feature is crucial to demonstrating the existence of a toroidal topology.

\begin{figure}
  \centering
  \begin{tabular}{ccc}
      {\includegraphics[width=0.33\columnwidth]{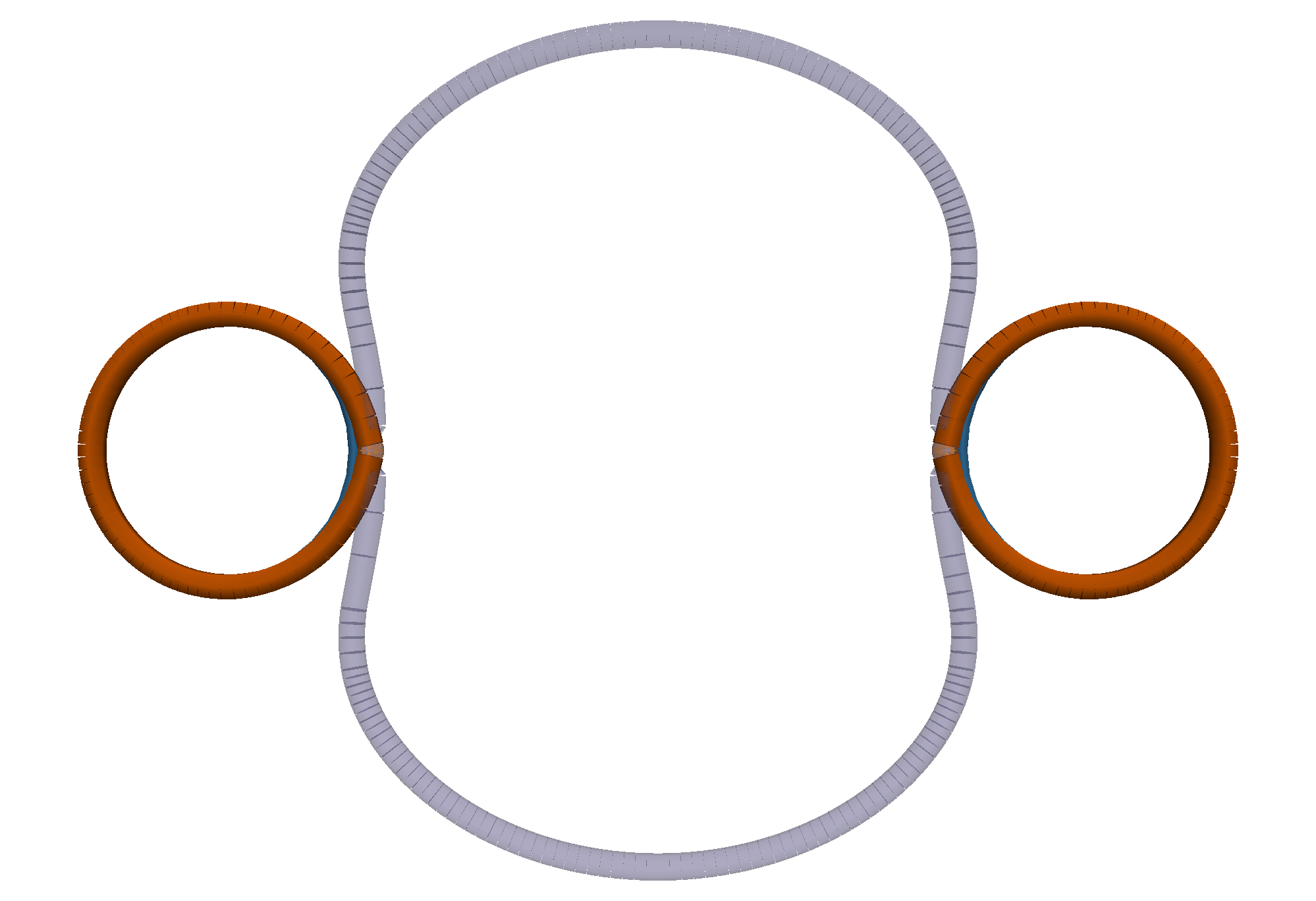}} &
      {\includegraphics[width=0.33\columnwidth]{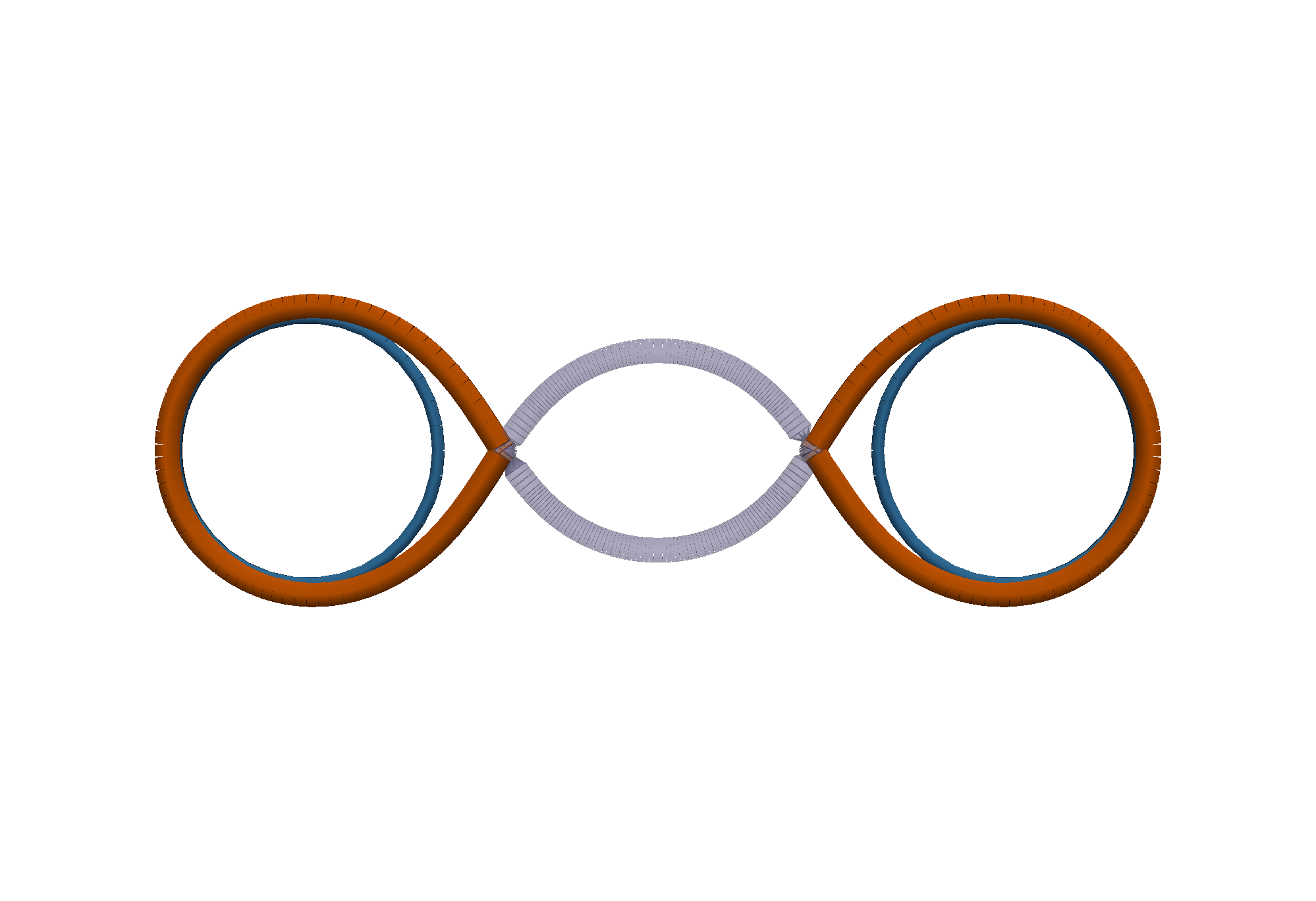}} &
      {\includegraphics[width=0.33\columnwidth]{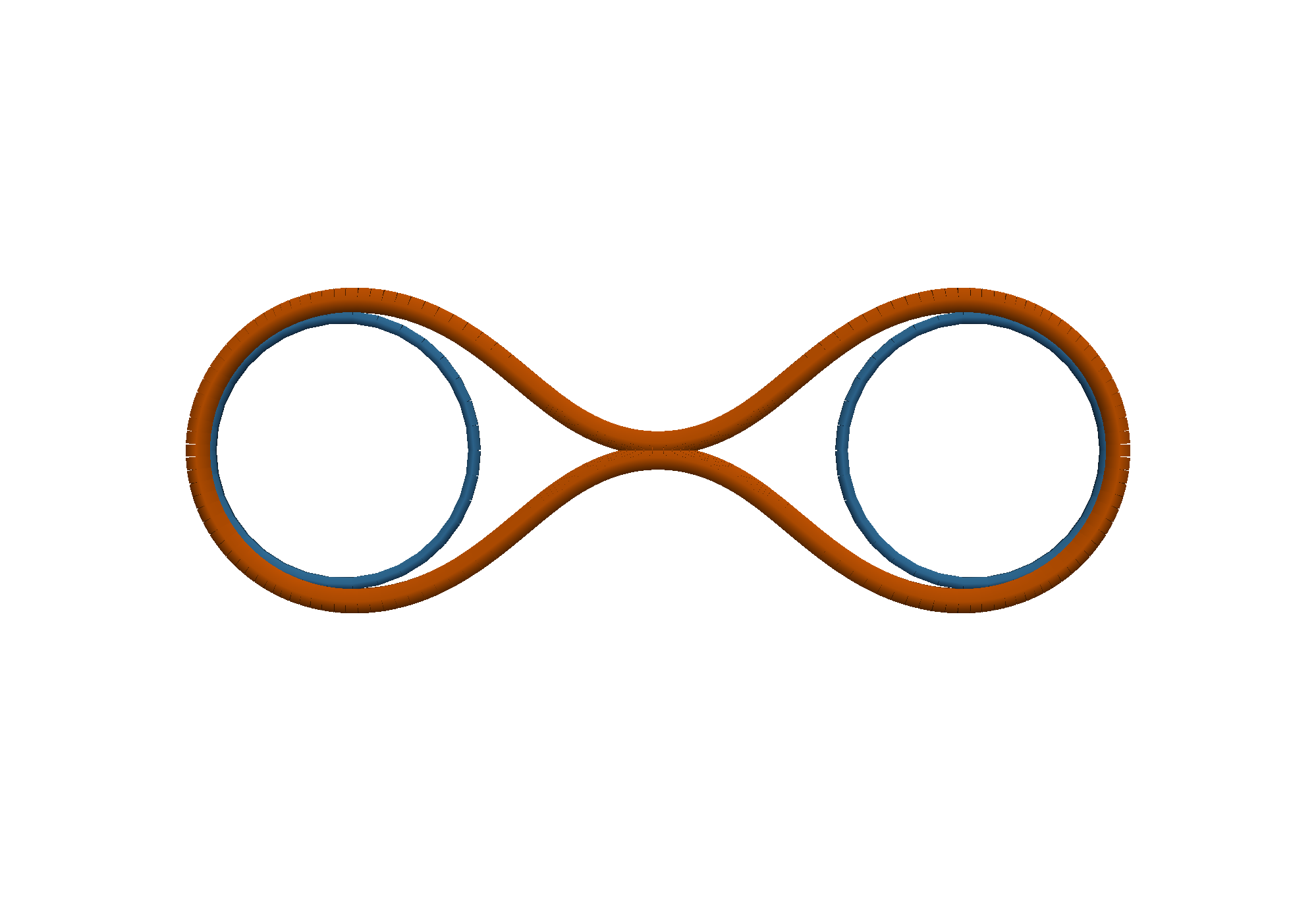}} \\
    (a) $t=414.000 M$ &
    (b) $t=416.500 M$ &
    (c) $t=417.500 M$ \\
      {\includegraphics[width=0.33\columnwidth]{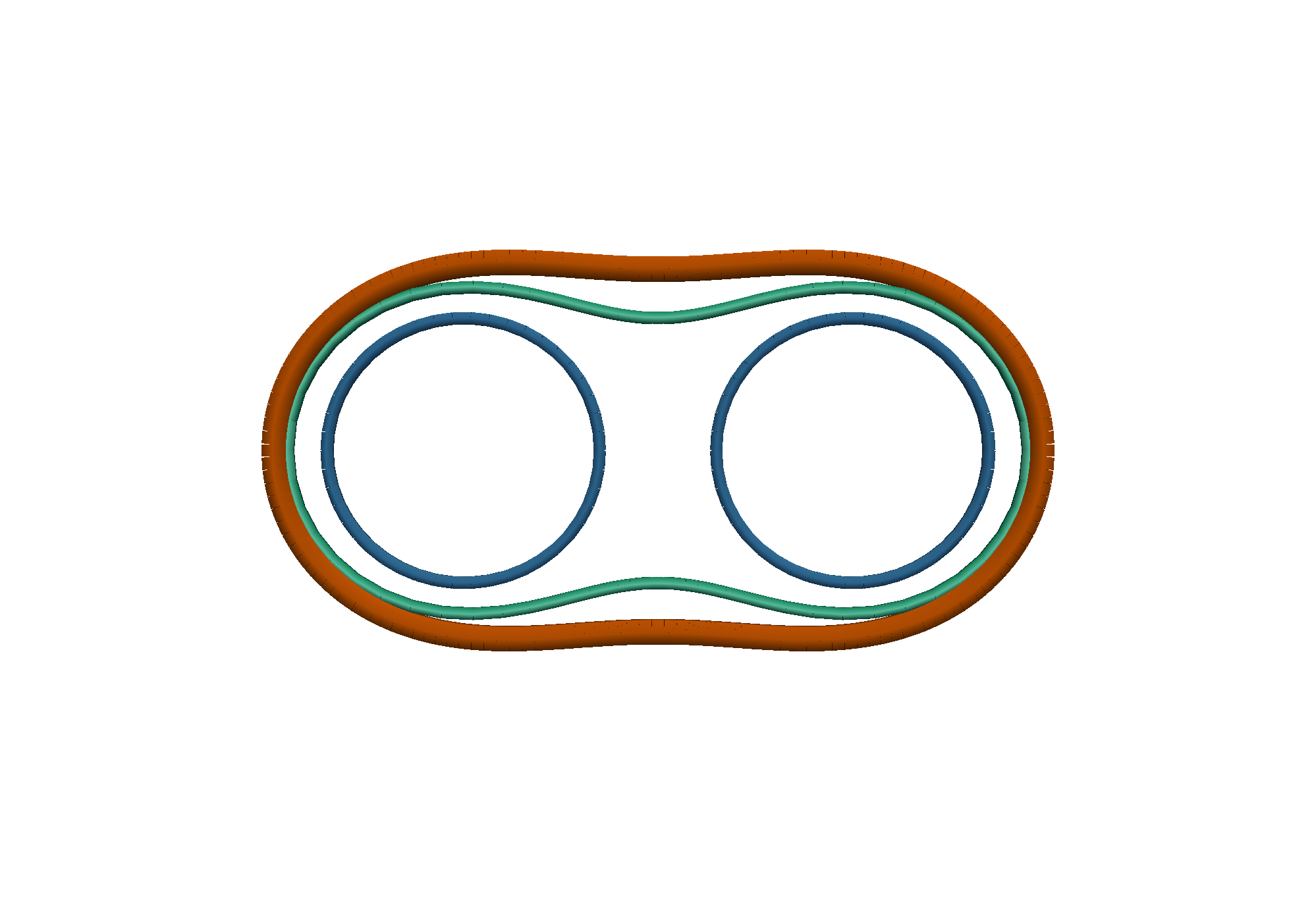}} &
      {\includegraphics[width=0.33\columnwidth]{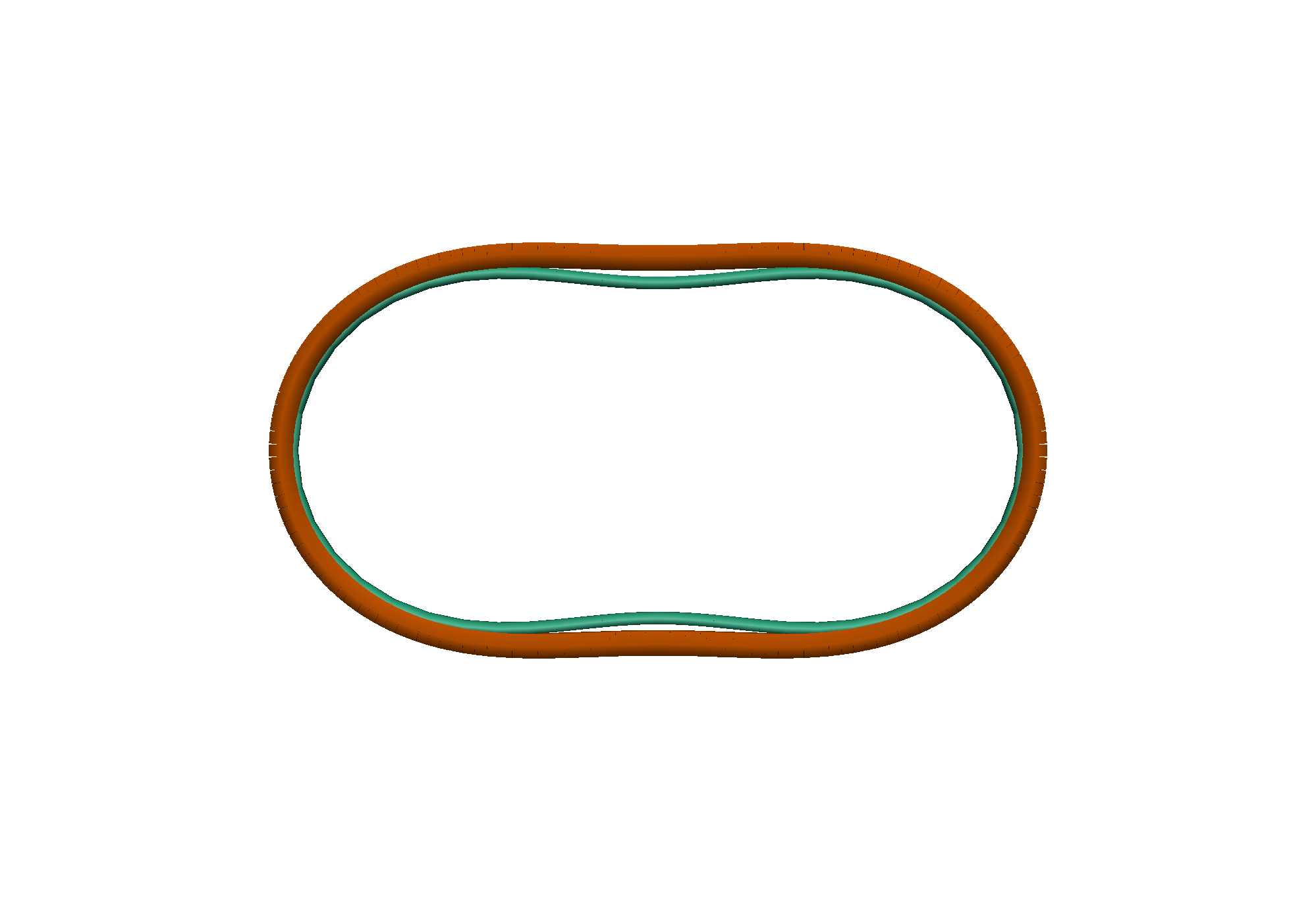}} &
      {\includegraphics[width=0.33\columnwidth]{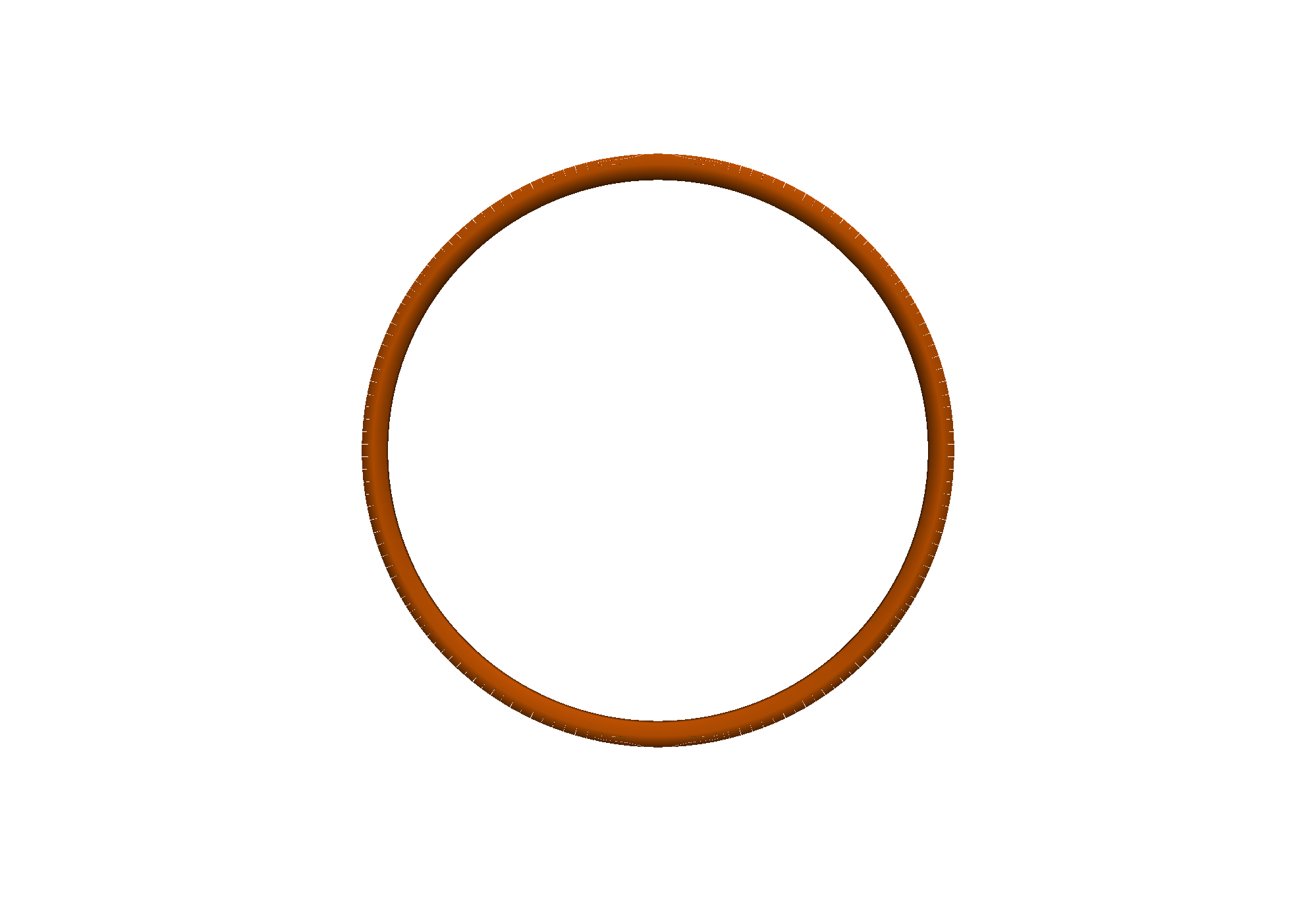}} \\
    (d) $t=420.003 M$ &
    (e) $t=420.266 M$ &
    (f) $t=470.639 M$ \\
  \end{tabular}
  \caption[Cross-sections of horizons for a head-on BBH merger]{
    Cross-sections through apparent horizons and the locus of event horizon
    generators
    for a head-on BBH merger.
    Shown in translucent purple are future generators of the horizon that
    continuously merge onto the event horizon, shown in orange, until the
    merger in panel~(c).
    Shown as blue curves in panels~(a-d) are apparent horizons associated
    with the two individual black holes, and shown as a green curve in
    panels~(d-f) is a common apparent horizon.
  }
  \label{figThorneSuggestion}
\end{figure}

To make the discussion concrete, consider a head-on equal mass binary black
hole merger, shown in \cref{figThorneSuggestion}.
We see a spatial cross-section of
apparent horizon surfaces shown blue or green, event horizon surfaces
shown in orange, and the future
generator surface shown in translucent purple.
In panel~(a), sufficiently long before the merger, the event horizon
surfaces lie almost on top of the blue apparent horizon
surfaces, which are hardly visible at this time.
The future generator surface is comprised of future generators that will
join onto the event horizon surface in the future.
When rotating this panel about the rotational axis of symmetry, the union
of the event horizon surfaces and future generator surface forms a smooth
$\mathcal{S}^2$.
In panel~(b), shortly before the merger, the future generator surface
is shrinking because some of the future generators have joined the event horizon
between this time and the time of the previous panel.
We can see the difference between the AH and EH
surfaces increases as we get closer to the merger.
There are no more future generators in panel~(c) since they have all
joined the event horizon surface, and therefore the event horizon surface
must be $\mathcal{S}^2$~\cite{Browdy1995}.

In panel~(d), a common apparent horizon shown in green has formed around the
two interior apparent horizons, and all three apparent horizons lie entirely
on or within the event horizon, as they should.
As time progresses to panels~(e)~and~(f), we stop tracking the blue inner
apparent horizons, the event horizon settles to a stationary state,
and the common apparent horizon in green approaches the event horizon until
the two surfaces eventually coincide.
With this picture in mind, the method is to evolve generators backwards in time
from panel~(f) toward panel~(a) which traces out the union of the
event horizon surface with the future generator surface.
Backwards in time, some generators ``leave'' the
event horizon surface as seen in panels~(b)~and~(a), so we must
be able to identify which generators leave the surface and when they leave.

One of the shortcomings of our previous event horizon finder was the lack of
flexibility to refine the distribution of event horizon
generators in certain regions of interest.
In the companion paper, we present
a new method of distributing and maintaining a set of
event horizon generators to address these issues.
In particular, we now have the ability to study in much greater detail
the region
where future generators join the event horizon surface.

\begin{figure}
  \centering
  \includegraphics[width=0.5\textwidth]{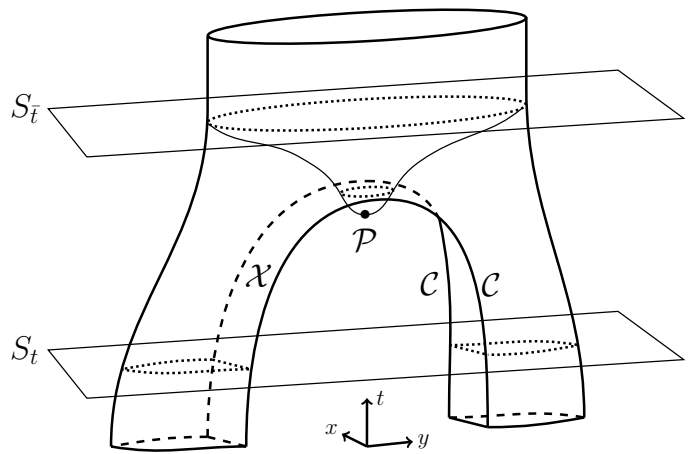}
  \caption[A ``pair of pants'' representation of merging black hole event
    horizons]{
    A $2+1$-dimensional ``pair of pants'' representation of slices of
    constant time
    $S_t$ and $S_{\bar{t}}$ through a BBH event horizon.
    The hypersurface $S_t$ is a slice of constant $t$ when
    the event horizon topology is two spheres, such as
    panel~(a) or panel~(b) of \cref{figThorneSuggestion}.
    $\mathcal{X}$ represents the spatial hypersurface of crossover points,
    which is surrounded on both sides by lines of caustics denoted by
    $\mathcal{C}$.
    The event horizon is toroidal on the spatial hypersurface
    $S_{\bar{t}}$, a slice of constant $\bar{t}$;
    the center of the hole in the torus is $\mathcal{P}$.
  }
  \label{figPairOfPants}
\end{figure}

In \cref{figPairOfPants}, we show a $2+1$ dimensional representation
of a BBH event horizon through merger.
The slice $S_t$ is a constant $t$ slice through the event horizon
at a time when the topology is two spheres, similar to
panels~(a)~and~(b) of \cref{figThorneSuggestion}.
At this time, event horizon generators are joining the event horizon
through, in general, both crossover points and caustics.
Connecting the crossover points together forms a spacelike hypersurface denoted
as $\mathcal{X}$, and connecting the caustic points forms
spacelike
hypersurfaces denoted as $\mathcal{C}$ that form
the boundary of the crossover
region.
Considering slices of constant $t$ in this example, the event horizon
topology is never toroidal.
However, a different spacelike slice $S_{\bar{t}}$ could dip through
$\mathcal{X}$ to form a toroidal event horizon with $\mathcal{P}$
a point in the middle of the hole.
In essence, we are looking for a slice where generators in the crossover
region are delayed near merger, similar to $S_{\bar{t}}$.

To accomplish this delay, we use a coordinate transformation of
the form
\begin{subequations}
\begin{align}
  \bar{x}^{i} &= x^i \\ 
  \bar{t} &= t + G(x^j, t),
\label{eqnCoordTransformation}
\end{align}
\end{subequations}
where $\bar{t}$ and $\bar{x}^{i}$ are the coordinates after the
transformation and $G(x^j, t)$ is some smooth function of position and time.
Equivalently, $t = \bar{t} - G(x^j, t)$, such that a slice of constant
$\bar{t}$ is associated with a smaller $t$ value where $G(x^j, t)$ is larger.
Therefore, the value of $G(x^j, t)$ controls how delayed generators
at $(x^j, t)$ are in the constant $\bar{t}$ slicing.
An example of an event horizon on a constant $\bar{t}$ slice
is shown in \cref{figLIGOTorus}, where the surface is colored
by the associated $t$ value and generators near
the hole in the event horizon correspond to earlier $t$ values.

The transformation has the Jacobian matrix
\begingroup
\renewcommand*{\arraystretch}{2}
\renewcommand*{\arraycolsep}{6pt}
\begin{equation}
  \label{eqnJacobian}
  \mathbf{J} = \\
    \frac{\partial{(\bar{t}, \bar{x}^{i})}}{\partial{(t, x^j)}} = \\
\begin{bmatrix} 1 + \partial_t{G} & \partial_{j}{G} \\
  0 & \delta\indices{^i_j} \end{bmatrix}.
\end{equation}
\endgroup
The normal to surfaces of constant $\bar{t}$ is given by
\begin{equation}
  \bar{n}_{\mu} = -\bar{\alpha} \nabla_{\mu} \bar{t},
\end{equation}
where $\bar{\alpha}$ is the lapse in the barred coordinates.
We can solve for $\bar{\alpha}$ from the normalization of the normal,
$\vec{n}\cdot\vec{n}=-1$, giving
\begin{equation}
\label{eqnNewLapse}
  \bar{\alpha}^2 = \frac{\alpha^2}{
    \left(1 + \partial_t{G} - \beta^k \partial_k G\right)^2 \\
  - \alpha^2 \gamma^{ij} (\partial_i G) (\partial_j G)},
\end{equation}
where $\beta^k$ is the shift vector and $\gamma^{ij}$ is the three metric.
The denominator of \cref{eqnNewLapse} must be greater than zero to obtain a
foliation of the spacetime with spacelike hypersurfaces, since
we know $\alpha^2$ is greater than zero.

\input{\tikzprefix/TikzSetup.tex}

\begin{figure}
  \centering
  \includegraphics[width=0.45\textwidth]{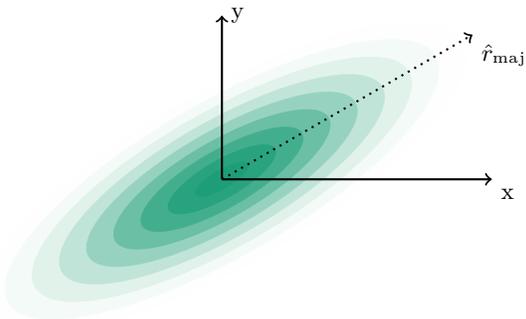}
  \caption[Gaussian function representation used for reslicing event horizons]{
    Representation of two spatial dimensions of the Gaussian function
    $G(t, x^j)$ from \cref{eqnCoordTransformation},
    where darker colored regions represent larger values of
    $G(t, x^j)$.
    $\hat{r}_{\rm{maj}}$ is an input parameter that specifies the major
    axis direction of the Gaussian.
    All directions in the plane perpendicular to $\hat{r}_{\rm{maj}}$ are
    treated equally.
  }
  \label{figGaussianEllipse}
\end{figure}

For the function $G(t, x^j)$, we choose a three-dimensional ellipsoidal
Gaussian,
with one dimension in time, one along a specified major axis, and the other
in the minor plane perpendicular to the major axis.
This gives $10$ free parameters to be specified: the amplitude ($A$),
the time center and time width
($t_0$ and $\sigma_{t}$), the spatial center ($\vec{r}_0$),
the major axis direction ($\hat{r}_{\rm{maj}}$), and the major and minor
widths ($\sigma_{\rm{maj}}$ and $\sigma_{\rm{min}}$).
A two-dimensional example is shown in \cref{figGaussianEllipse}, where
the time dimension has been omitted, and the plane perpendicular to
$\hat{r}_{\rm{maj}}$ has been projected down
into one dimension.
The function $G(t, x^j)$ has the form
\begin{equation}
\begin{split}
  G(&t, x^j) = \\
            & A \exp{\left[-\left(t-t_0\right)^2 / \left(2 \sigma_t^2\right)\right]} \\
            & \times \exp{\left[-[\hat{r}_{\rm{maj}} \cdot \left(\vec{x} - \vec{r}_0\right)]^2 /
              \left(2 \sigma_{\rm{maj}}^2\right)\right]} \\
            & \times \exp{\left[-\left(\left(\vec{x}-\vec{r}_0\right)^2 - [\hat{r}_{\rm{maj}} \cdot
              \left(\vec{x} - \vec{r}_0\right)]^2\right)/\left(2 \sigma_{\rm{min}}^2\right)\right]},
\end{split}
\end{equation}
where the first exponential localizes the Gaussian to the time of merger, the
second preferentially modifies geodesics along some major axis, and the third
limits the range in the plane perpendicular to the major axis.
The major axis is chosen in the thinnest direction of the small neck connecting
the two black holes just after merger, which we will analyze in
\cref{secEMI}.
This choice produces time slices that cut through the spacelike crossover
surface
arising during the merger, as illustrated in
\cref{figLIGOTorus,figPairOfPants}.
After finding Gaussian parameters that yield a toroidal event horizon on
at least one constant $\bar{t}$ slice,
it is sufficient to verify that the new lapse is positive and real using
\cref{eqnNewLapse}.

To reslice the event horizon in practice,
we first trace a set of generators to locate the EH in
the generalized harmonic coordinate system used to merge the binary in \SpEC{},
as detailed in the companion paper~\cite{BohnMethods2016}.
During the generator evolution, we record the generator locations
at a set of times that are finely spaced as the event horizons merge and
coarsely spaced after the merger.
Using \cref{eqnCoordTransformation}, we then calculate $\bar{t}$ for each
generator at each of these times.
We want the locations of the generators on constant $\bar{t}$ slices, and
we accomplish this with a $3$\ts{rd}-order Lagrange interpolation polynomial
in $\bar{t}$.
The spacetime location where an EH generator joins the horizon is a spacetime
event, so we simply apply the coordinate transformation to determine when
the generator joins the horizon in the barred coordinate system.

\section{Discussion}
\label{secTori}

Previous studies of merging event horizons
infer the possibility of a toroidal event horizon by
studying the distribution of caustics and crossover points during the
merger.
As discussed in \cref{secReslicing},
the set of crossover points is known to live on a spacelike hypersurface
that becomes asymptotically null as the surface approaches a set of
caustics~\cite{Shapiro1995}.
There should therefore exist a spacelike foliation of the spacetime
that cuts a hole out of the spacelike surface of crossover points, resulting
in a short-lived toroidal event horizon.
In this section, we are interested in explicitly finding such a reslicing where
the event horizon has a toroidal topology.

It is useful to first study null
hypersurfaces in flat space,
where the distribution of caustics and crossover points is known
analytically.
We will use these wavefronts as model horizons and refer to them as
``horizons'' for convenience
in the spherical model in \cref{secSphericalModel} and in the ellipsoidal
model in \cref{secEllipsoidalModel}.
These models were introduced in
Shapiro~\textit{et al.}~\cite{Shapiro1995} and also studied in
Siino~\cite{Siino1998a}.

All of the systems in this discussion section can be found on the
SXS collaboration website~\cite{SXSWebsite} at the
page~\cite{EHTopologiesWebsite}.

\subsection{Spherical model}
\label{secSphericalModel}

\begin{figure}
  \centering
  \includegraphics[width=0.3\textwidth]{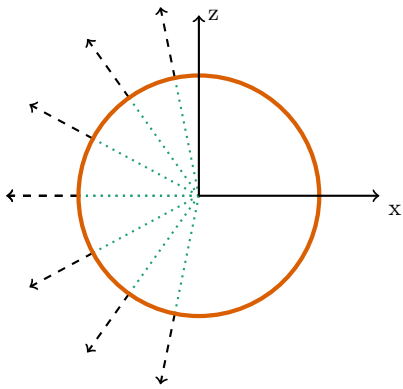}
  \caption[Initial data configuration for spherical model horizon in flat
    space]{
    Initial data configuration for the spherical model horizon
    in flat space.
    The orange circle, when rotated about the $z$-axis, forms the sphere
    used as initial data for the generator tracing.
    Along this surface, we place null geodesics normal to the surface
    as described in the companion paper~\cite{BohnMethods2016}, illustrated as
    black dashed arrows.
    The green dashed lines show where the generators came from
    earlier in coordinate time, and that the trajectories all met
    at the origin at the same time in the past.
  }
  \label{figS0InitialData}
\end{figure}

We trace generators for a spherical wavefront backwards
in time through the Minkowski spacetime until all the generators
leave the horizon through a caustic or a crossover point.
These points are identified using the same algorithms as used for
binary black hole event horizons, described in the methods
paper~\cite{BohnMethods2016}.
The initial data for this model horizon is a sphere of radius $1$
at $t=0$,
shown in
\cref{figS0InitialData}, where the $z$-axis is an axis of rotation.
Generators are placed on the sphere pointing perpendicular to the surface
outward and evolved backwards in time through flat space,
where the black dashed arrows denote some generators of the horizon
and the dotted teal lines show the corresponding generator trajectories.
The generators begin the simulation
on the surface and we search for caustics or crossover points to determine
if and when generators leave the horizon backwards in time.

Because of the symmetry of the system,
all future generators must join onto the horizon
at the same location and time through a caustic, since all the generators
meet together at the origin.
The code properly labels all of the generators as joining through caustics
and we do not find a surface of crossover points, as expected.
The lack of a crossover surface makes this model illustrative for the
head-on merger of equal mass black holes as featured
in the following section.

Since there is no crossover surface, which would form a spacelike hypersurface,
we should not expect to be able to find a slicing of the spacetime
that yields a toroidal surface,
so this provides a good test of our reslicing algorithm.
Using the coordinate transformation in \cref{secReslicing} with
a flat metric and $\sigma_t$ set large enough to keep the transformation
independent of
time, the new lapse from \cref{eqnNewLapse} simplifies to
\begin{equation}
  \bar{\alpha}^2 = \frac{1}{1 - \delta^{ij}(\partial_i G)(\partial_j G)}.
\end{equation}
We must therefore keep the spatial gradients of $G(x^j, t)$ small to maintain
a spacelike foliation.
However, we know that any coordinate transformation will
preserve events.
In particular, the caustic event where all the generators meet at the origin
of the coordinate system will be preserved, meaning all the generators
will join the horizon at the same time in all foliations of the spacetime.

\begin{figure}
  \centering
  \begin{tabular}{ccc}
    \includegraphics[trim={20cm 0 20cm 0},clip,
      width=0.31\columnwidth] {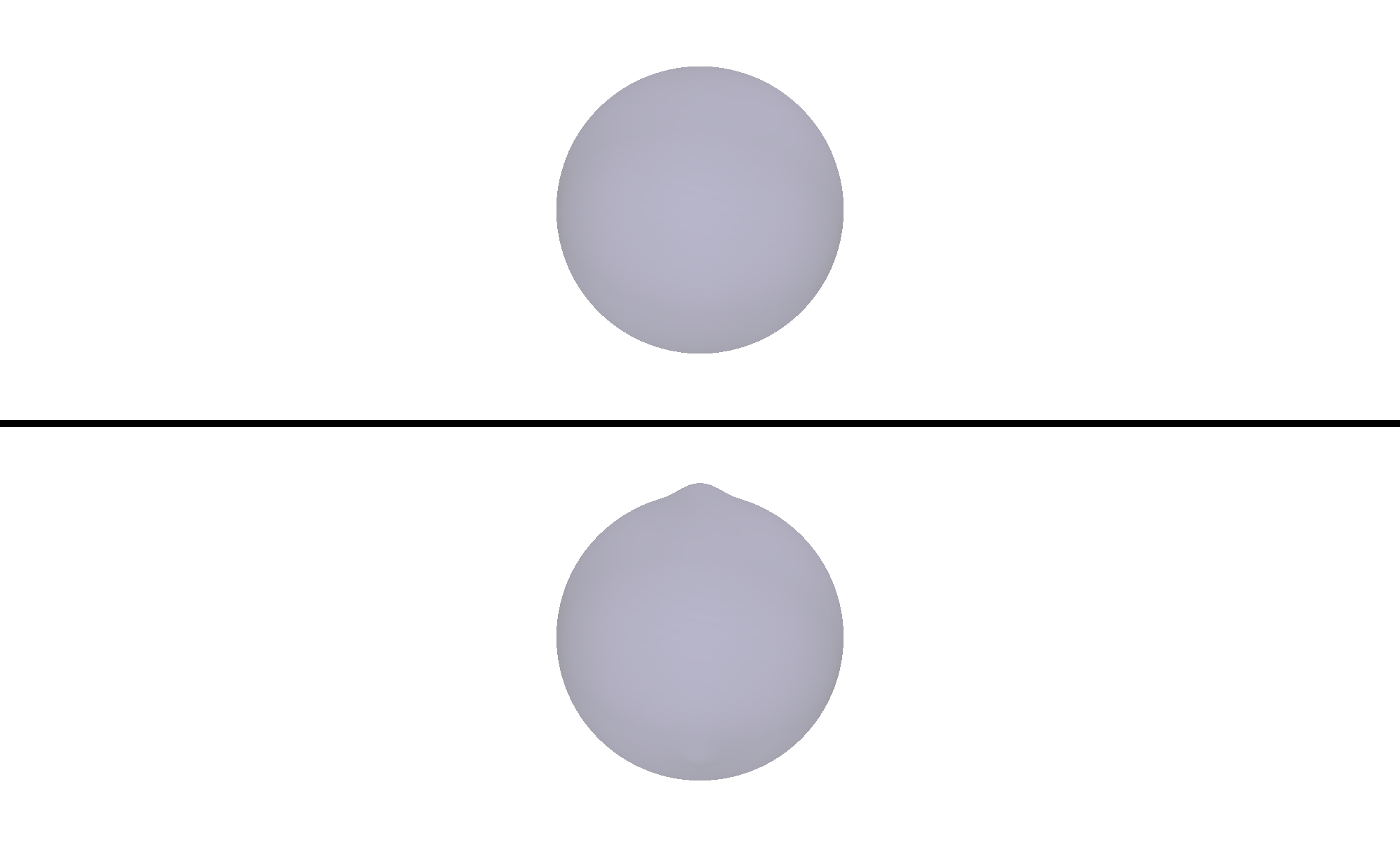} &
    \includegraphics[trim={20cm 0 20cm 0},clip,
      width=0.31\columnwidth]{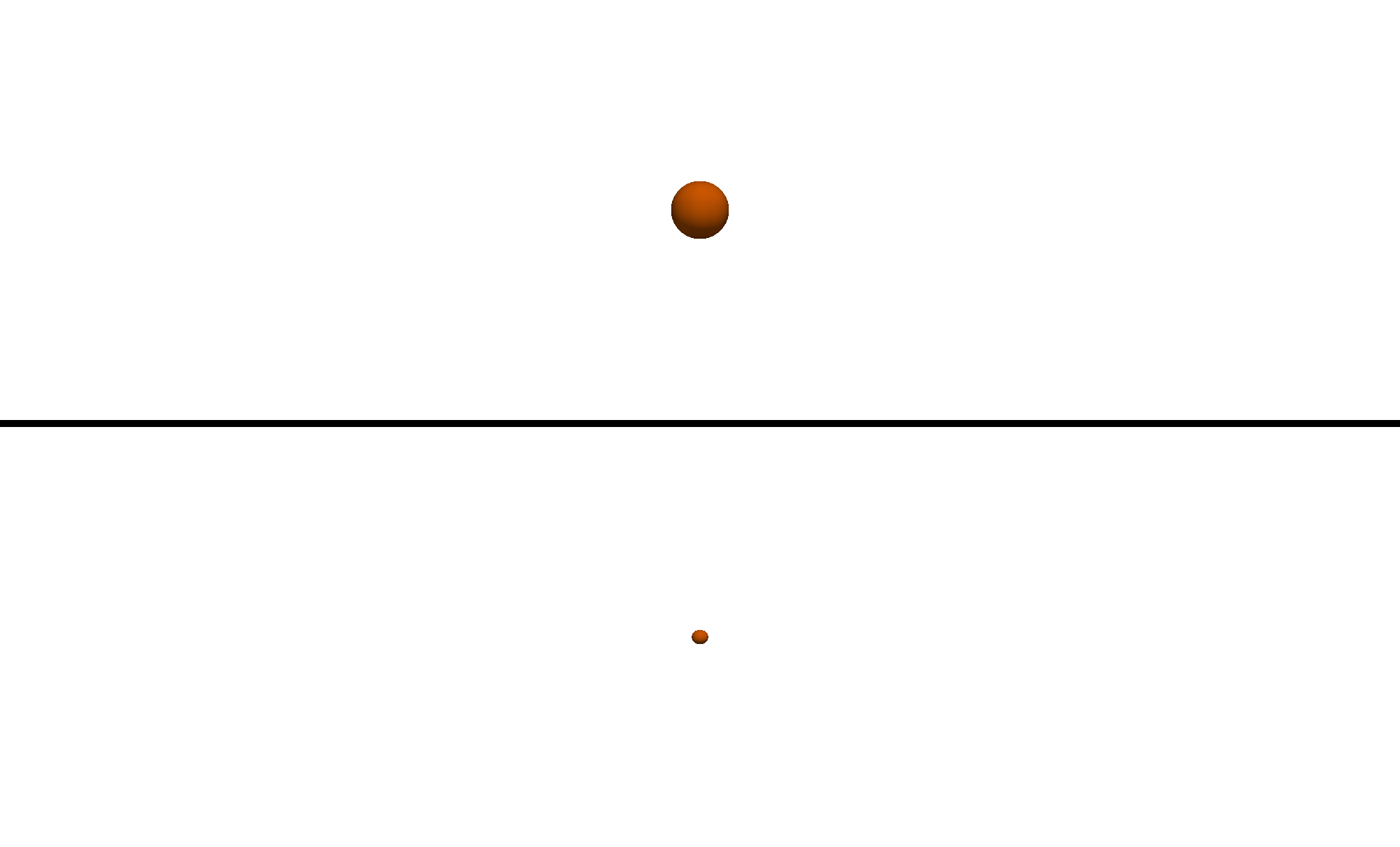} &
    \includegraphics[trim={20cm 0 20cm 0},clip,
      width=0.31\columnwidth]{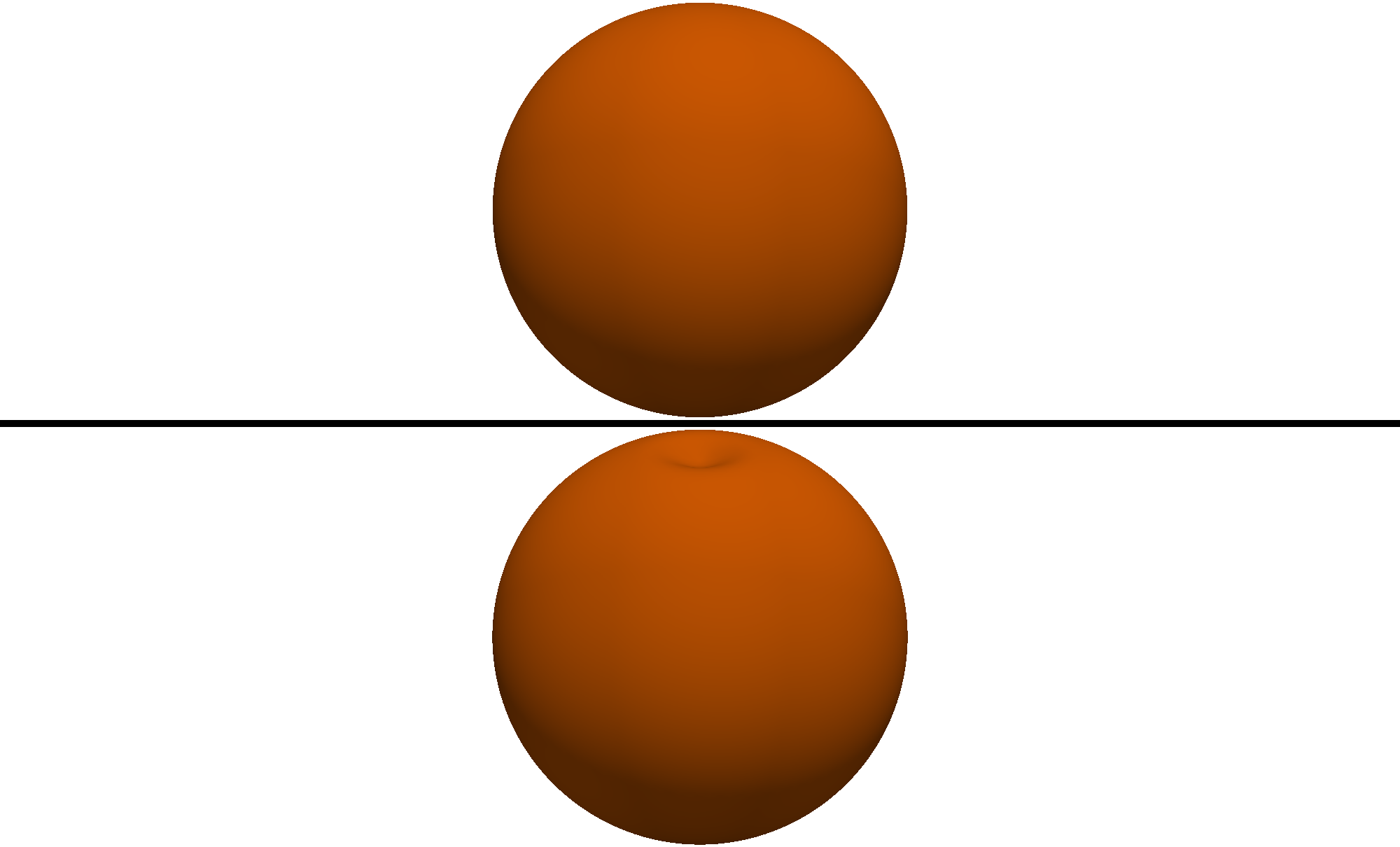} \\
    (a) $-1.351 M$ &
    (b) $-0.935 M$ &
    (c) $-0.550 M$
  \end{tabular}
  \caption[Generator surface for spherical horizon model]{
    Generator surface for the spherical horizon model in
    \cref{secSphericalModel}, shown in two different coordinate systems.
    The top row shows a slice of constant $t$ coordinate, which is the original
    coordinate system of the spherical model,
    and the bottom row shows a slice of constant $\bar{t}$ coordinate
    after using the
    transformation in \cref{eqnCoordTransformation}.
    Regions of the surface colored in translucent purple denote areas
    of future generators that are not currently part of the
    horizon surface,
    and orange denotes areas where generators are on the horizon surface.
    $M$ is the unit of time in this coordinate system, where the speed of
    light is $1$.
  }
  \label{figToyS0Surface}
\end{figure}

\Cref{figToyS0Surface} shows this surface in two
foliations of the spacetime,
where the top row shows the original slicing with spherical initial data,
and the bottom row shows the resliced horizon.
It is important to reiterate that we will show horizons going forwards
in time from left to right, but the generator evolution is performed
backwards in time from right to left in these figures.
Therefore, the initial data for the spherical model is in the
top row of the rightmost panel.
The bottom row is an attempt at a coordinate transformation into a new slicing
of the spacetime to look for a torus.

Going along the top row from left to right, all the generators are
initially
future generators of the horizon,
as indicated by the translucent
purple color.
As coordinate time progresses forward, all the generators meet
together at a single point in time just before panel~(b),
where they join the
surface through a caustic at the origin.
The surface continues to expand linearly through panel~(c)
until reaching the unit sphere.

\begin{figure}
  \centering
  \begin{tabular}{cccc}
    \includegraphics[trim={20cm 0 20cm 0},clip,
      width=0.24\columnwidth]{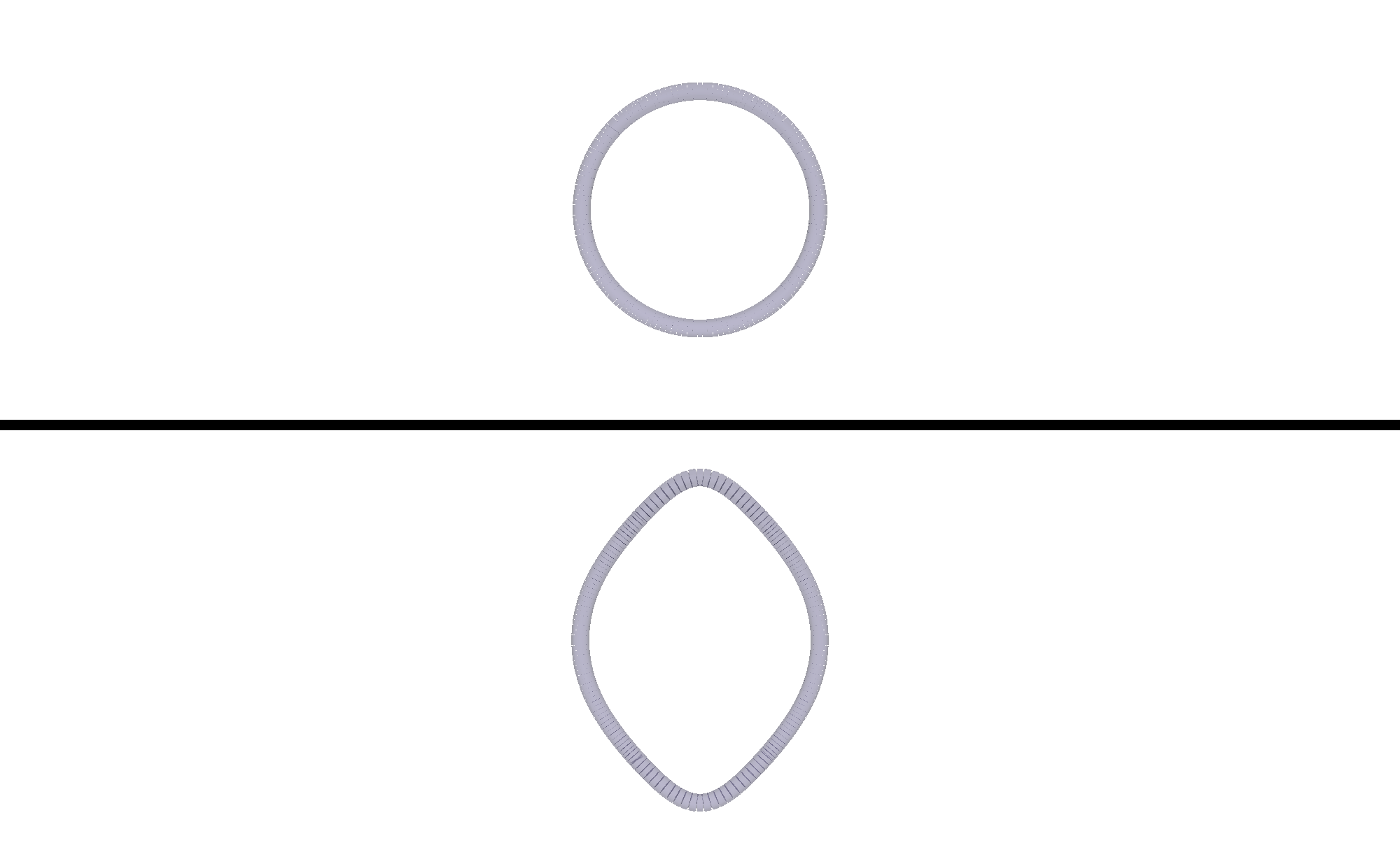} &
    \includegraphics[trim={20cm 0 20cm 0},clip,
      width=0.24\columnwidth]{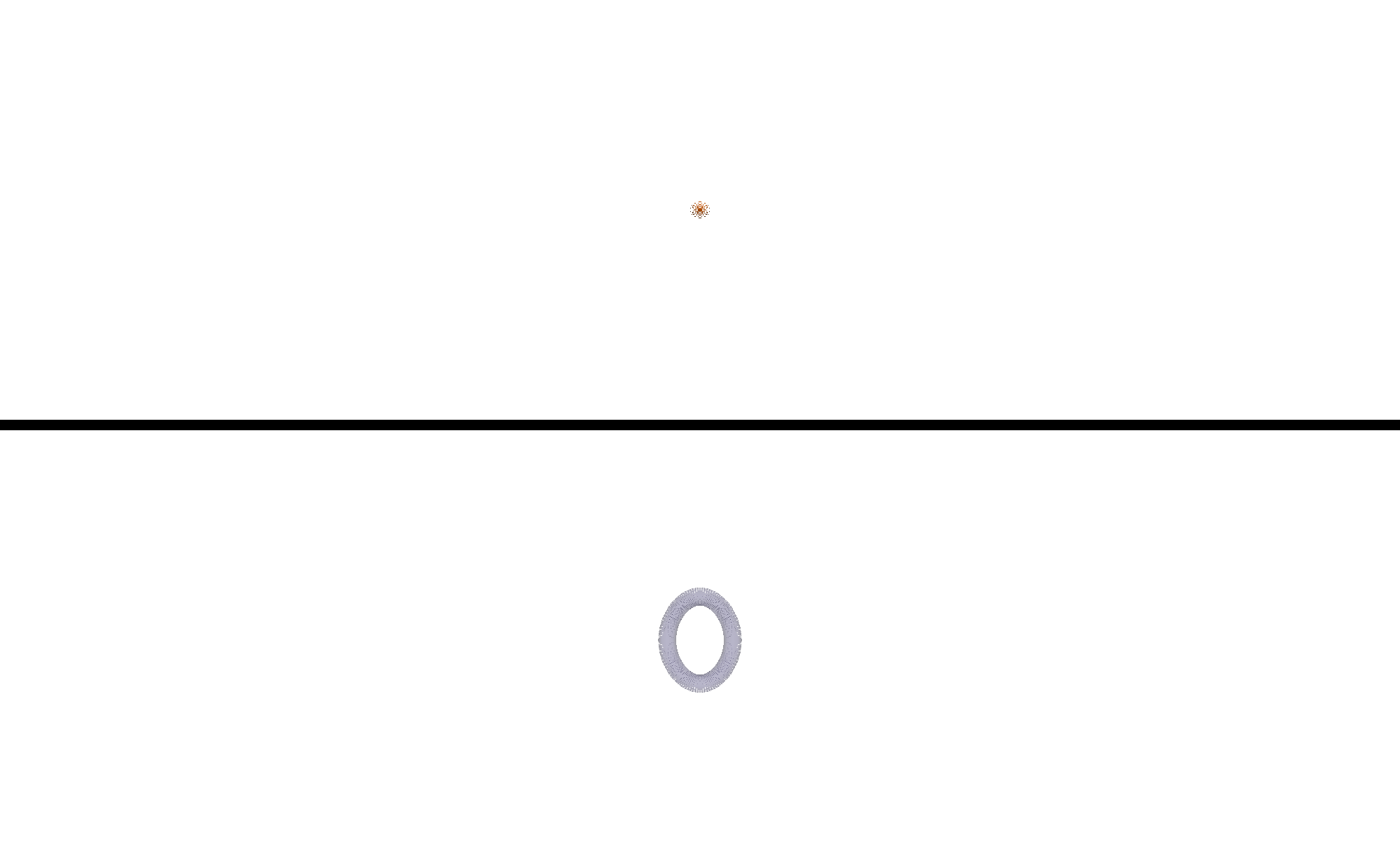} &
    \includegraphics[trim={20cm 0 20cm 0},clip,
      width=0.24\columnwidth]{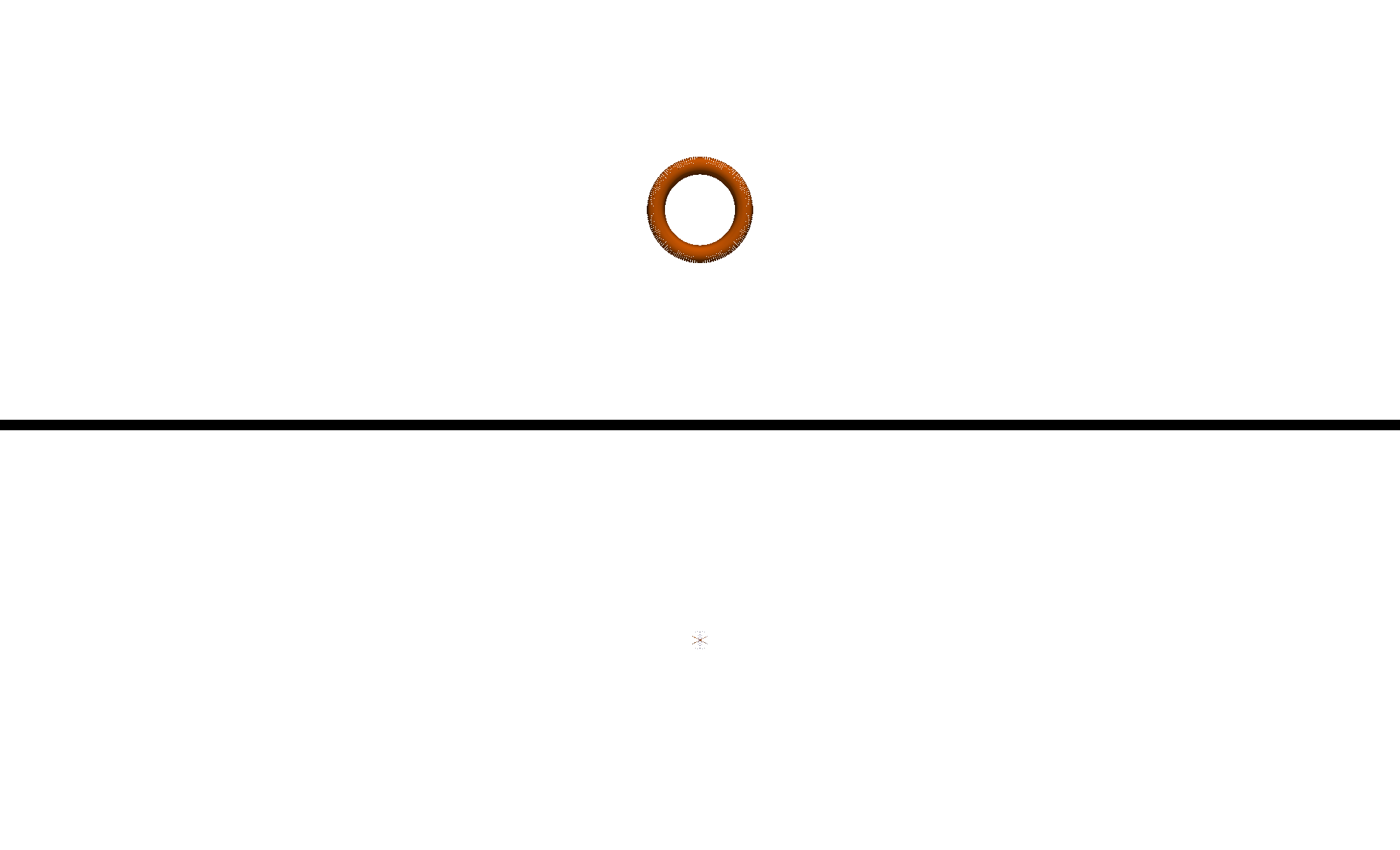} &
    \includegraphics[trim={20cm 0 20cm 0},clip,
      width=0.24\columnwidth]{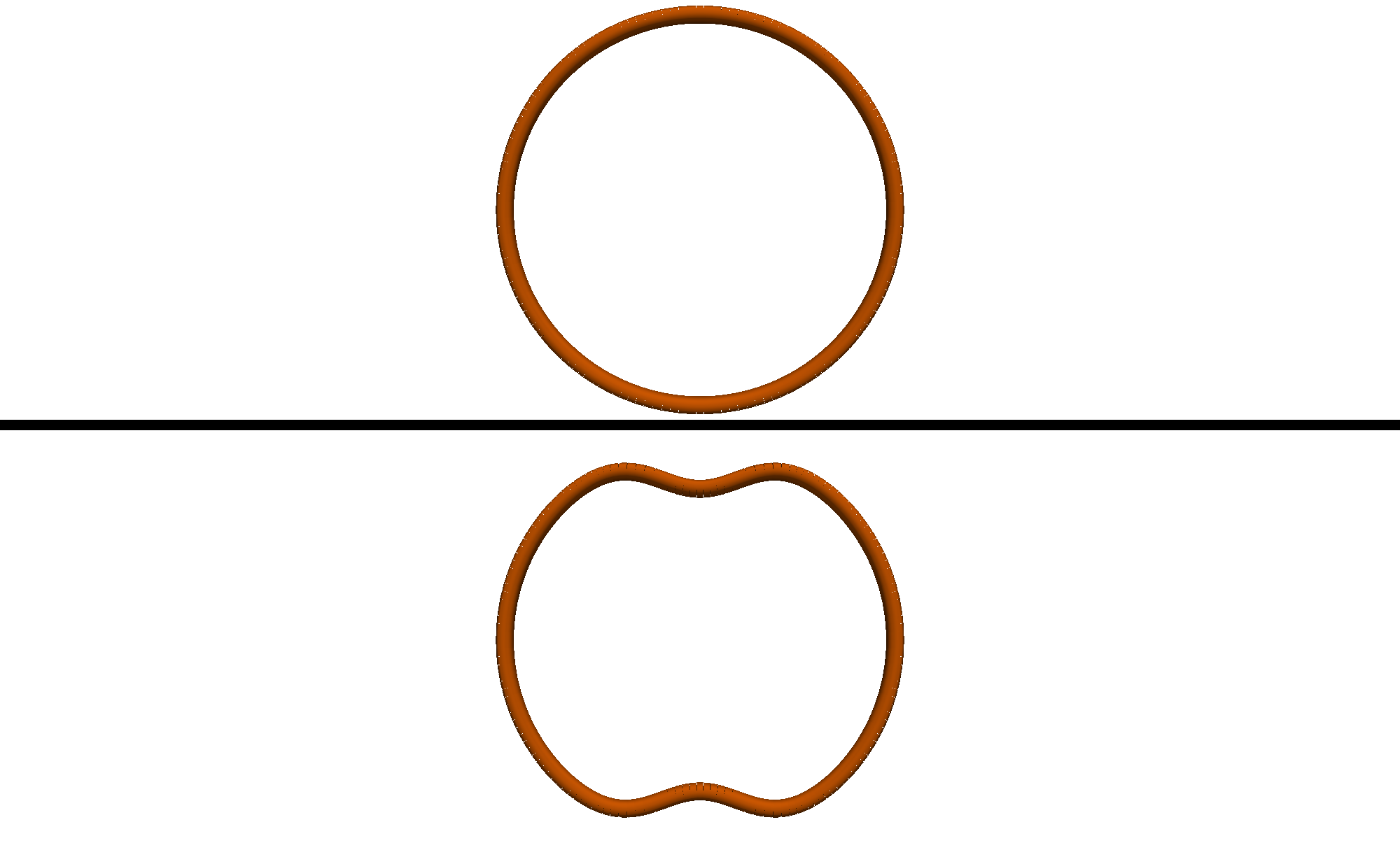} \\
    (a) $-1.065 M$ &
    (b) $-0.999 M$ &
    (c) $-0.952 M$ &
    (d) $-0.873 M$
  \end{tabular}
  \caption[Spatial cuts through spherical model horizon]{
    Zoomed in slices of the spherical horizon in flat space, covering a
    small duration of time near panel~(b) in \cref{figToyS0Surface}.
    The full surface is generated by rotating these slices around the vertical
    direction of the figure.
    The color scheme and coordinate systems are the same as
    in \cref{figToyS0Surface}.
  }
  \label{figToyS0Slice}
\end{figure}

The bottom row paints a very similar picture, where
we have applied the coordinate transformation in
\cref{eqnCoordTransformation} to search for a toroidal
topology.
We tried a variety of parameters with similar results,
but show the values from Case A of \cref{figResliceParameters} for
these figures.
Just as in the original slicing, all the generators join at the same time
through a caustic just before panel~(b).
The coordinate transformation changes the shape of the horizon, but leaves
the topology unaffected.

\begin{table*}
  \setlength{\tabcolsep}{6pt}
  \renewcommand{\arraystretch}{1.5}
  \centering
  \begin{tabular}{ c c c c c c c c }
    \hline
    \hline
    Case & $A$ & $\vec{r}_0$ & $t_0$ & $\sigma_t$ & $\hat{r}_{\rm{maj}}$ & $\sigma_{\rm{maj}}$ & $\sigma_{\rm{min}}$ \\
    \hline
    A & $5\times10^{-2} M$ & $\vec{0}$ & $0$ & $\infty$ & $\hat{z}$ & $1 M$ & $5\times10^{-2} M$ \\
    B & $3\times10^{-2} M$ & $\vec{0}$ & $417.424 M$ & $1 M$ & $\hat{z}$ & $1 M$ & $2\times10^{-2} M$ \\
    C & $5\times10^{-2} M$ & $\vec{0}$ & $7540.018 M$ & $3 M$ & $\frac{\sqrt{2}}{2} (-\hat{x}+\hat{y})$ & $1 M$ & $2\times10^{-2} M$ \\
  \hline
  \hline
\end{tabular}
  \caption[Sets of parameters supplied to the Gaussian coordinate
    transformation]{
    Sets of parameters supplied to the Gaussian coordinate transformation
    in \cref{eqnCoordTransformation}, used in different circumstances throughout
    this paper.
    The unit $M$ is the unit of the corresponding coordinate system, where it is
    the total mass of the black holes for BBH simulations.
  }
  \label{figResliceParameters}
\end{table*}

It is instructive to simplify horizons by taking a slice through the surface.
In \cref{figToyS0Slice}, we take a slice through the spherical model
horizon along the major axis of the Gaussian coordinate
transformation, such that a rotation of the slice produces the full surface
in both coordinate systems.
To analyze exactly how generators join the horizon, we have magnified
the spatial and temporal scales relative to \cref{figToyS0Surface}.
Note that the rows show slices of constant time in different coordinate
systems, so we do not expect events such as the joining of generators
onto the horizon to align.
In the top row, the generators join the horizon in panel~(b)
simultaneously at a single point, and similarly for the bottom row in panel~(c).

Though the surfaces appear different in the two coordinate systems,
we see clearly
that the caustic event is preserved under coordinate transformation.
Therefore the horizon of this model
instantaneously
transitions from not existing to having a spherical topology independent
of the slicing as expected.

\subsection{Equal mass head-on merger}
\label{secHOMerger}

The simplest binary black hole merger to study is the
head-on merger of equal mass non-spinning black holes.
The system we consider has black holes initially at rest centered
at $\pm 25 M \hat{y}$,
where $M$ is the total mass of the black holes.
This binary has rotational symmetry about the $y$-axis connecting
the two black holes as well as a mirror symmetry about the $xz$-plane
halfway between the black holes.
The expectation for the topology of this event horizon is
two spheres before the merger that transition
to a single sphere, with
no toroidal phase in any slicing of the
spacetime~\cite{MassoEtAl:1999, Lehner1999, Husa-Winicour:1999}.

Straightforward symmetry arguments show that the event horizon
topology must be composed of only spheres, as we now show.
In this system, the resultant black hole after the merger settles down to
a static Schwarzschild horizon since there is no angular momentum in the
system about the origin.
The initial data
for the event horizon simulation is therefore a spherically
symmetric surface.
Consider the event horizon generators at the intersection
between the EH and the $y=0$ mirror plane, forming a ring.
The generators on this ring should initially look exactly like those in the
spherical
model shown in \cref{figS0InitialData}.
These generators must remain in this plane for the entire simulation
owing to the mirror symmetry.
Furthermore,
the spacetime is axisymmetric about the $y$-axis, and so the generators
must respect this symmetry and remain in a circle in
this coordinate system.
We can see from these symmetries that the generators
in the mirror plane must
all join simultaneously through a caustic at the origin,
identical to the spherical
model horizon in \cref{secSphericalModel}.
When considering planes where $y \neq 0$,
the rotational symmetry still enforces that the intersection
of the plane and the event horizon always remains circular,
where all the generators in a circle similarly join the EH through a caustic
along the $y$-axis.
We can parameterize all the future generators into rings by where
along the $y$-axis they join the EH.
In any coordinate system, the generators in a given ring
are either all future generators
at a given time or all true generators of the EH.
Because generators never cross after joining the EH surface,
it is therefore impossible for a torus to form in any coordinate slicing of
the head-on merger.
Changing the number of $\mathcal{S}^2$ EH surfaces is however possible with
certain coordinate transformations that change the relative times when
neighboring rings join the EH, as we will see in \cref{secBabies}.

Another way to state the argument is based on the lack of a crossover surface.
The $2+1$-dimensional event horizon hypersurface is null everywhere
except for where future generators join the EH through caustics or crossover
points, where it is spacelike.
Using coordinate transformations, we can only cut a hole through the event
horizon hypersurface where it is spacelike, along the inseam
of the pair of pants in \cref{figPairOfPants}.
We already argued that there are only caustics (and so no crossover points)
in the coordinate system where the BBH system is axisymmetric,
and that coordinate transformations preserve these caustics.
The inseam of the pair of pants is thus $1$-dimensional and
composed of only caustics, and the rest of the event horizon hypersurface is
null,
therefore there is no $2$-dimensional spacelike hypersurface through which to
cut a hole in the EH.

\begin{figure*}
  \centering
  \begin{tabular}{ccc}
    \includegraphics[
      width=0.33\textwidth]{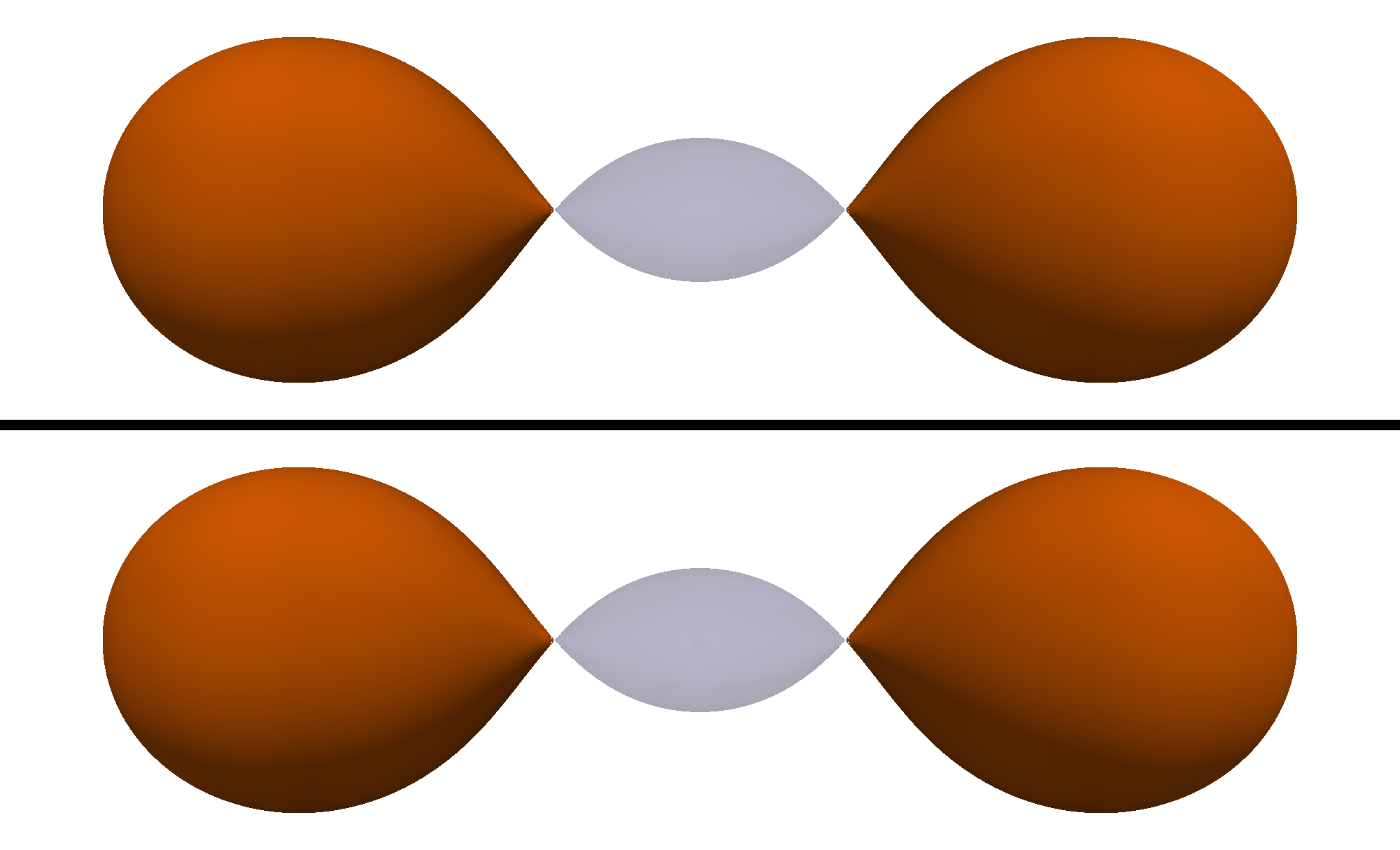} &
    \includegraphics[
      width=0.33\textwidth]{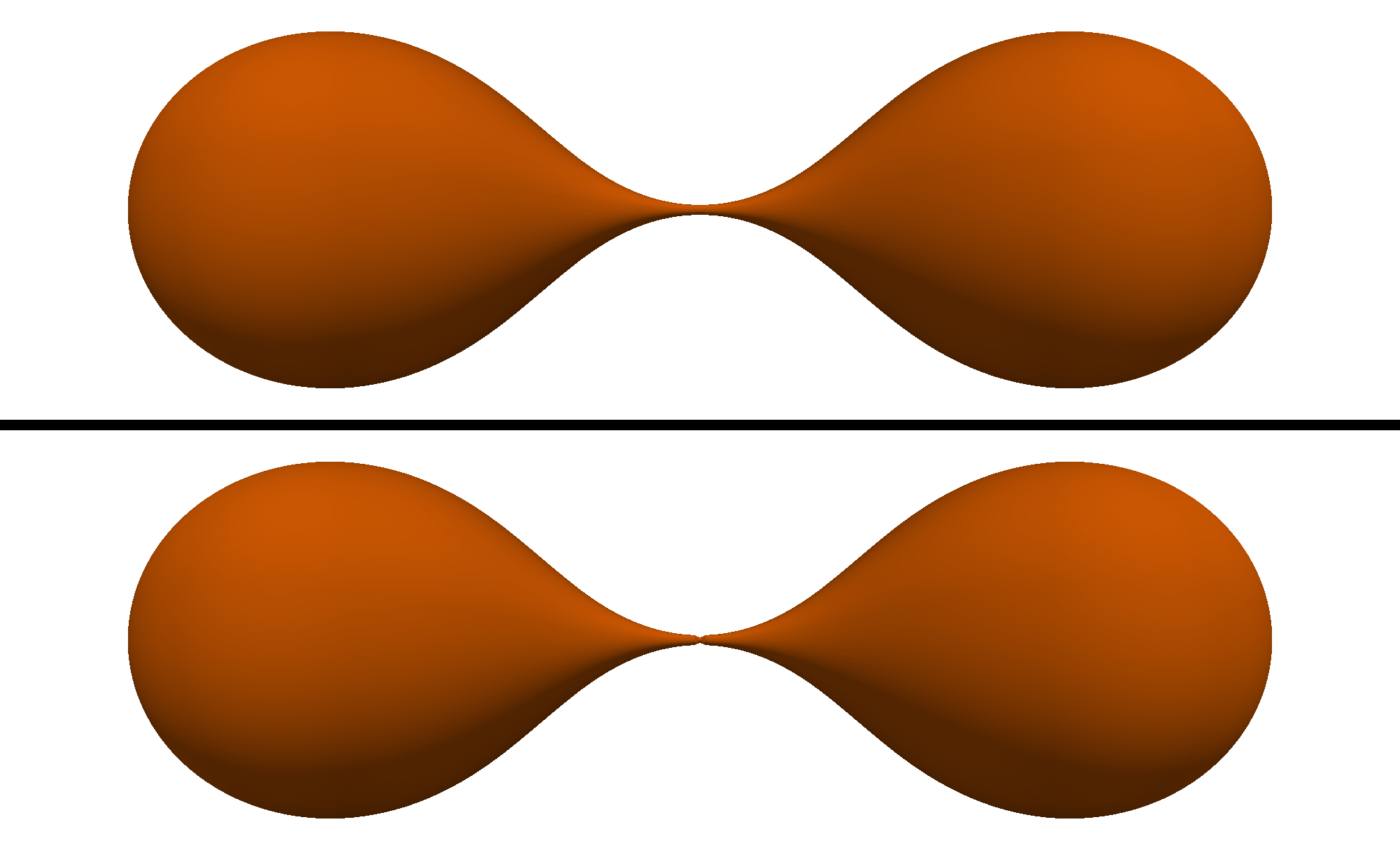} &
    \includegraphics[
      width=0.33\textwidth]{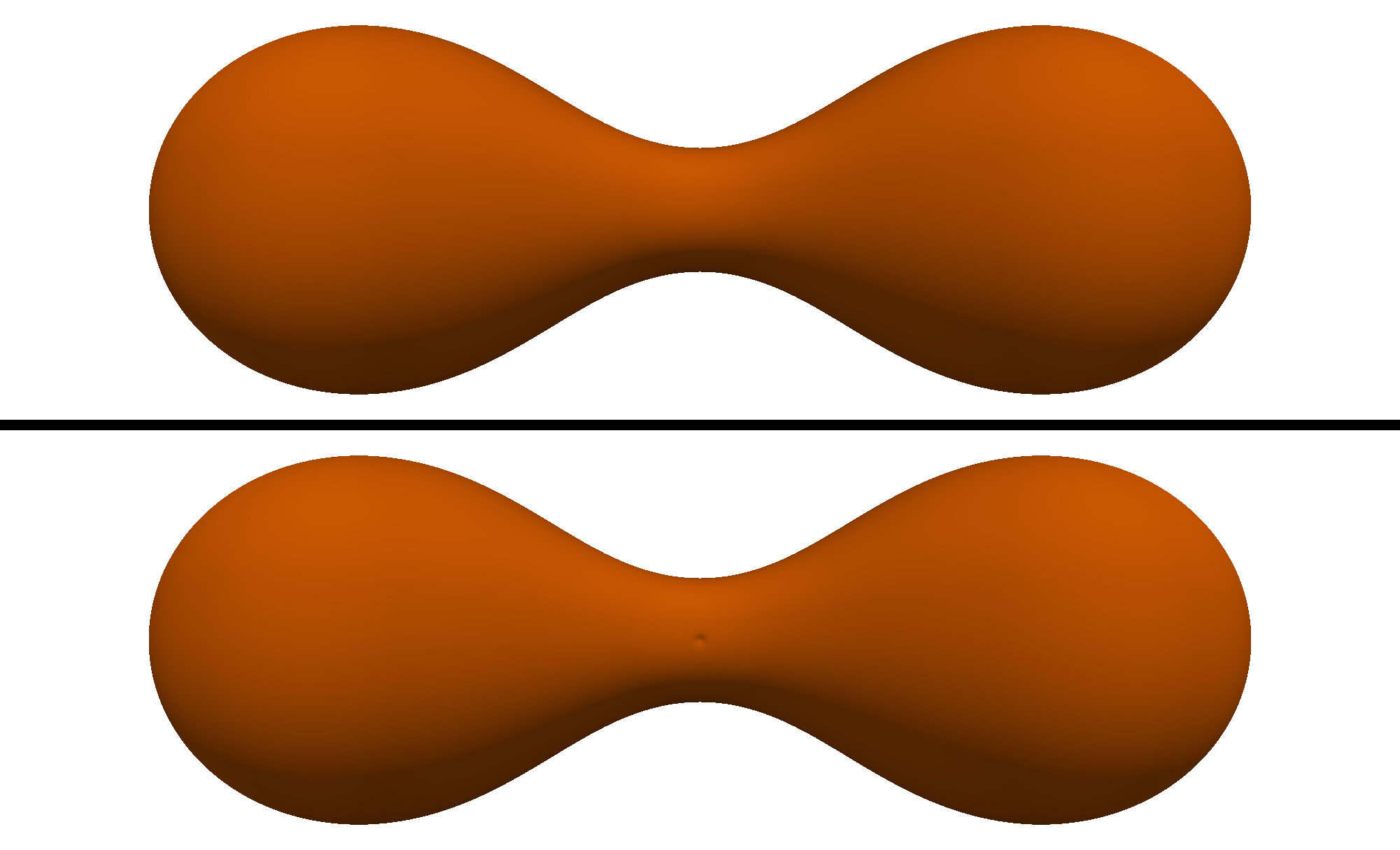} \\
    (a) $416.800 M$ &
    (b) $417.460 M$ &
    (c) $418.000 M$
  \end{tabular}
  \caption[Event horizon generator surfaces for the equal mass head-on binary]{
    Event horizon generator surfaces for the equal mass head-on binary.
    The $t$ slicing in the top row is almost identical to the $\bar{t}$ slicing
    in the bottom row, because of the small size of the Gaussian parameters
    relative to the
    horizon scale.
  }
  \label{figHOSurface}
\end{figure*}

\Cref{figHOSurface} shows the event horizon surface before, during, and
after the black hole merger.
The parameters of the coordinate transformation are
labeled Case B in \cref{figResliceParameters}.
The event horizon in these two coordinate systems looks virtually
indistinguishable
because the spatial scale of the coordinate transformation is
small compared to the scale of the figure.
Topologically,
both coordinate systems are identical.
We have one spherical surface for each event horizon
($2\times \mathcal{S}^2$) in panel~(a).
After all the future generators join the EH, the horizon transitions
into a single $\mathcal{S}^2$ shown in panel~(b) and remains that way.

\begin{figure}
  \centering
  \begin{tabular}{cccc}
    \includegraphics[trim={20cm 0 20cm 0},clip,
      width=0.24\columnwidth]{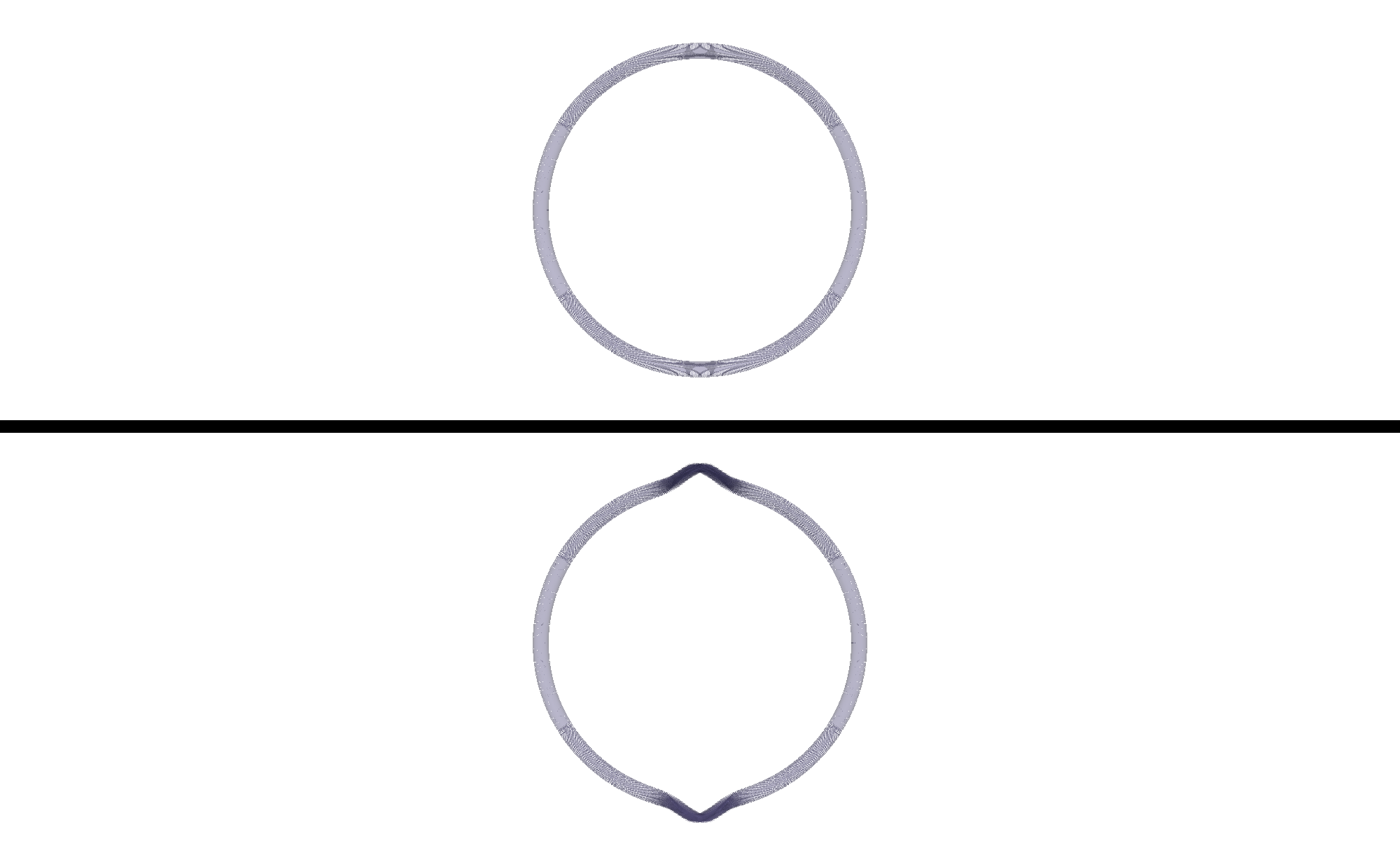} &
    \includegraphics[trim={20cm 0 20cm 0},clip,
      width=0.24\columnwidth]{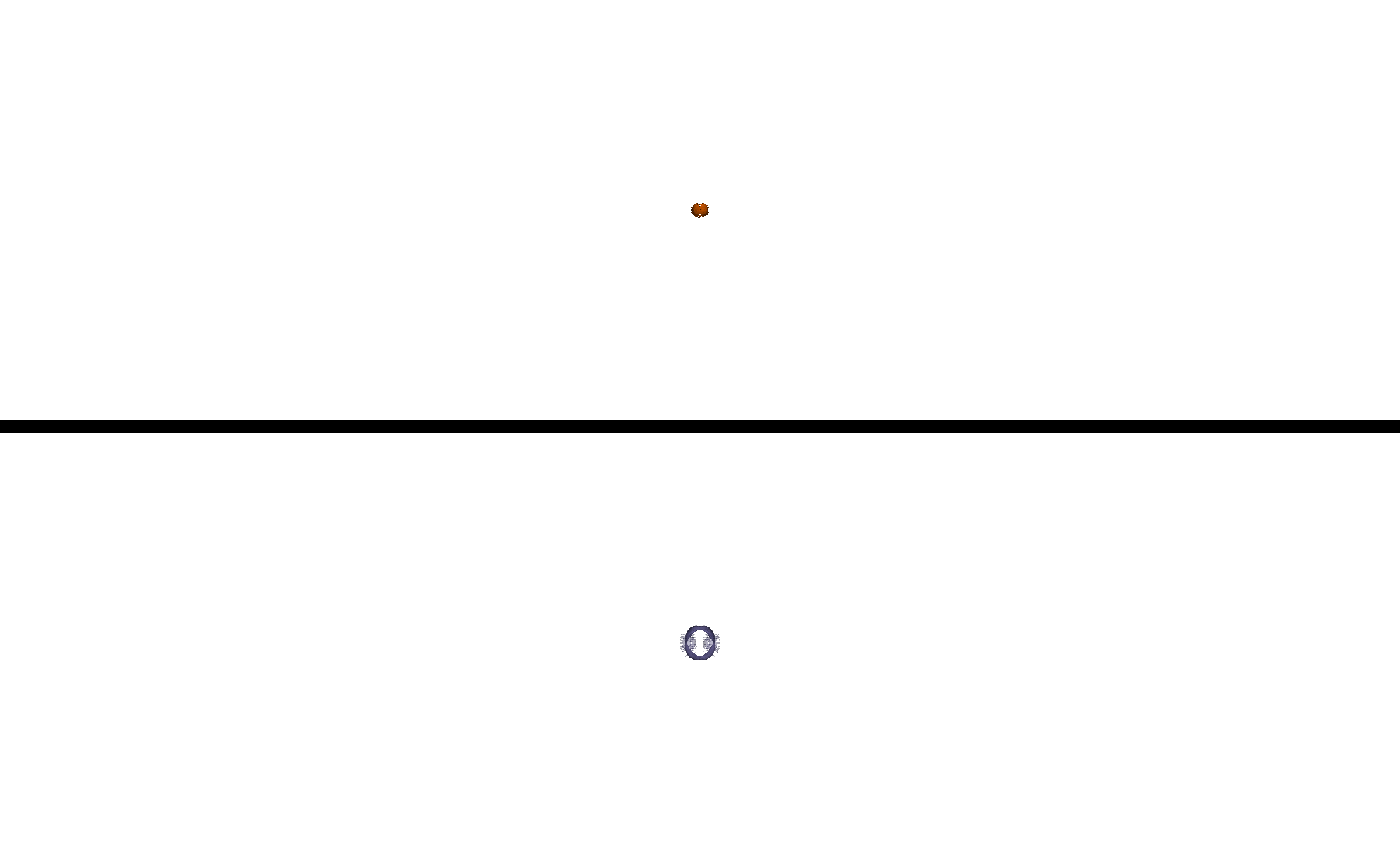} &
    \includegraphics[trim={20cm 0 20cm 0},clip,
      width=0.24\columnwidth]{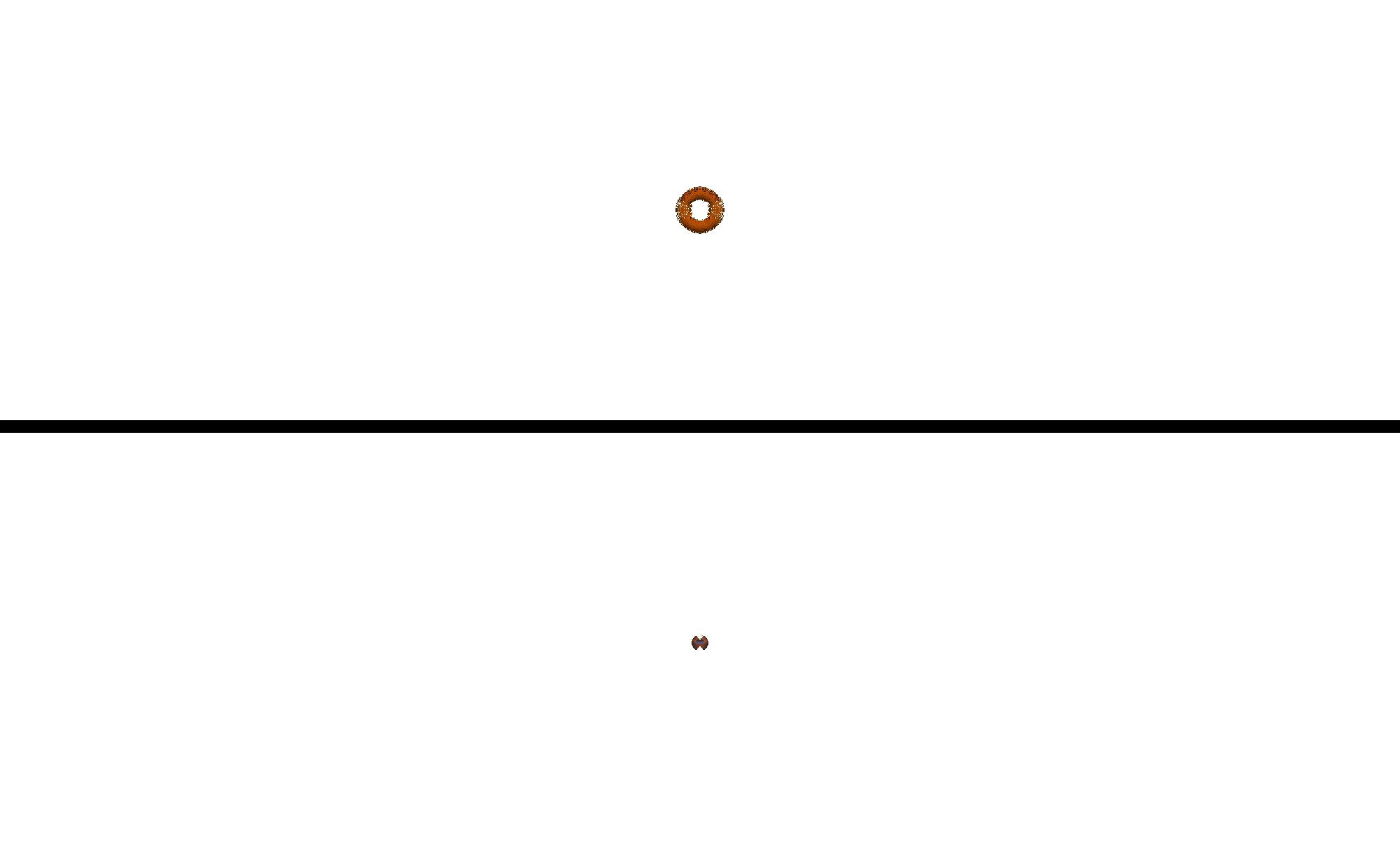} &
    \includegraphics[trim={20cm 0 20cm 0},clip,
      width=0.24\columnwidth]{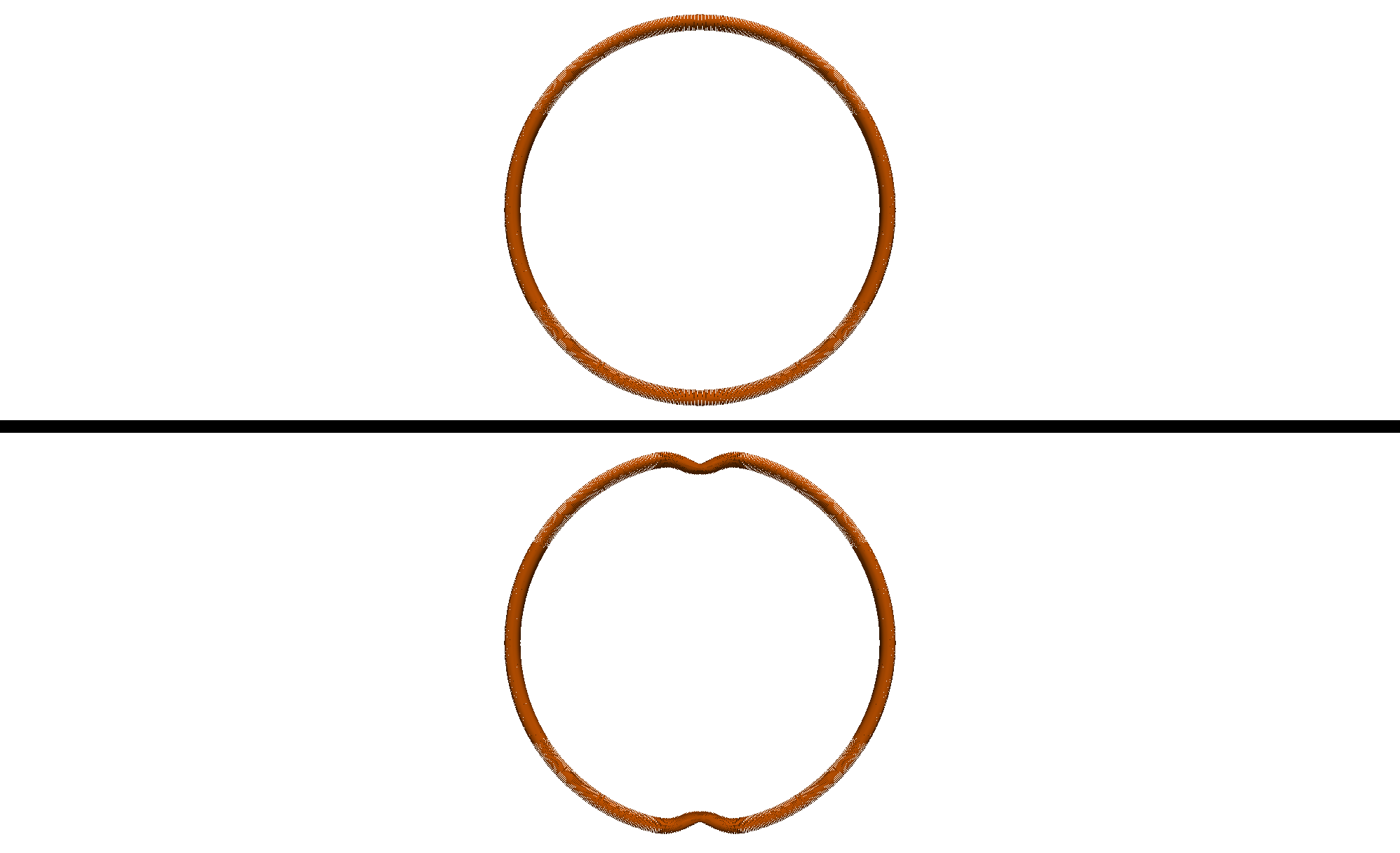} \\
    (a) $417.160 M$ &
    (b) $417.433 M$ &
    (c) $417.460 M$ &
    (d) $417.773 M$
  \end{tabular}
  \caption[Spatial cuts through the head-on BBH event horizon]{
    Slices in the mirror symmetry plane
    of \cref{figHOSurface}, near the time the EHs merge.
    The generators join the EH simultaneously through a caustic in
    both coordinate systems.
  }
  \label{figHOSlice}
\end{figure}

\Cref{figHOSlice} shows spatial slices through the mirror symmetry plane.
The top row shows the event horizon in the slicing used for the
\SpEC{} BBH spacetime evolution and the bottom row the transformed slicing.
These slices look similar to slices of the spherical model shown
in \cref{figToyS0Slice}, where the $t$ coordinate slice in the top row
remains a circle and generators on the circle join the
horizon simultaneously through a caustic.

Just as in the spherical model, we cannot alter the relative timing of
when the generators join the horizon in this slice, since these generators
meet at a single event in spacetime, and coordinate transformations preserve
events.
We perform a reslicing anyway to illustrate the point and to test our
code.
In the bottom row of \cref{figHOSlice}, we see small scale deformations
along the top and bottom of the ring.
Because of the coordinate transformation in \cref{eqnCoordTransformation},
generators in regions where $G(x^i, t)$ is relatively large
are delayed in the $\bar{t}$ slicing, causing the small bumps in
panels~(a)~and~(d).
The caustic event where generators join the horizon
occurs in panel~(c), showing that the caustic is preserved by the
coordinate transformation.
No hole in this event horizon could possibly exist because of the lack of
a crossover surface.

Independent of the slicing of the spacetime, the head-on binary
starts as a set of spheres and
transitions to a single sphere.
These results are consistent with the findings in~\cite{MassoEtAl:1999,
Lehner1999, Husa-Winicour:1999},
as well as the spherical model in \cref{secSphericalModel}.
The highest resolution of the \SpEC{} BBH evolution was used for these figures,
but the topological structure is the same in all three resolution levels
of the \SpEC{} evolution.

\subsection{Ellipsoidal model}
\label{secEllipsoidalModel}

The prediction of Siino~\cite{Siino1998b} and
Husa and Winicour~\cite{Husa-Winicour:1999} is that
toroidal event horizons should appear in generic BBH mergers, where
there is no axis of symmetry.
We analyze in this section an ellipsoidal wavefront, identical
to the oblate spheroid model in~\cite{Shapiro1995, Siino1998a},
that provides a more generic caustic and crossover distribution
than the spherical wavefront model in~\cref{secSphericalModel}.
The appearance of both caustics and crossovers makes this model illustrative
for generic BBH mergers, such as the equal mass inspiral featured in the
following section.

\begin{figure}
  \centering
  \includegraphics[width=0.3\textwidth]{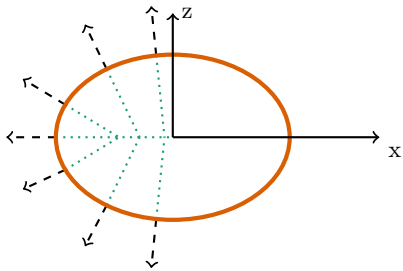}
  \caption[Initial data configuration for ellipsoidal model horizon in flat
    space]{
    Initial data configuration for the ellipsoidal model horizon
    in flat space.
    Similar to \cref{figS0InitialData}, but the initial data surface is
    an ellipsoid rather than a sphere.
    The green dashed lines show where the generators came from
    earlier in coordinate time, and that the trajectories met at
    different locations in the past.
  }
  \label{figS1InitialData}
\end{figure}

\begin{figure}
  \centering
  \begin{tabular}{cccc}
    \includegraphics[trim={20cm 0 20cm 0},clip,
      width=0.24\columnwidth]{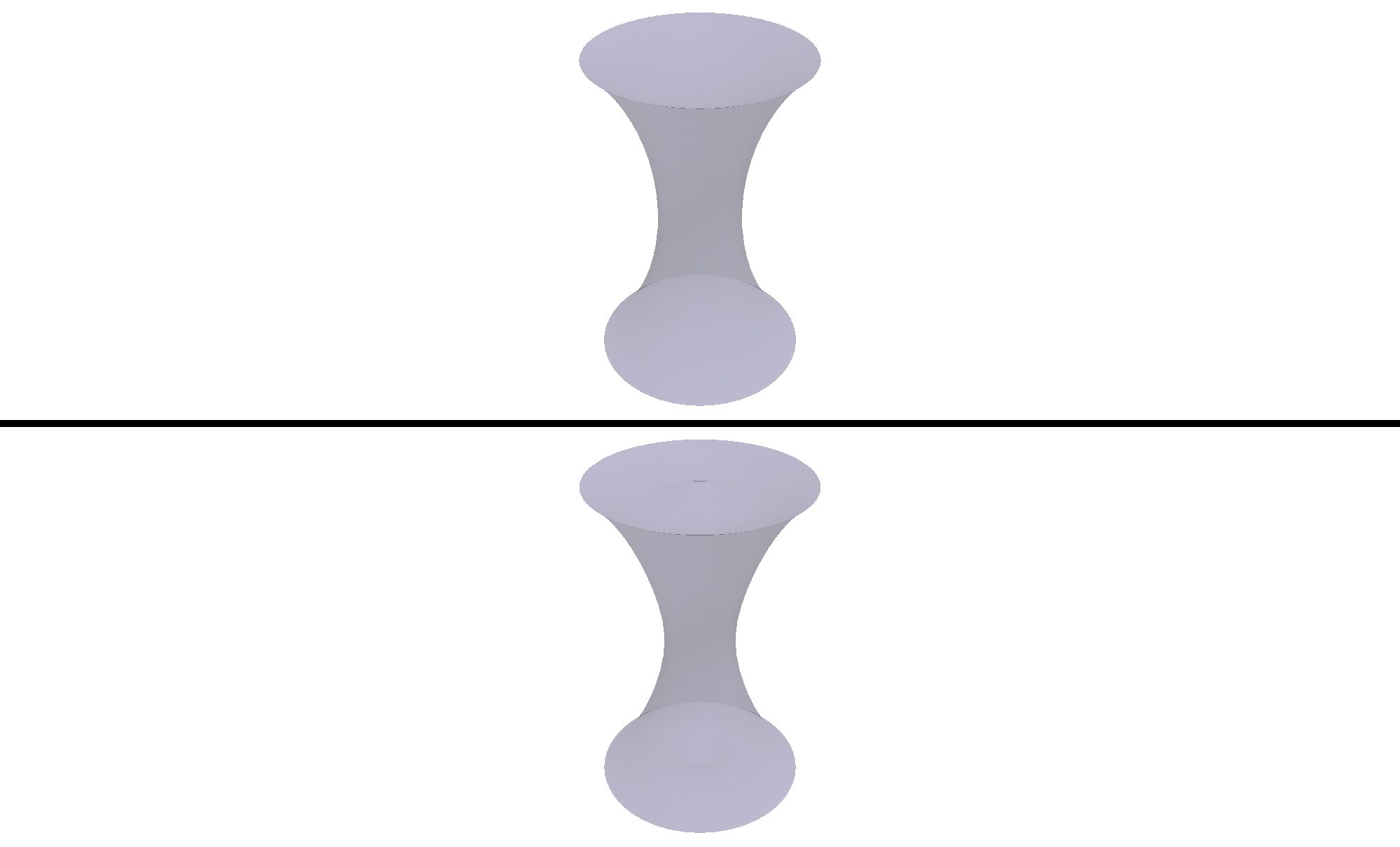} &
    \includegraphics[trim={20cm 0 20cm 0},clip,
      width=0.24\columnwidth]{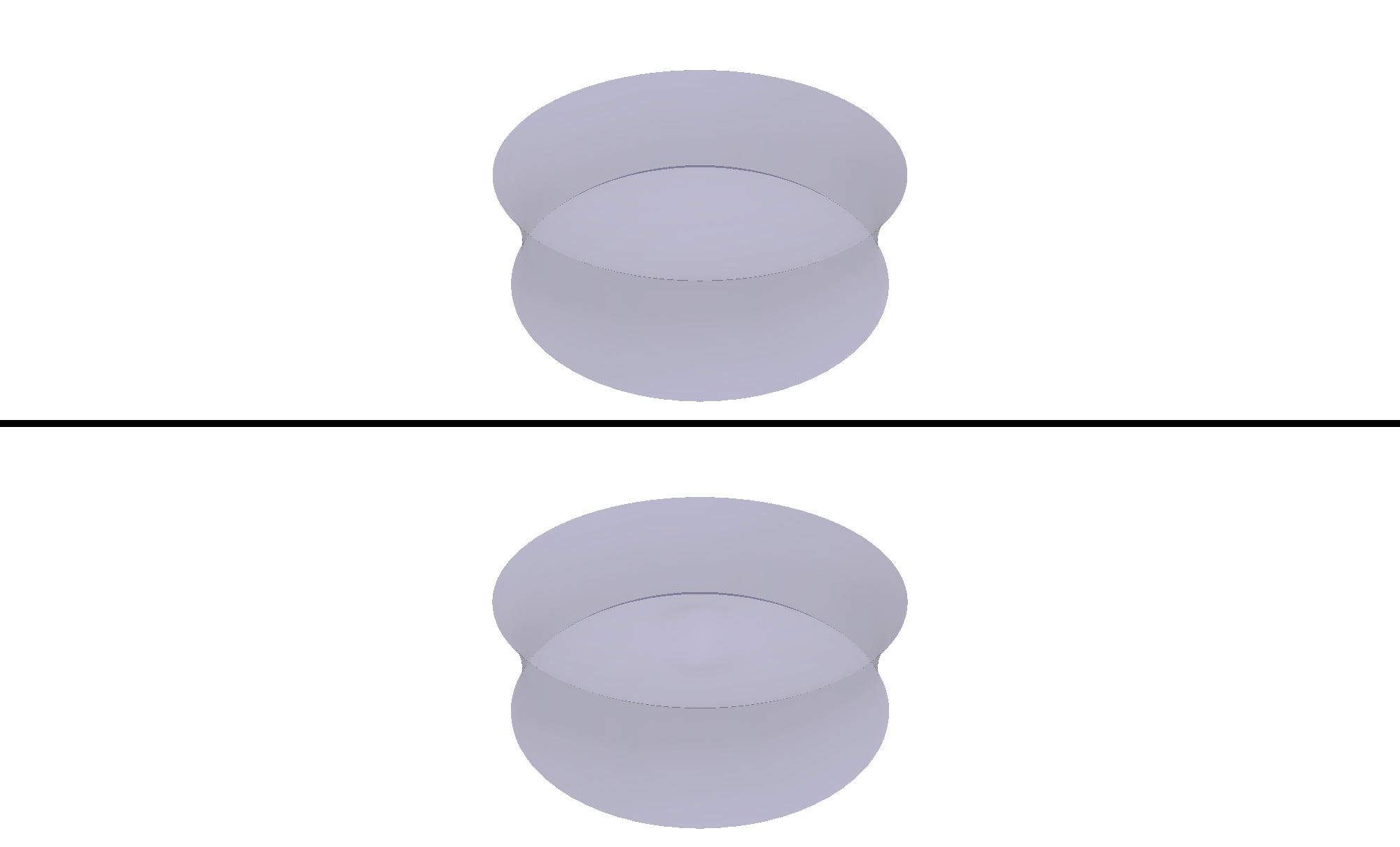} &
    \includegraphics[trim={20cm 0 20cm 0},clip,
      width=0.24\columnwidth]{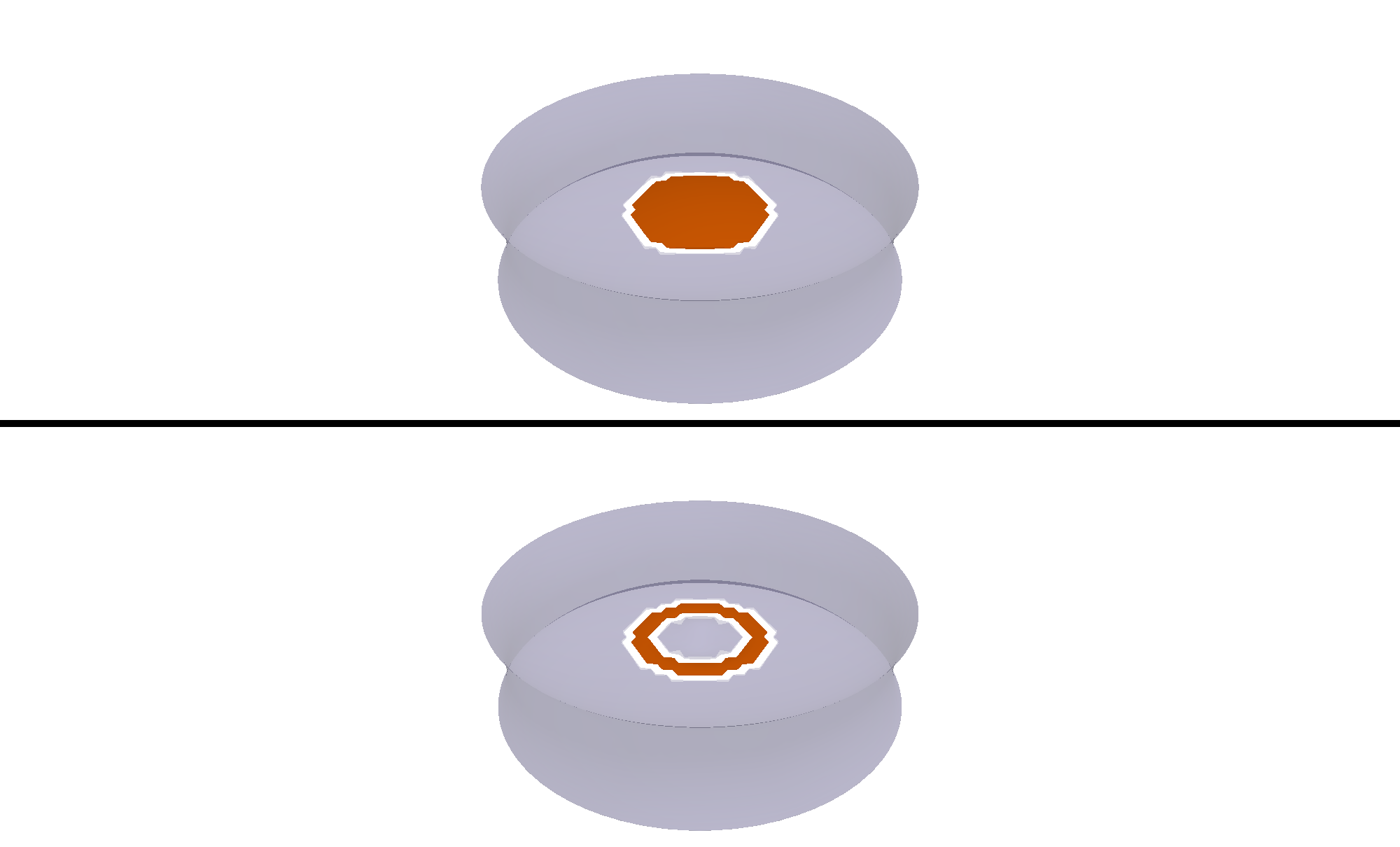} &
    \includegraphics[trim={20cm 0 20cm 0},clip,
      width=0.24\columnwidth]{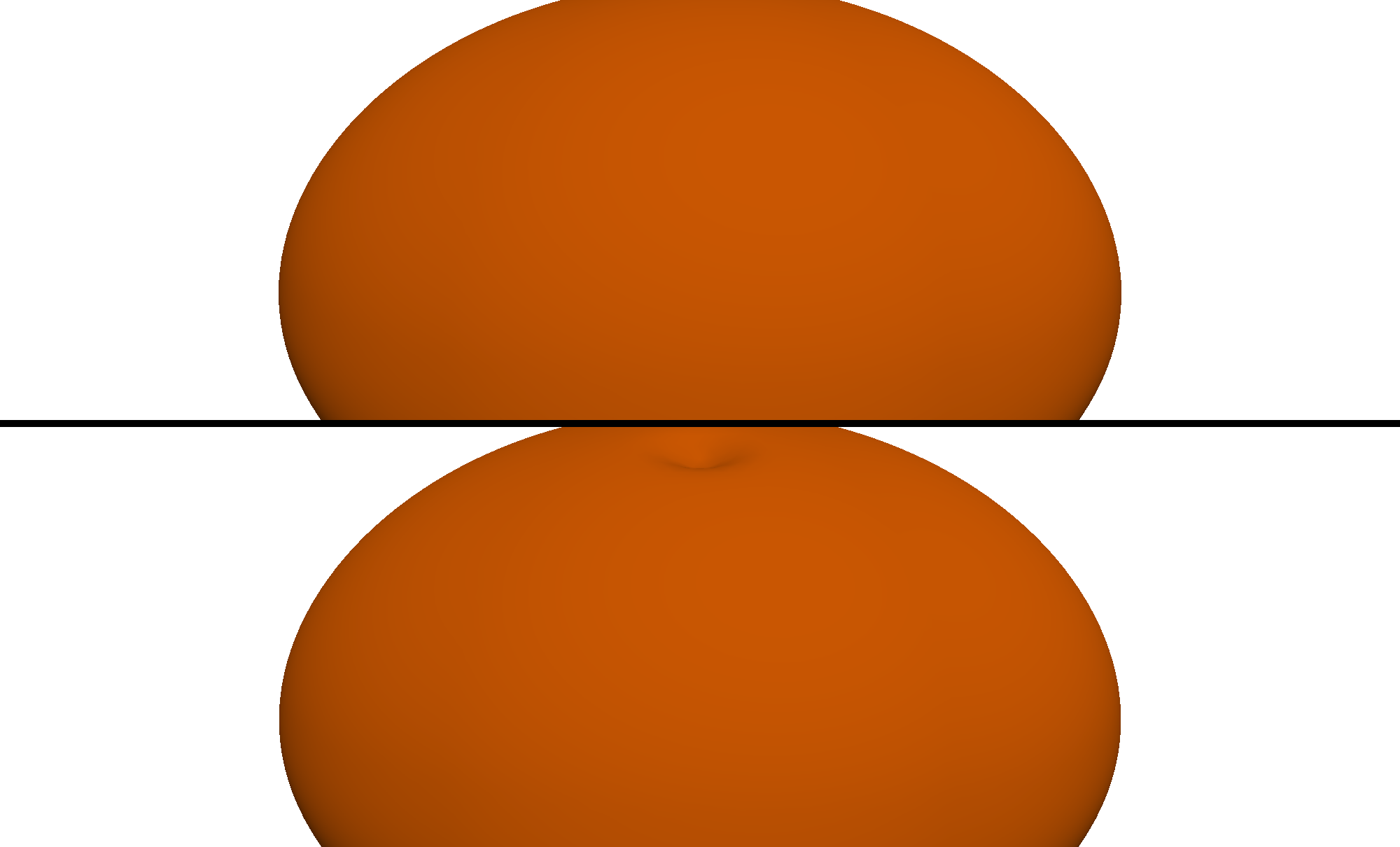} \\
    (a) $-1.307 M$ &
    (b) $-1.038 M$ &
    (c) $-0.988 M$ &
    (d) $-0.571 M$
  \end{tabular}
  \caption[Generator surface for the ellipsoidal model horizon]{
    Similar to \cref{figToyS0Surface}, but with an ellipsoidal
    horizon used as initial data.
    The figures are zoomed in to show the small scale features that arise
    as generators
    join the horizon.
  }
  \label{figToyS1Surface}
\end{figure}

The initial data for the generator evolution is similar to the spherical
model, but we place generators normal to the ellipsoid
\begin{equation}
  \frac{x^2 + y^2}{2} + z^2 = 1,
  \label{eqnS1ID}
\end{equation}
shown in \cref{figS1InitialData}.
\Cref{figToyS1Surface} shows this ellipsoidal horizon on a few time slices
using the same color scheme and layout as \cref{figToyS0Surface}.
In agreement with Shapiro~\textit{et al.}~\cite{Shapiro1995},
the first generators to join the horizon
join at the origin through crossover points in the top row of panel~(c).
The horizon is smooth everywhere,
apart from a one-dimensional ring around the outside of the horizon
where generators continue to join through crossover points.
If we connect these crossover events to form a surface,
we obtain a two-dimensional
spacelike hypersurface in the equatorial plane (the $xy$-plane).
Much later, the last future generators join the horizon along the outside ring
in the equatorial plane,
forming a one-dimensional ring of caustic events.
This slicing therefore shows only a spherical topology.

The bottom row of \cref{figToyS1Surface} shows the horizon
after the coordinate transformation in \cref{eqnCoordTransformation}
with parameters identical to those used in the spherical model
(case A of \cref{figResliceParameters}).
While applying coordinate transformations will ensure that spacetime events
such as caustic or
crossover points are preserved,
the relative time between neighboring
caustic or crossover points can be altered.
This coordinate transformation is sufficient to obtain a horizon
that initially appears with a toroidal topology as shown in the bottom row of
panel~(c).
The horizon is smooth
apart from two
one-dimensional rings where crossover generators continue to join
the surface.
One ring is on the outside of the torus and the other is on the inside.
Shortly after the torus forms,
the hole in the horizon closes, leaving the same spherical
topology as seen in the top row of panel~(c).

\begin{figure*}
  \centering
  \begin{tabular}{cccc}
    \includegraphics[trim={6cm 0 6cm 0},clip,
      width=0.235\textwidth] {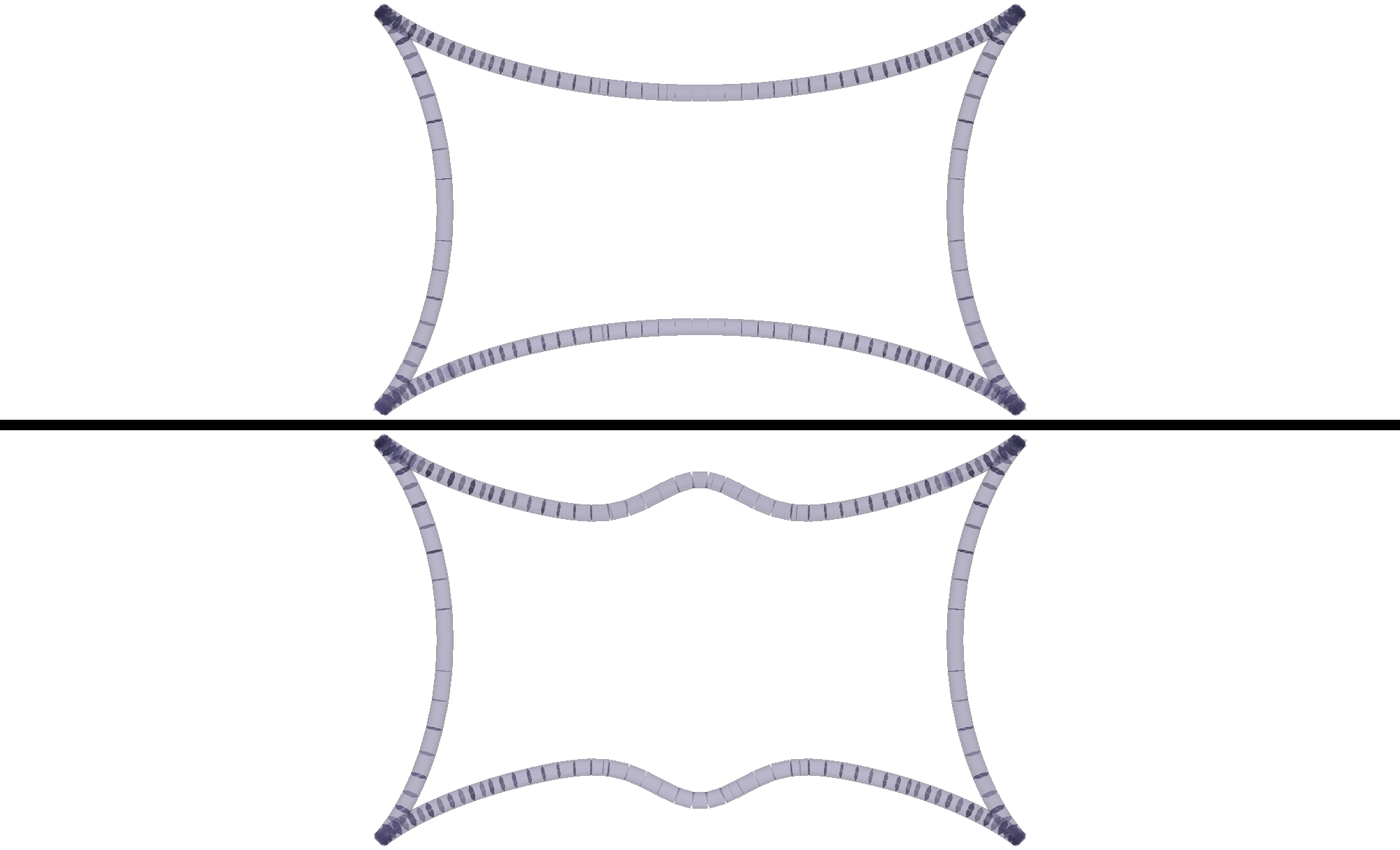} &
    \includegraphics[trim={6cm 0 6cm 0},clip,
      width=0.235\textwidth]{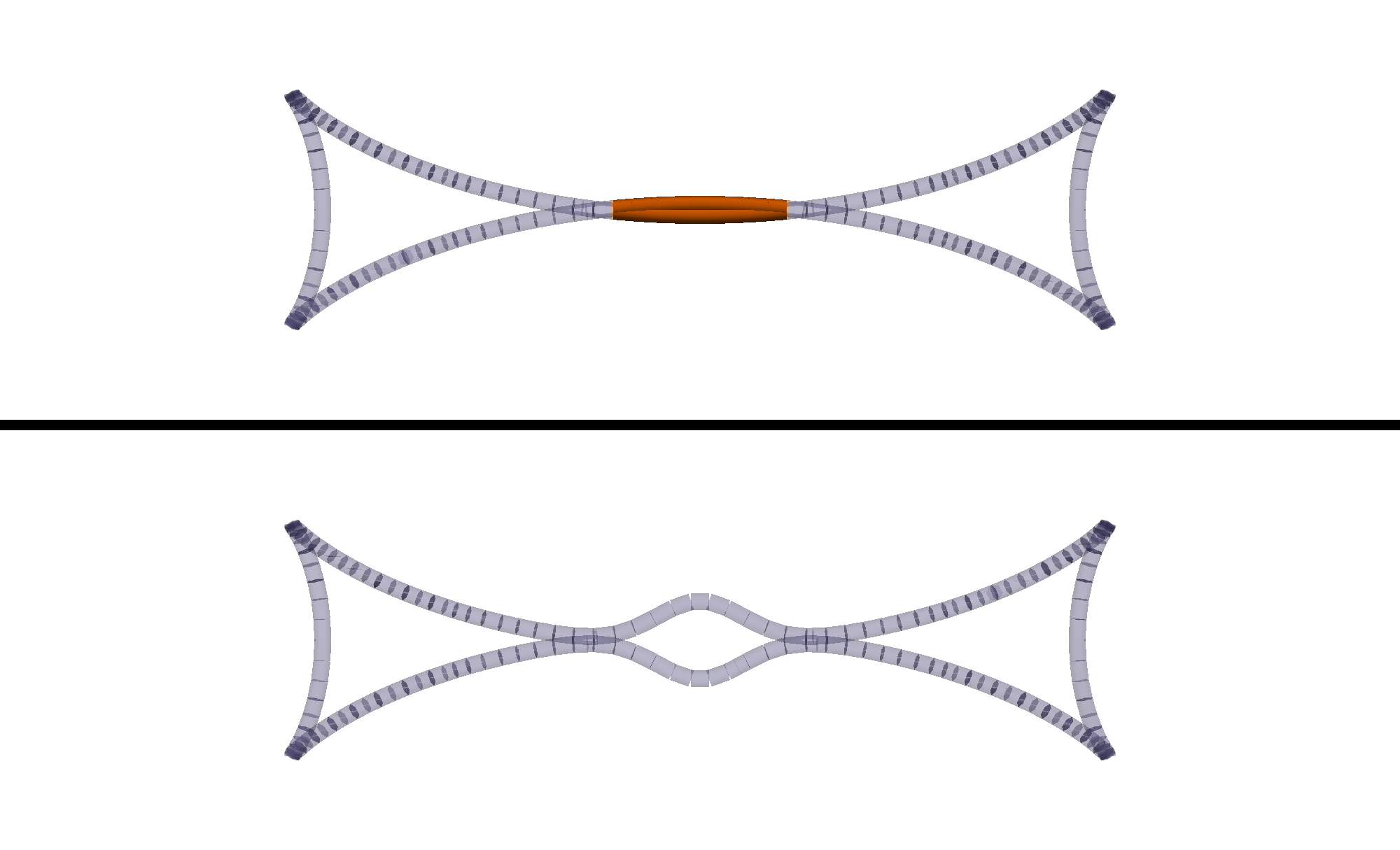} &
    \includegraphics[trim={6cm 0 6cm 0},clip,
      width=0.235\textwidth]{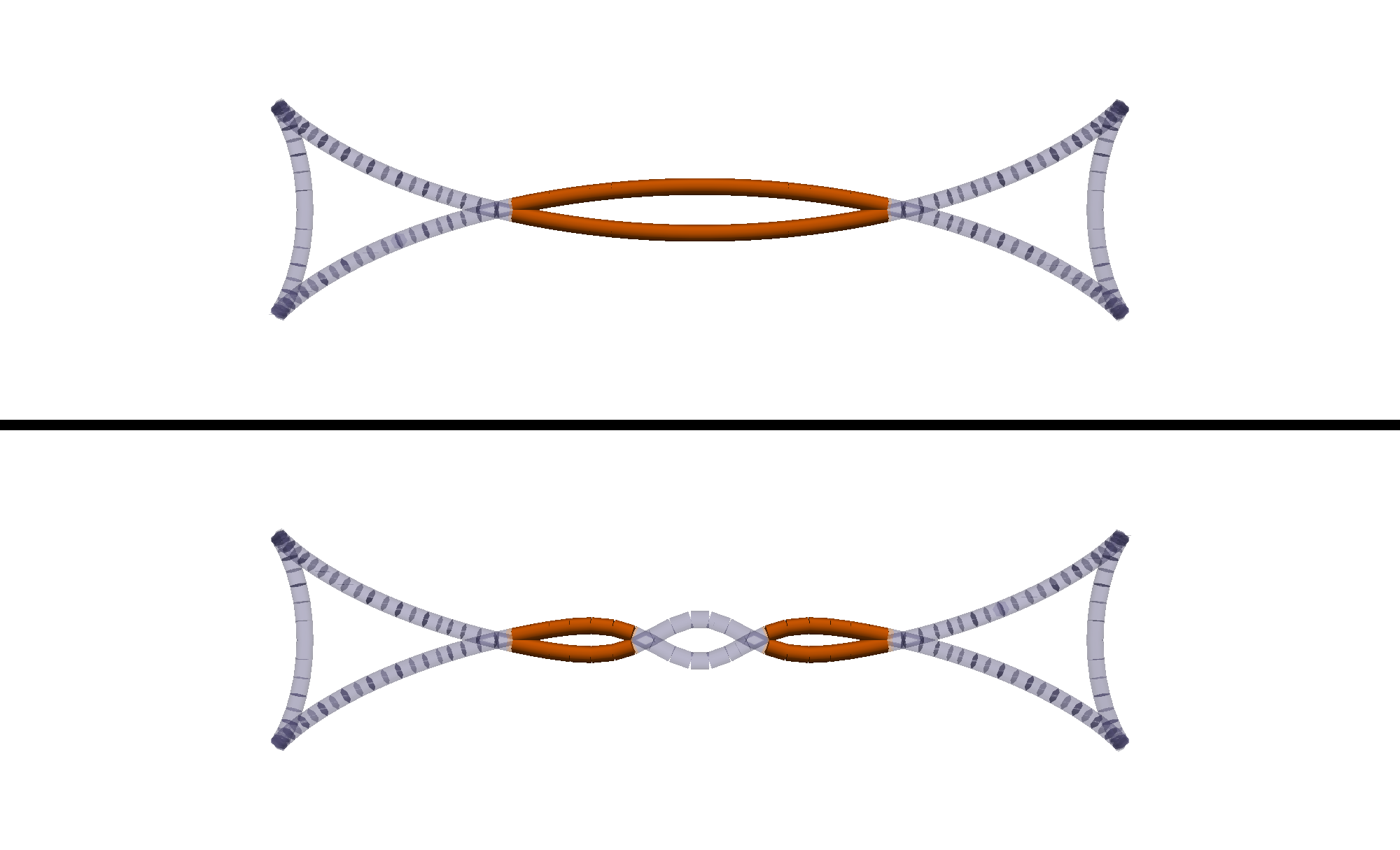} &
    \includegraphics[trim={6cm 0 6cm 0},clip,
      width=0.235\textwidth]{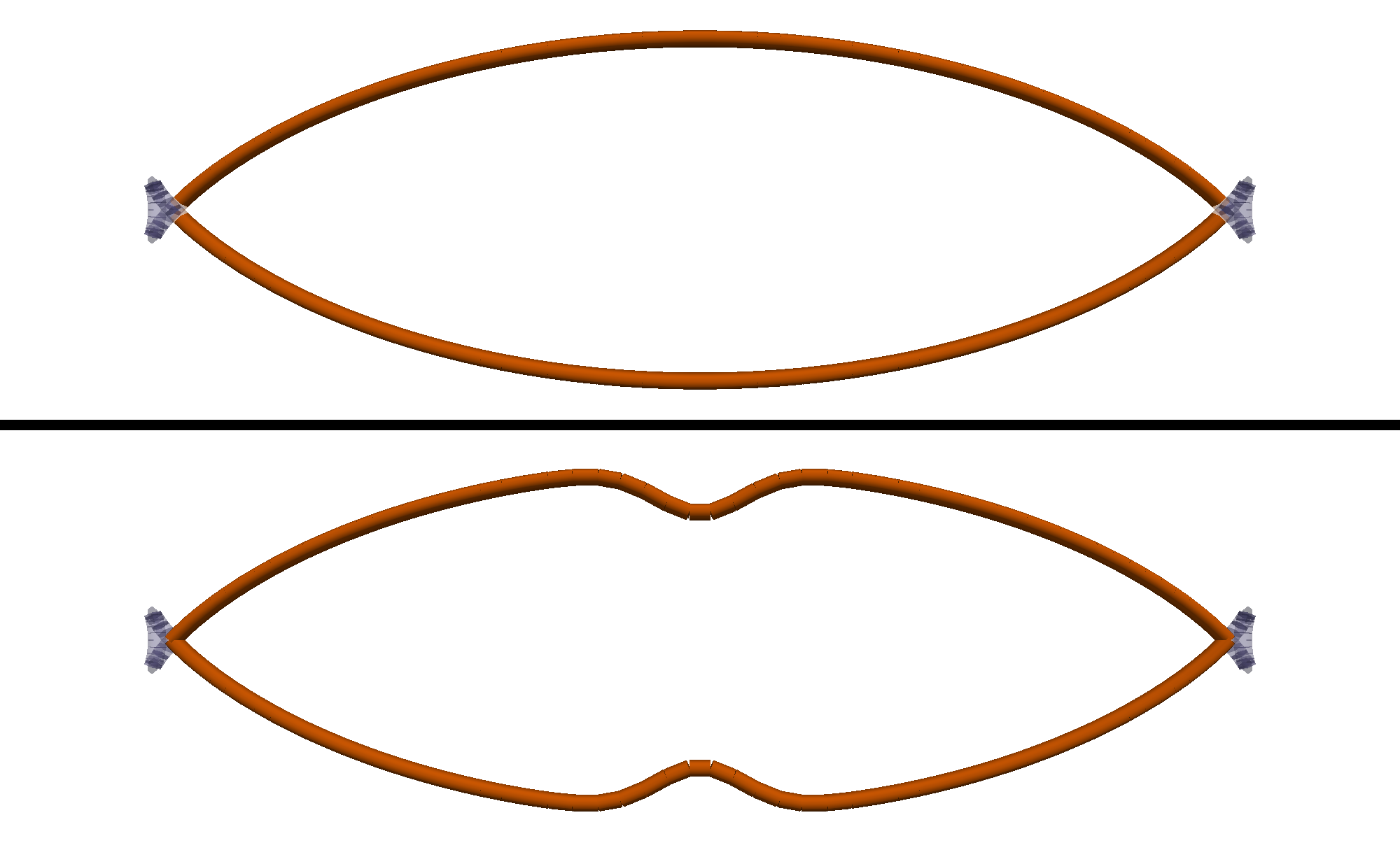} \\
    (a) $-1.131 M$ &
    (b) $-0.998 M$ &
    (c) $-0.978 M$ &
    (d) $-0.811 M$
  \end{tabular}
  \caption[Spatial cuts through the ellipsoidal model horizon]{
    Zoomed in slices of the ellipsoidal
    horizon shown in \cref{figToyS1Surface}, covering a small
    duration of time near panel~(c).
    The setup is identical to \cref{figToyS0Slice}.
  }
  \label{figToyS1Slice}
\end{figure*}

As we did for the spherical model,
in \cref{figToyS1Slice}
we take a slice through the horizon along the $z$-axis
to learn why it was possible to apply a coordinate transformation and obtain a
torus.
The spatial and temporal scales are magnified in this figure compared to
\cref{figToyS1Surface} to showcase how the generators join
the surface in both coordinate systems.

Panel~(a) of \cref{figToyS1Slice} shows a slice of future generators
with a quite different shape compared to what is seen in the spherical
model.
In the top row of panel~(b), generators begin to join the horizon
through crossover points,
where generators from the top half of the slice meet the
bottom half.
The horizon instantaneously appears as an $\mathcal{S}^2$.
In the $\bar{t}$ slicing of the bottom row,
the generators in the middle of the slice are
delayed relative to their neighbors because of the positive Gaussian
in the coordinate transformation.
The delay is sufficient to cause the first generators that join the horizon
to be spatially separated on the slice as seen in panel~(c).
After rotating about the vertical axis of symmetry, the
surface initially appears with a toroidal topology.
Finally, in the bottom row of panel~(d), the interior region has closed
to yield an $\mathcal{S}^2$ topology.
We have thus found a coordinate transformation that cuts a hole out of the
spacelike crossover surface along the inseam.

\subsection{Equal mass inspiral}
\label{secEMI}

The primary reason that the equal mass head-on merger did not yield a toroidal
event horizon is the rotational symmetry of the system causing all the
future generators to join the horizon through caustics.
A binary black hole system in a quasi-circular orbit
removes this rotational symmetry.
We expect to see a more generic distribution of caustics and crossover
points similar to the ellipsoidal model, enabling us to reslice
the EH into a torus.
For simplicity, we analyze a pair of non-spinning black holes, initially
in a quasi-circular orbit with a
separation of $17 M$.

\begin{figure*}
  \centering
  \begin{tabular}{ccc}
    \includegraphics[
      width=0.33\textwidth]{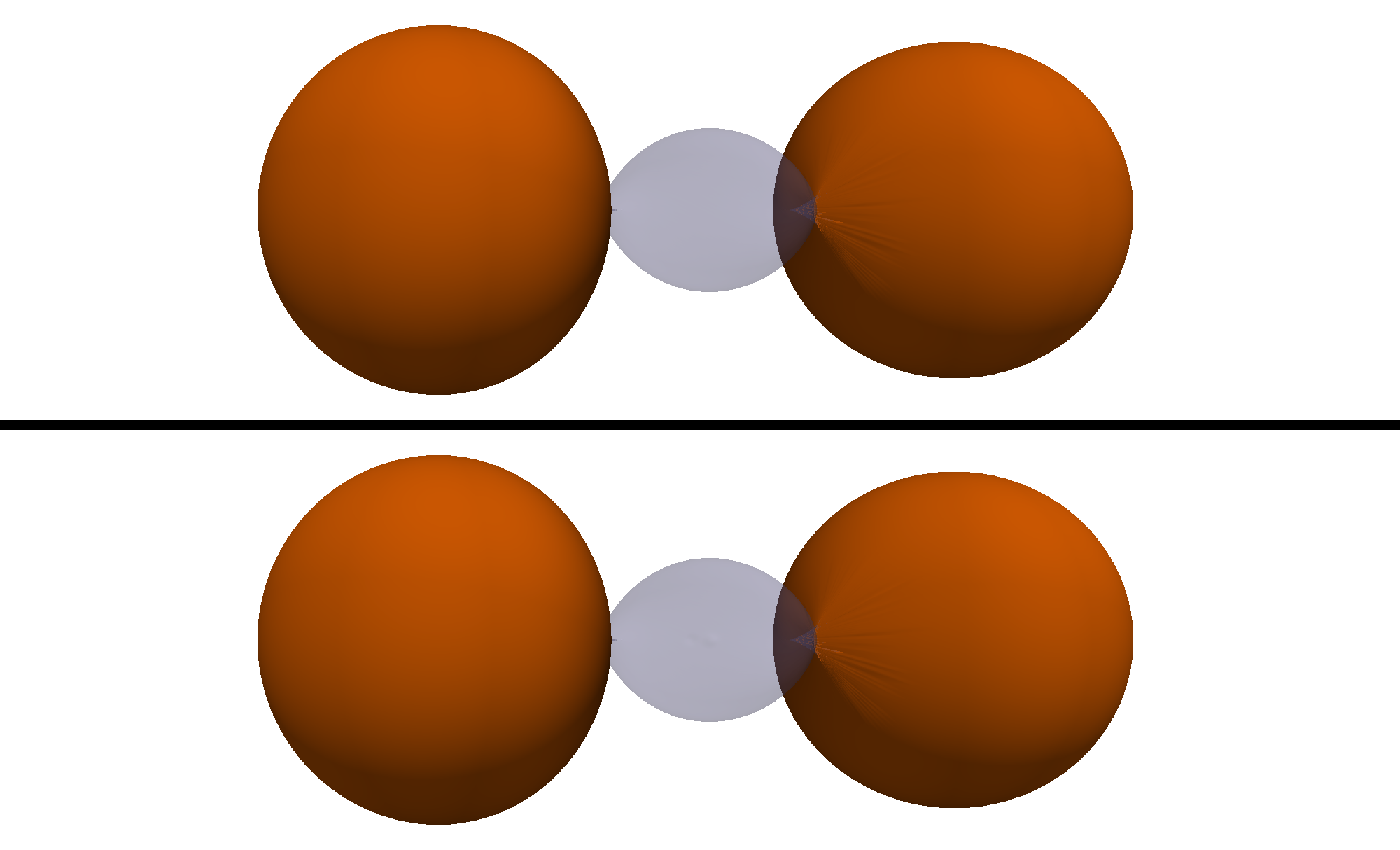} &
    \includegraphics[
      width=0.33\textwidth]{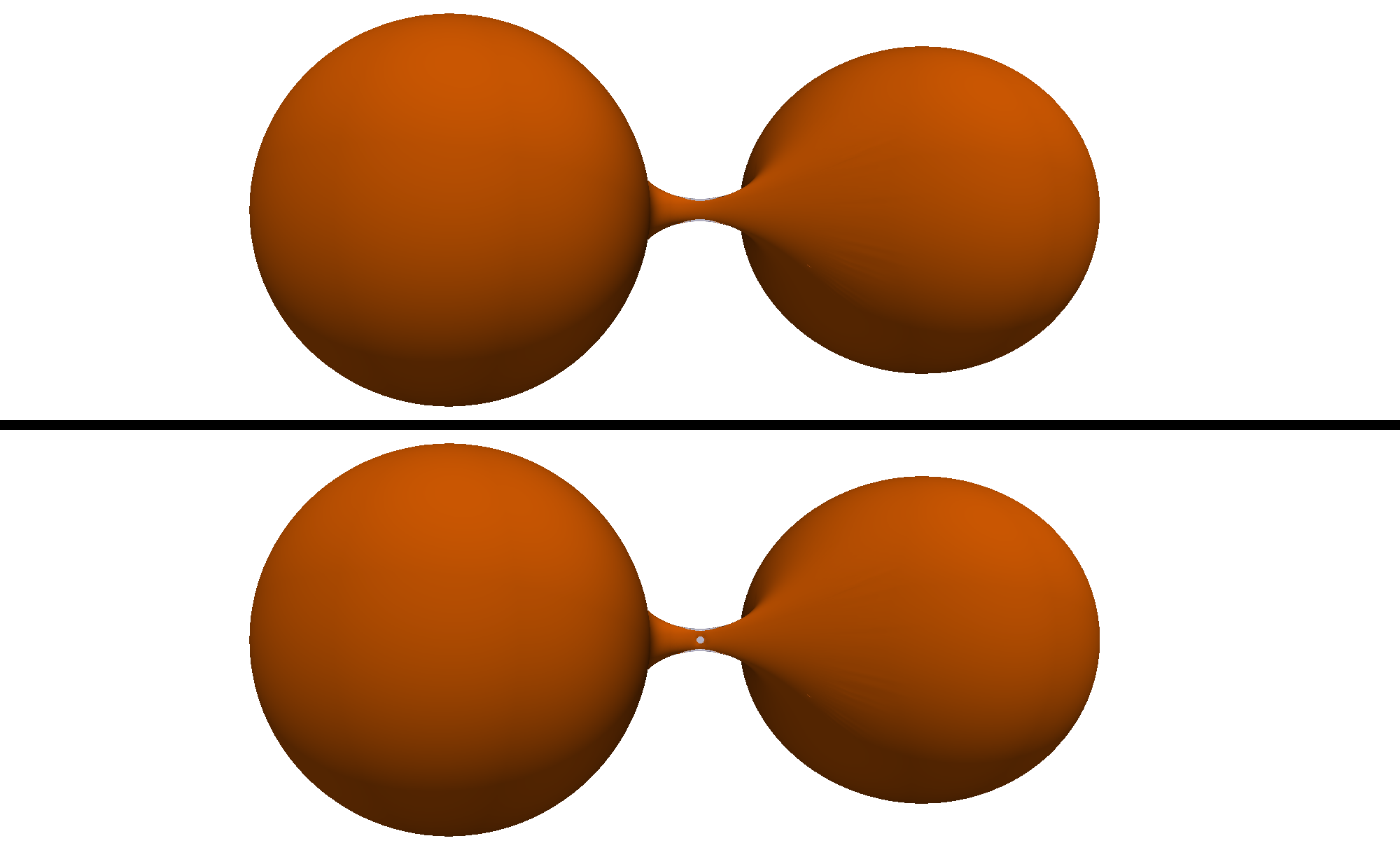} &
    \includegraphics[
      width=0.33\textwidth]{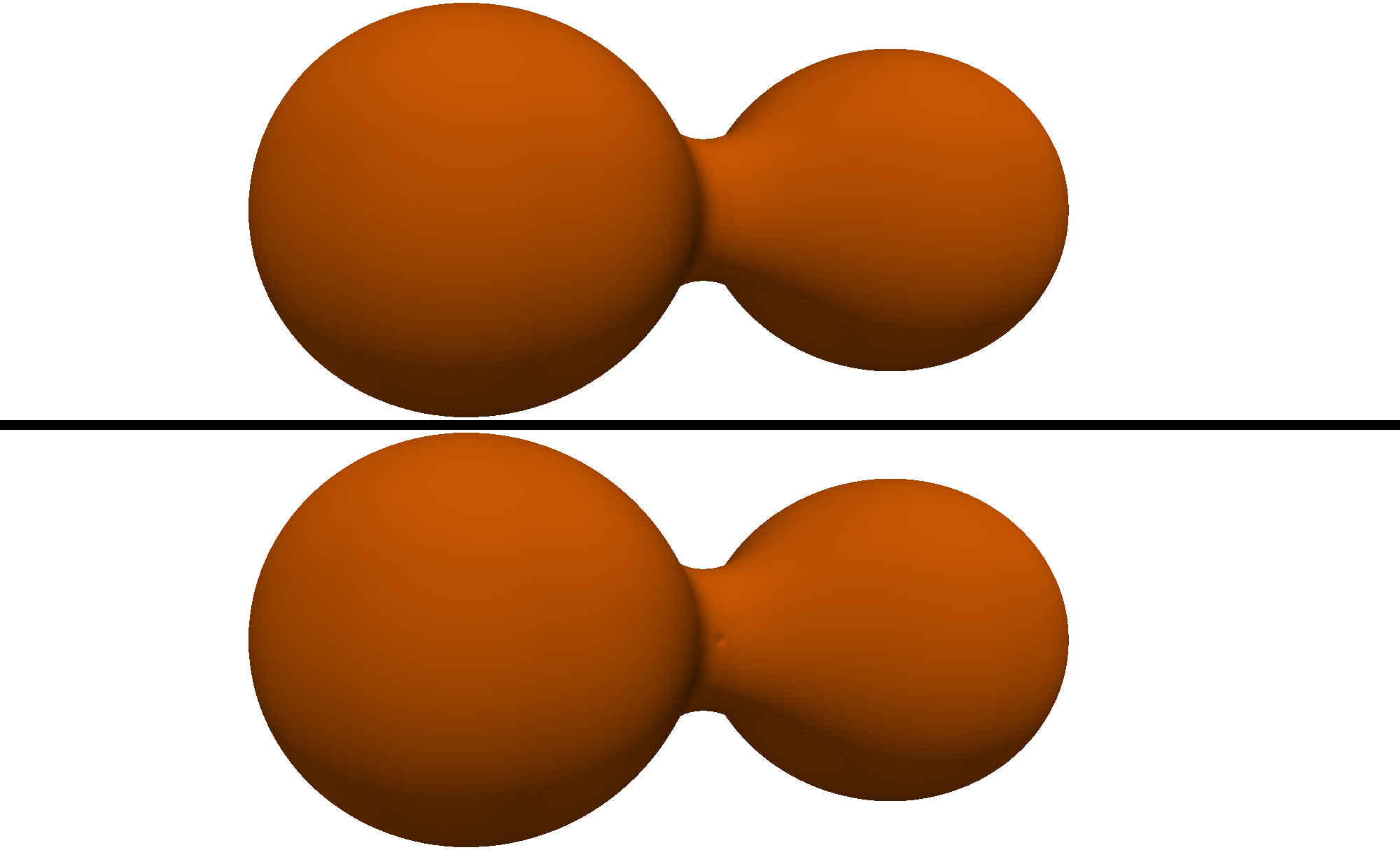} \\
    (a) $7539.011 M$ &
    (b) $7539.947 M$ &
    (c) $7540.786 M$
  \end{tabular}
  \caption[Generator surface for the equal mass inspiral]{
    EH generator surface for the equal mass inspiral, with
    the orbital angular momentum of the system pointing upward.
    A slice of the neck will be analyzed in more detail in \cref{figEMSlice}
    and a close-up is seen in \cref{figTORUS}.
  }
  \label{figEMSurface}
\end{figure*}

We show the event horizon surfaces in \cref{figEMSurface}, where
the camera is in the orbital plane and the orbital
angular momentum of the system is pointing up.
The coordinate transformation uses parameters with the label Case~C in
\cref{figResliceParameters}, and the amplitude is yet again quite small
compared to the figure size.
The time and space centers, $t_0$ and $\vec{r}_0$, are chosen to coincide
with the location where the event horizons first meet.
In this BBH, the neck joining the event horizons
has an elliptical shape, similar to what was seen in the slices of the
ellipsoidal
model horizon.
We learned from the ellipsoidal model that the direction
of the major axis $\hat{r}_{\rm{maj}}$ should be chosen roughly
along the direction in which the crossover generators were
traveling as they joined the horizon.
The final parameter that is important to tune is
the width of the Gaussian perpendicular to $\hat{r}_{\rm{maj}}$,
$\sigma_{\rm{min}}$, such that it is smaller
than the width of the neck connecting the horizons.

\begin{figure*}
  \centering
  \begin{tabular}{cc}
    \includegraphics[width=0.4\textwidth]{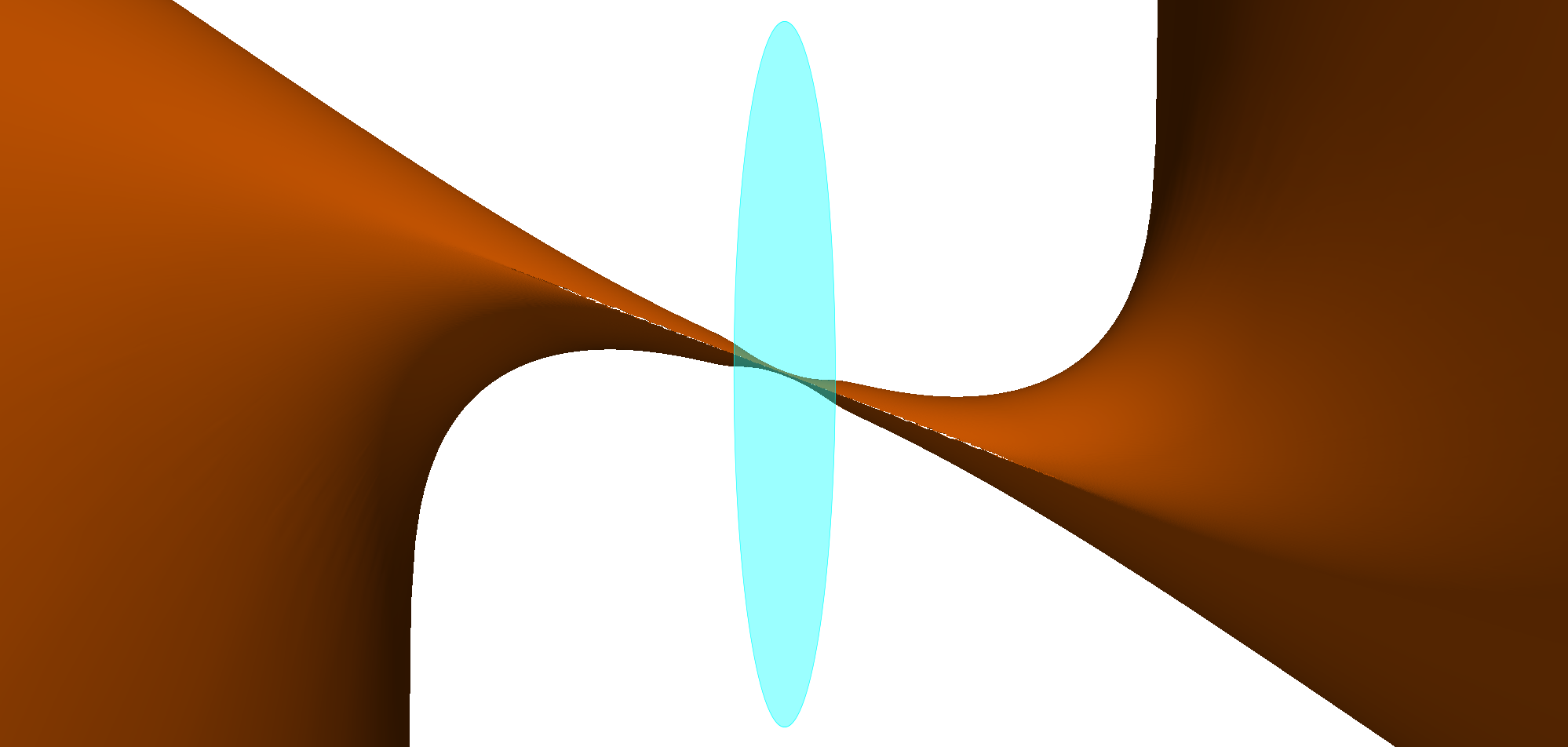} &
    \includegraphics[width=0.4\textwidth]{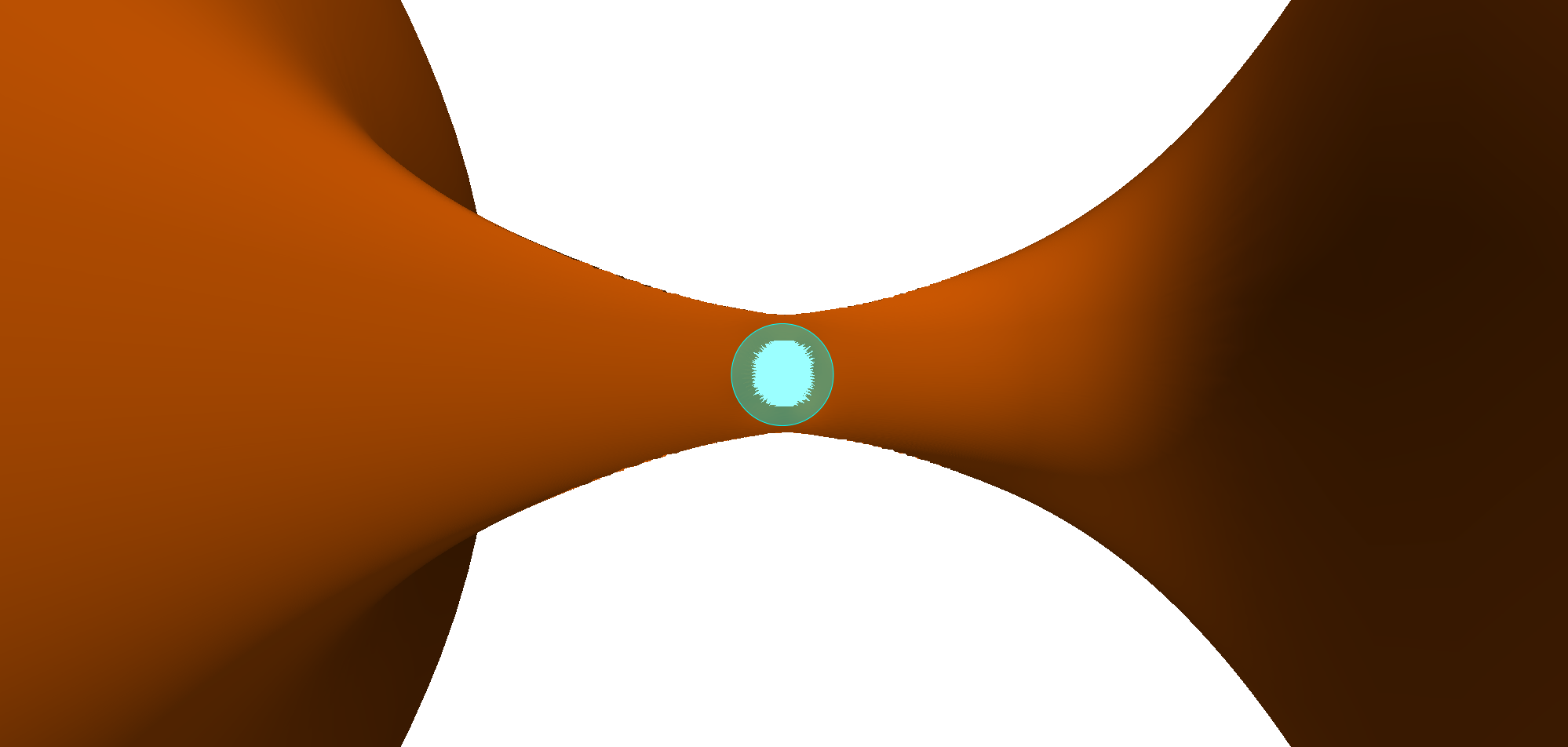} \\
    (a) $\hat{r}_{\rm{maj}}$ pointing upward &
    (b) $\hat{r}_{\rm{maj}}$ pointing out of the page
  \end{tabular}
  \caption[Overlaying the Gaussian ellipse over the equal mass inspiral]{
    At $\bar{t} = 7539.943 M$, visualizing the Gaussian ellipse
    on top of the equal mass inspiral surface shown in the barred
    coordinate system.
    In panel~(a), $\hat{r}_{\rm{maj}}$ is pointing upward and is pointing
    out of the page in panel~(b).
    The minor axis width $\sigma_{\rm{min}}$ is on the same spatial scale
    as the width of the neck causing a pinching of the neck in
    panel~(a) and causing a hole in the horizon surface to appear in panel~(b).
  }
  \label{figLapseOverlay}
\end{figure*}

\Cref{figLapseOverlay} shows a cartoon illustration of this coordinate
transformation overlaid on the event horizon in barred coordinate system.
The camera viewpoints are chosen such that in panel~(a), $\hat{r}_{\rm{maj}}$
is pointing up, and in panel~(b), $\hat{r}_{\rm{maj}}$ is pointing into the
page.
The major axis Gaussian width $\sigma_{\rm{maj}}$ is not shown to scale in this
figure,
but the precise value of $\sigma_{\rm{maj}}$ has little effect on the coordinate
transformation once it is sufficiently large.
In panel~(a) the effect of the coordinate transformation is only
to pinch the neck in the $\bar{t}$ coordinate system
in the region where the Gaussian is different than zero.
The minor axis Gaussian width $\sigma_{\rm{min}}$
has most of the control over the size of the
hole, where a smaller width causes a smaller (and thus harder to resolve
numerically)
hole.
Smaller values of $\sigma_{\rm{min}}$ also result in sharper gradients
of the function $G$, which can cause the new lapse in \cref{eqnNewLapse}
to become imaginary.
However, a minor axis width that is too large gives a shallower
gradient of the function
$\bar{t} = t + G(x^i, t)$, which
could result in the lack of a toroidal horizon.

\begin{figure*}
  \centering
  \begin{tabular}{cccc}
    \includegraphics[trim={6cm 0 6cm 0},clip,
      width=0.235\textwidth] {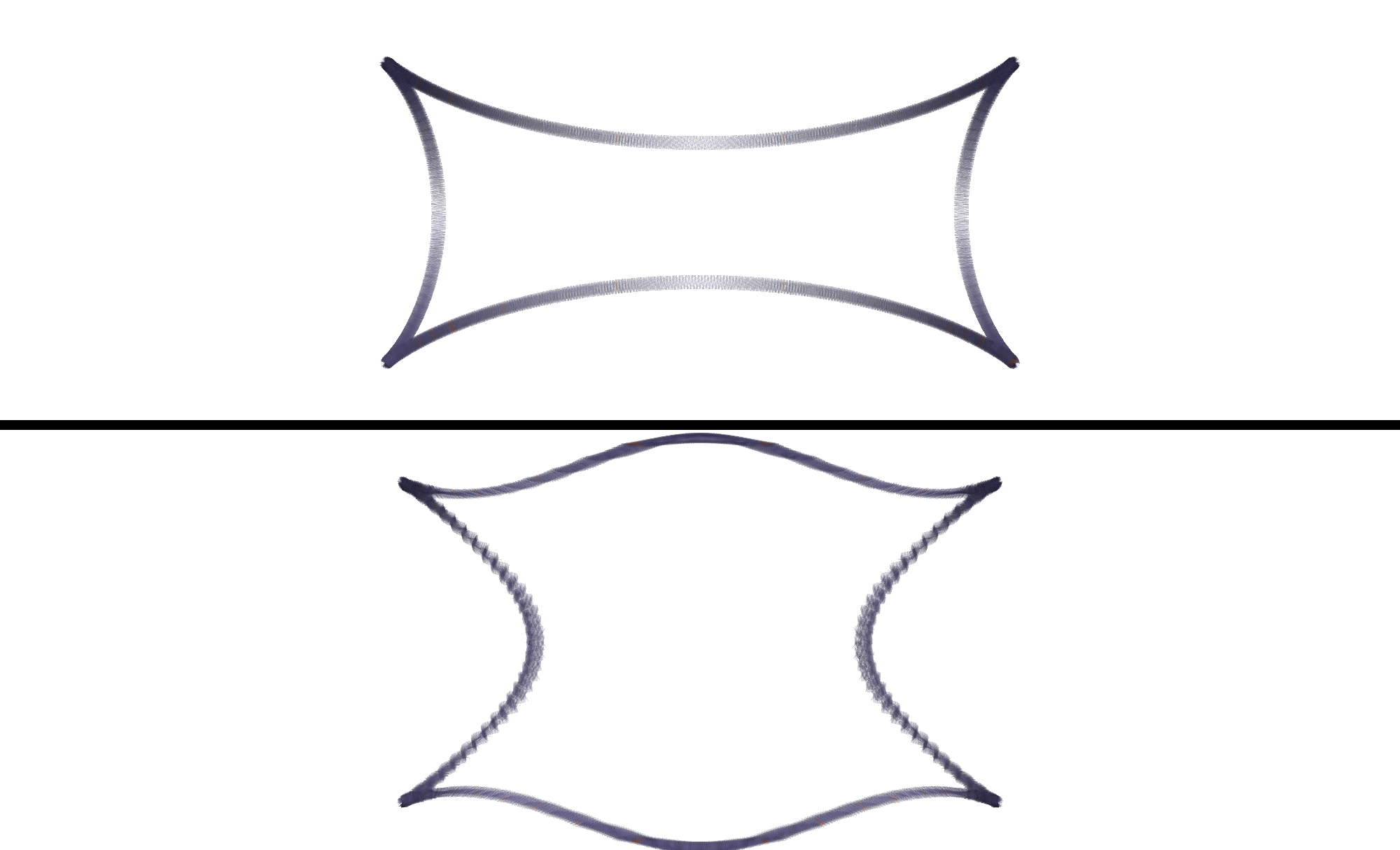} &
    \includegraphics[trim={6cm 0 6cm 0},clip,
      width=0.235\textwidth]{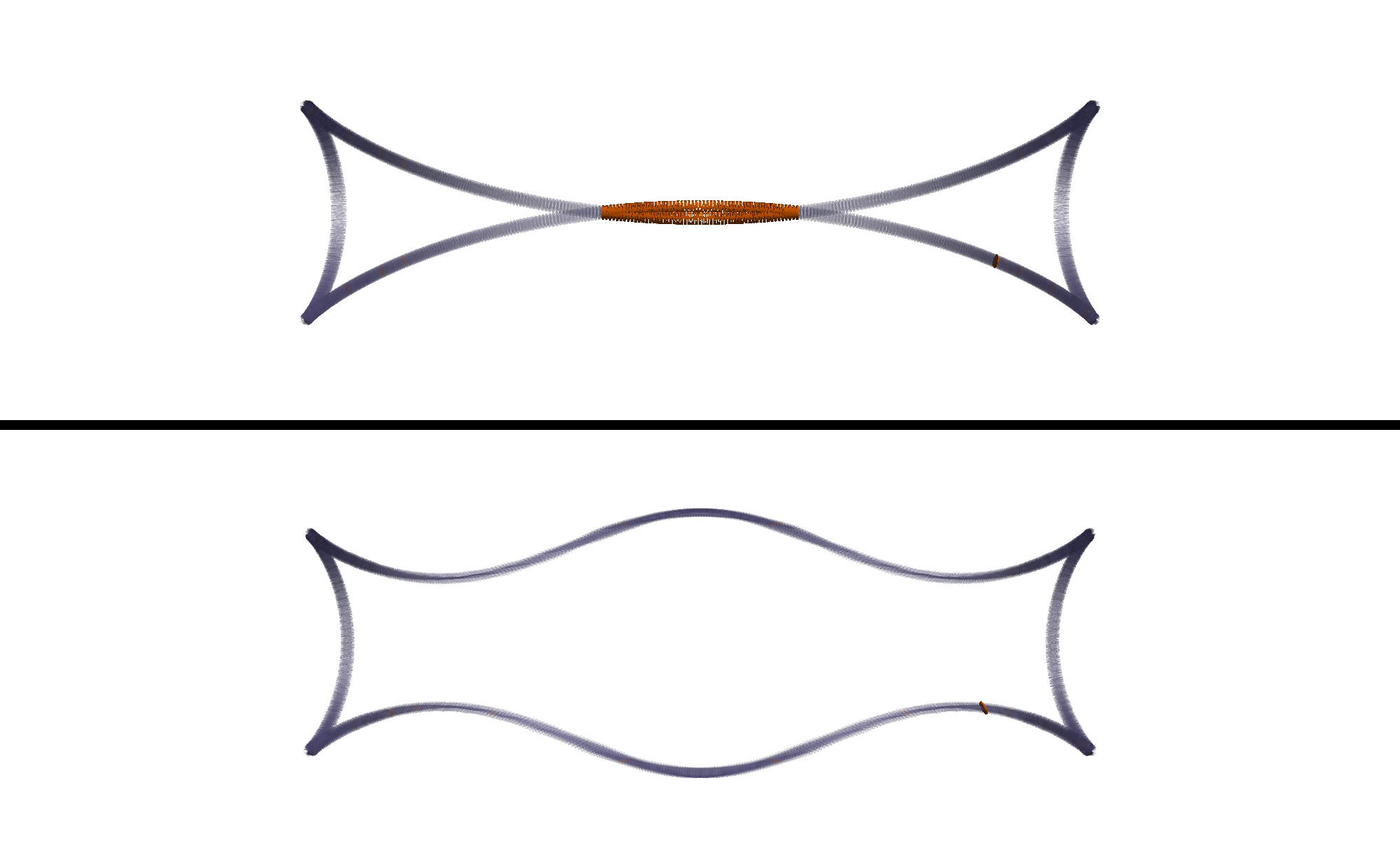} &
    \includegraphics[trim={6cm 0 6cm 0},clip,
      width=0.235\textwidth]{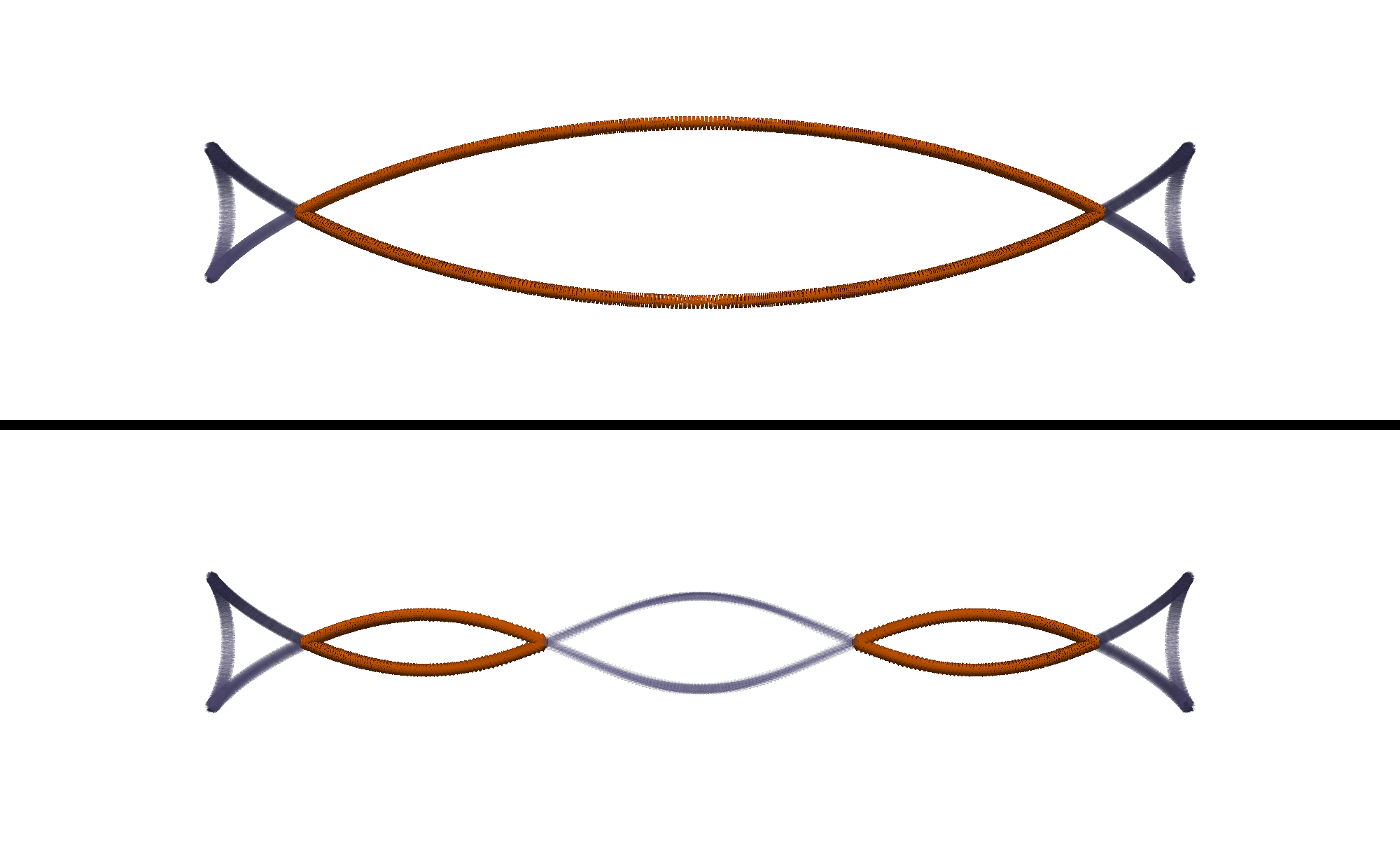} &
    \includegraphics[trim={6cm 0 6cm 0},clip,
      width=0.235\textwidth]{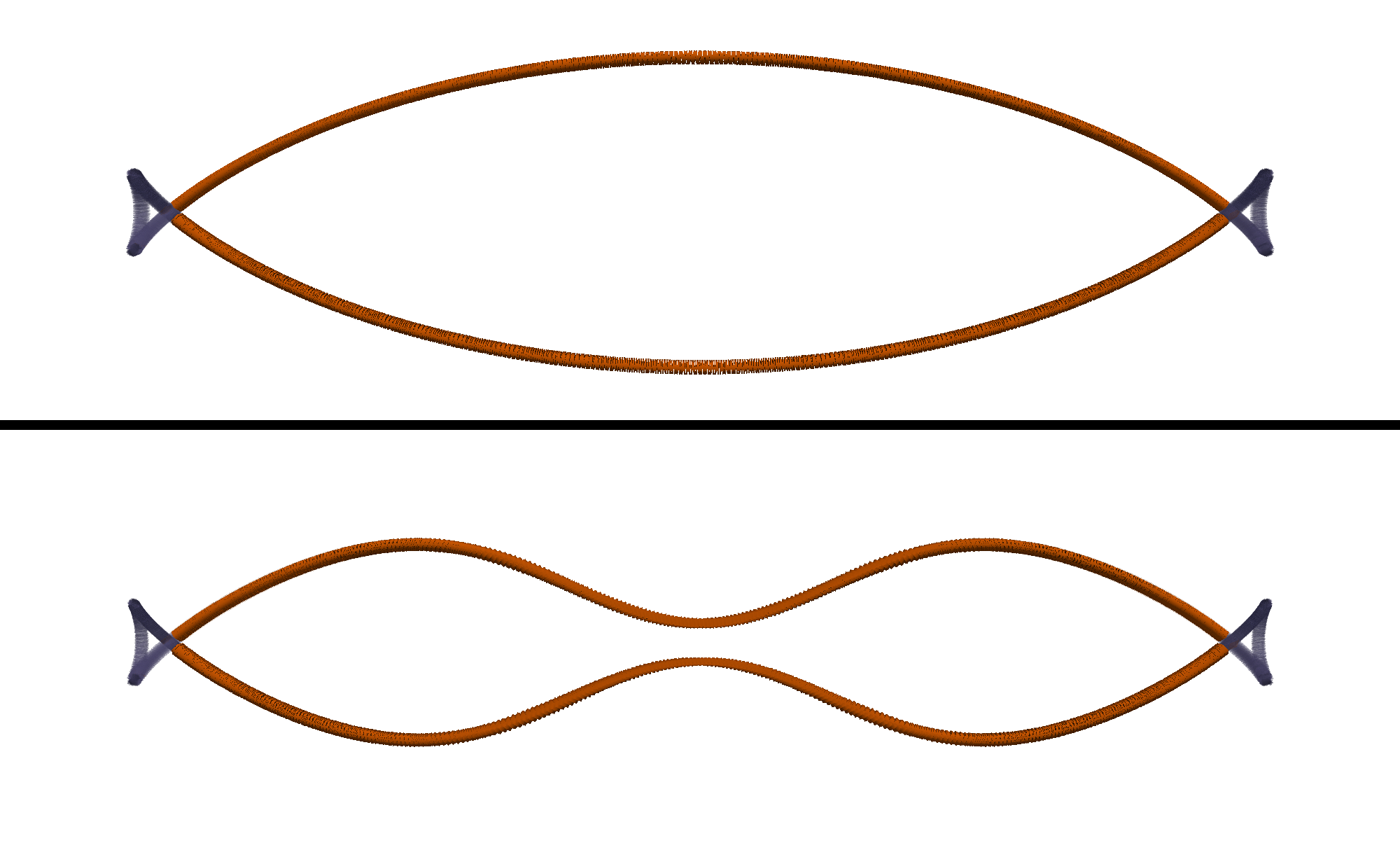} \\
    (a) $7539.891 M$ &
    (b) $7539.918 M$ &
    (c) $7539.948 M$ &
    (d) $7540.971 M$
  \end{tabular}
  \caption[Spatial cuts of the equal mass inspiral during the merger]{
    Slices of the equal mass inspiral during the merger of
    \cref{figEMSurface}, where the vertical direction
    in the figure is parallel to $\hat{r}_{\rm{maj}}$, and the slice
    is taken through the hole in the EH.
    The slices have the same character as those in \cref{figToyS1Slice}.
  }
  \label{figEMSlice}
\end{figure*}

The torus is illuminated more clearly by taking
spatial cuts through the EH surface in both coordinate systems
as shown in \cref{figEMSlice}.
The vertical direction in the figure is parallel to $\hat{r}_{\rm{maj}}$.
The slices in this figure bear a remarkable resemblance
to the ellipsoidal model slices in \cref{figToyS1Slice}, suggesting that
the future generators join the horizon in a similar manner.
In panel~(b), the first generators to join the EH in the constant
$t$ slicing join through crossover points.
We are able to delay these generators such that the first generators
to join the horizon in the constant $\bar{t}$ slicing are spatially
separated in the slice in panel~(c).
As time progresses forward, generators continue to join at the
interfaces
between future generators and
event horizon regions in the $\bar{t}$ slicing.
Finally in panel~(d), the two pieces of the horizon connect after all the
remaining generators in the gap join the horizon.

\begin{figure}
  \centering
  \includegraphics[width=1.00\columnwidth]{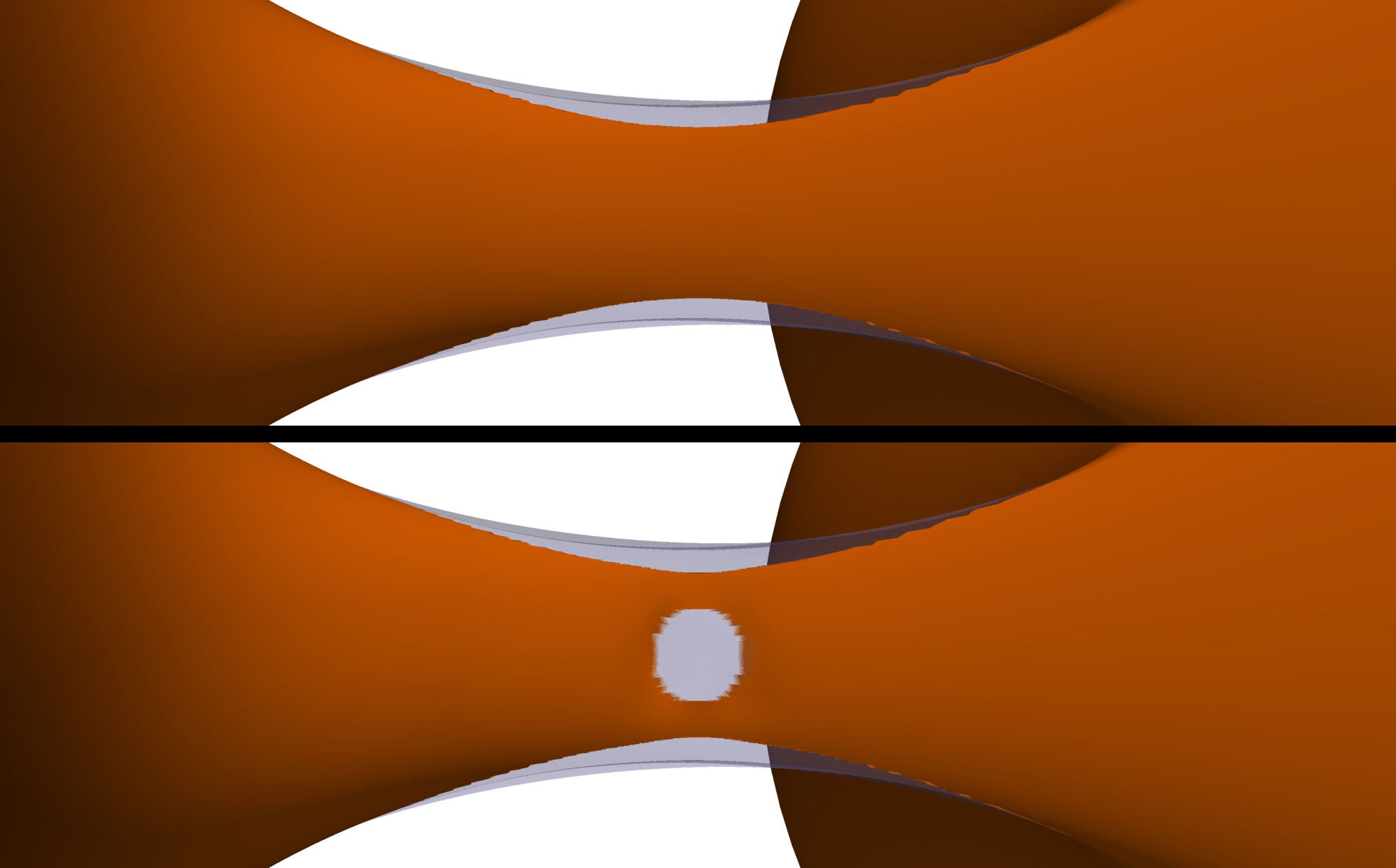}
  \caption[Toroidal event horizon in the equal mass inspiral binary]{
    Zoomed in figure of the hole in the horizon.
    The full event horizon generator surface including future generators
    are shown at time $7539.948 M$ corresponding to panel~(c) of
    \cref{figEMSlice}.
    This toroidal event horizon and the other systems in the discussion section
    \cref{secTori} at the webpage~\cite{EHTopologiesWebsite}.
  }
  \label{figTORUS}
\end{figure}

\Cref{figTORUS} shows up close what the hole in the horizon
looks like.
The top and bottom rows are constant $t$ and constant $\bar{t}$ slices.
We are showing both the full generator surface as well as the same spatial slice
as seen in panel~(c) of \cref{figEMSlice}.
The constant $\bar{t}$ slice shows clearly that there is a hole
in the event horizon surface, so the EH has a toroidal topology.
For the hole in the horizon, the EH surface pinches off along a
one-dimensional non-smooth ring where event horizon generators
will continue to join through crossover points.
The left and right edges of the event horizon surface shown in orange
are also not smooth, where
generators continue to join through crossover points.
The final generators to join the event horizon surface do so
through caustic events, just as seen in the ellipsoidal model
(\cref{figToyS1Slice}).
This torus is seen in all three refinement levels of the \SpEC{} BBH evolution.

The coordinate transformation used does not guarantee
that constant $\bar{t}$ hypersurfaces are spacelike.
We therefore must check that the new lapse $\bar{\alpha}$ is well behaved
by evaluating \cref{eqnNewLapse} in the region where $\bar{t}$ differs
from $t$, that is, where $G(x^i, t)$ is non-negligible.
We construct a grid of points centered about $\vec{r}_0$ and $t_0$ to evaluate
the new lapse in the range of
\begin{subequations}
\begin{align}
  t &= t_0 \pm 4 \sigma_t \\
  \vec{x} &= \vec{r}_0 \pm 4 \sigma_{\rm{maj}} \hat{r}_{\rm{maj}}
    \pm 4 \sigma_{\rm{min}} \hat{r}_{\rm{min}1}
    \pm 4 \sigma_{\rm{min}} \hat{r}_{\rm{min}2},
\label{eqnLapseCheckBoundaries}
\end{align}
\end{subequations}
where $\hat{r}_{\rm{min}1}$ and $\hat{r}_{\rm{min}2}$ are unit vectors
perpendicular to
each other and perpendicular to $\hat{r}_{\rm{maj}}$.
Beyond this range, the Gaussian function is vanishingly small
($G(x^i, t) < e^{-8} = \mathcal{O}(10^{-4})$)
for our purposes.

We use a grid of points with $N_{\rm{pts}}$ points distributed in each dimension
of the
four-dimensional space defined by \cref{eqnLapseCheckBoundaries} to calculate
the new lapse $\bar{\alpha}$ and check that it is real.
Because the new lapse is a function of the metric in the \SpEC{} coordinate
system, we must interpolate the metric $g_{\mu \nu}$ to the location in space
and time where $\bar{\alpha}$ is to be calculated.
These interpolations are performed the same way as is done during the
generator evolution, described in the companion paper~\cite{BohnMethods2016}.

\begin{figure}
  \centering
  \includegraphics[width=1.0\columnwidth,trim=3cm 1cm 1cm 2cm,clip]
    {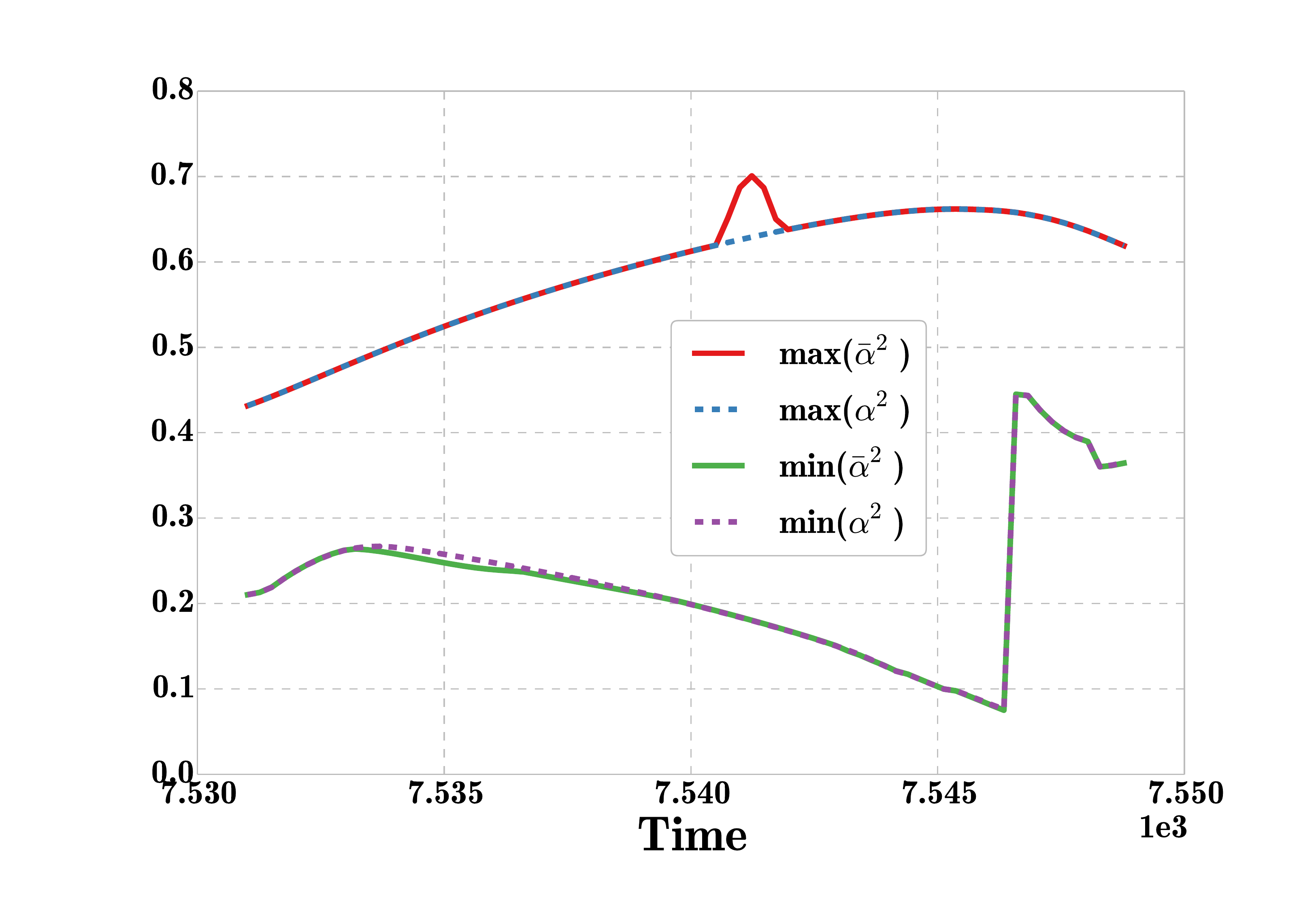}
  \caption[Confirmation that the lapse function is well behaved]{
    Confirmation that the lapse is well behaved for both the $t$ and
    the $\bar{t}$ coordinate systems.
    The minimum and maximum values of $\alpha^2$ are plotted as a function of
    time.
    Note that the large jump in the minimum lapse squared is
    caused by the domain regrid as \SpEC{} transitions into the ringdown, and
    the coordinate transformation has no effect on the jump.
  }
  \label{figLapseCheck}
\end{figure}

\Cref{figLapseCheck} shows the lapse squared
in both the \SpEC{} coordinate system ($\alpha^2$) and in the new coordinate
system ($\bar{\alpha}^2$) using a grid with $74^4$ evenly distributed points
over the Gaussian.
At each of the $74$ times, we calculate the square of the lapse
on $74^3$ spatial points
and
plot the maximum and minimum found in both coordinate systems.
This plot shows that the constant $\bar{t}$ hypersurfaces are indeed spacelike,
because $\bar{\alpha}^2$ is positive at all times.
It should be noted that we could not check the lapse at all points on this wide
grid, since some of the points live off the \SpEC{} evolution domain
because of the excision region inside the black holes;
However, these
locations are guaranteed to be inside the event horizon and so do
not affect the event horizon.
All other points in the \SpEC{} domain and in the space defined by
\cref{eqnLapseCheckBoundaries} contribute to
\cref{figLapseCheck}.
The large spike in the minimum lapse squared in both coordinate systems is
an expected feature from how the excision surfaces in \SpEC{} change during
the BBH merger phase.

\subsection{Baby event horizons}
\label{secBabies}

To obtain toroidal event horizons, we used a positive amplitude
Gaussian in our coordinate transformation in \cref{eqnCoordTransformation}
to delay generators in a small region around where the event horizons
merge.
We now consider the effect of a negative amplitude Gaussian that
will advance generators in a small region.

\begin{figure*}
  \centering
  \begin{tabular}{ccc}
    \includegraphics[
      width=0.33\textwidth]{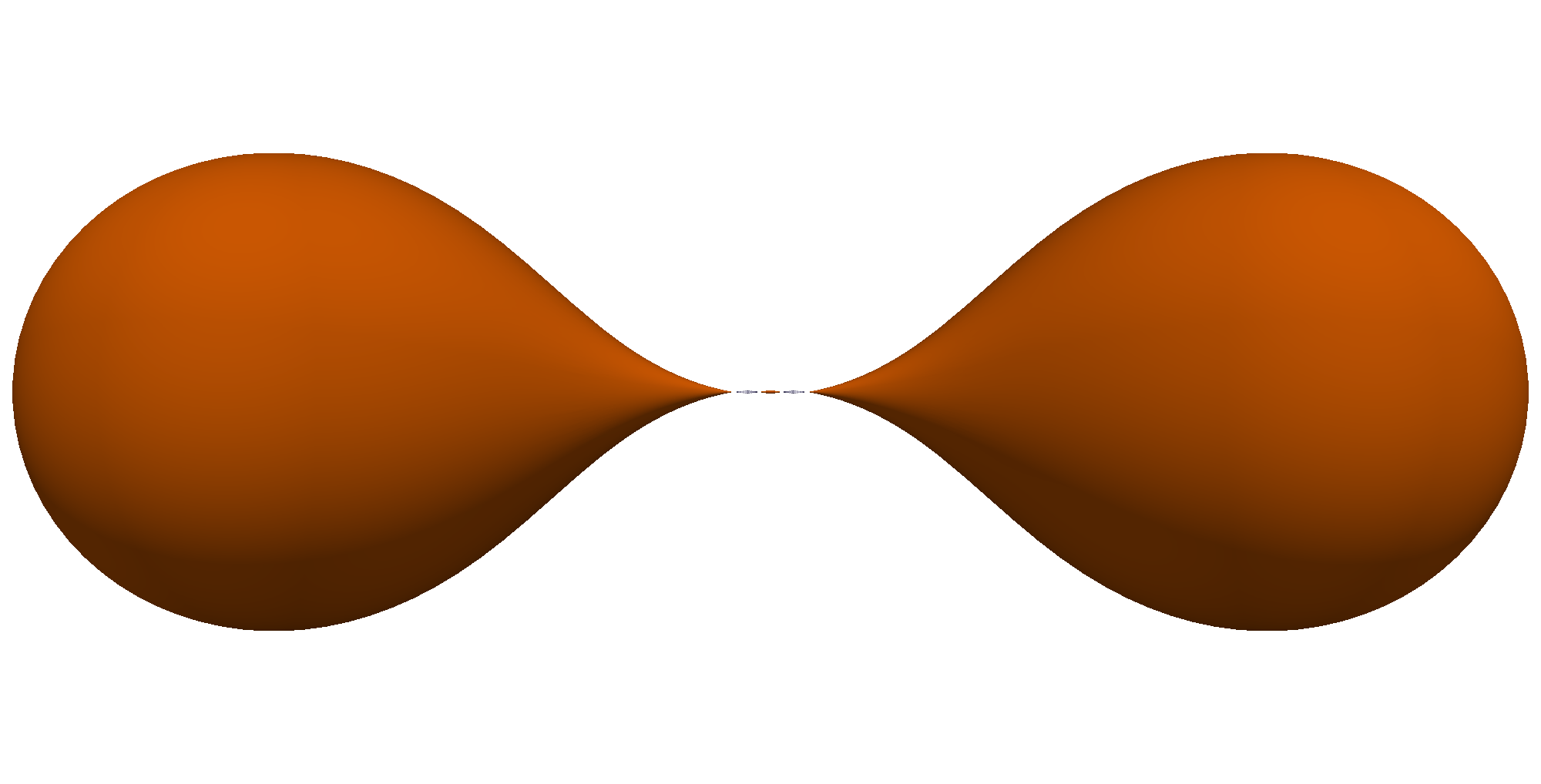} &
    \includegraphics[
      width=0.33\textwidth]{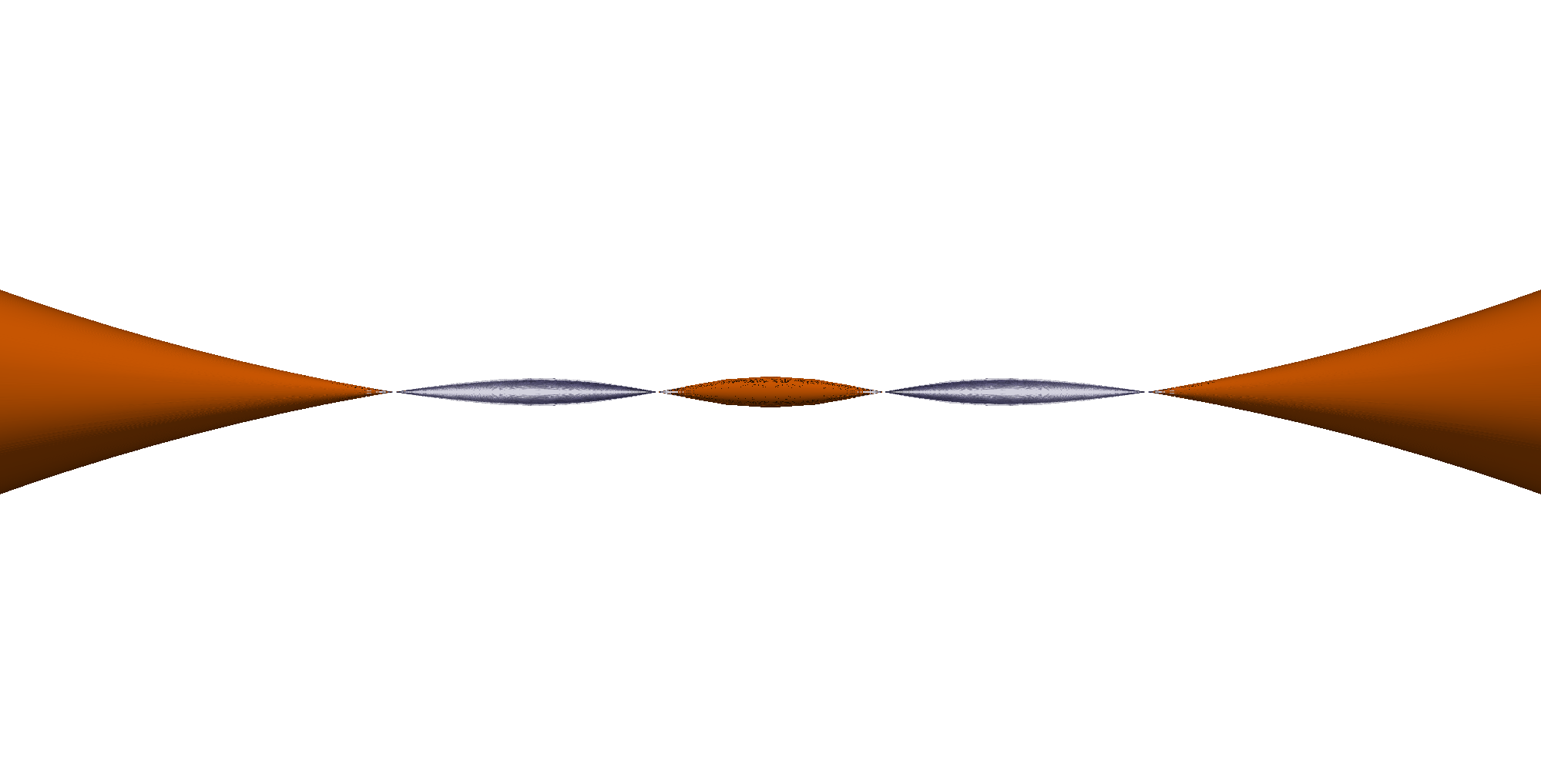} &
    \includegraphics[
      width=0.33\textwidth]{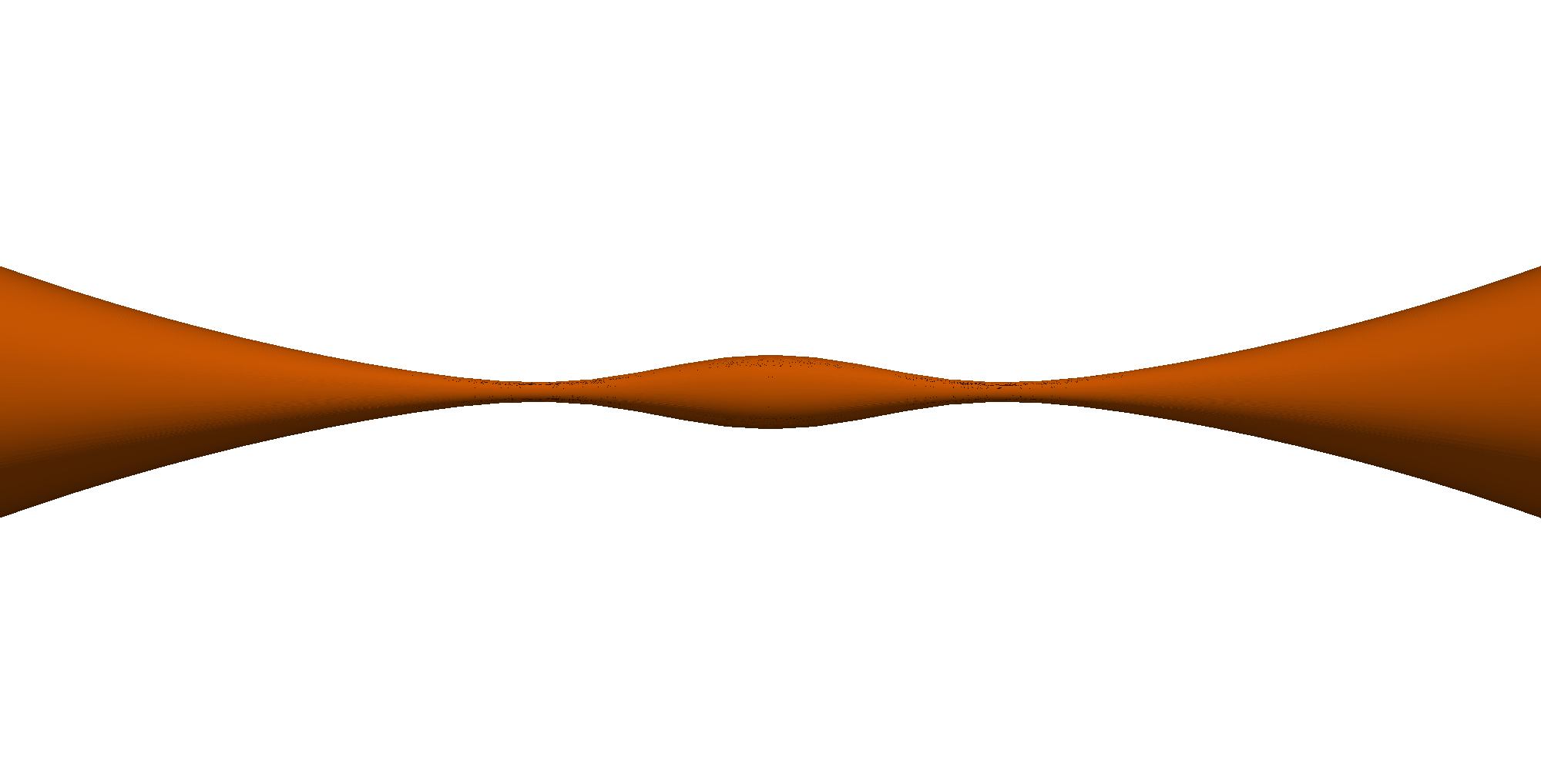} \\
    (a) $\bar{t}=417.407 M$: Zoomed out &
    (b) $\bar{t}=417.407 M$ &
    (c) $\bar{t}=417.422 M$
  \end{tabular}
  \caption[Generation of a baby event horizon in the head-on BBH system]{
    Event horizon of a head-on equal mass BBH merger, after
    performing the coordinate transformation with Case~B of
    \cref{figResliceParameters}, but with a negative amplitude.
    Panel~(a) is a zoomed out view of panel~(b), where the topology
    of the event horizon is three spheres $3\times \mathcal{S}^2$, before
    merging into one sphere in panel~(c).
    This baby event horizon and the other systems in the discussion section
    \cref{secTori} at the webpage~\cite{EHTopologiesWebsite}.
  }
  \label{figHOBaby}
\end{figure*}

The head-on BBH event horizon from \cref{secHOMerger} has all the future
generators joining through caustics that form a one-dimensional
spacelike line along the inseam of the pair of pants diagram.
If we advance generators in a small region near this line, we can
push the time slice across this spacelike line in a small region.
The event horizon on the new time slice would have the topology of three spheres
$3\times \mathcal{S}^2$ instead of $2\times \mathcal{S}^2$ before the merger.
In theory, we could make our time slicing cross the spacelike line of caustics
as many times as we would like to create a topology of $n\times \mathcal{S}^2$,
a possibility proved by Siino~\cite{Siino1998b} in corollary III.8.
This is directly demonstrated in \cref{figHOBaby}.
We have also created an additional ``baby'' event horizon in more generic
mergers such as the binary in \cref{figLIGOTorus}, where there are
not only caustics but also crossover points.

Similarly, when we can reslice an event horizon to produce a torus with one
hole, we can reslice into a torus with $n$ holes.
The crossover surface is spacelike, so we can construct a slicing
that intersects this crossover surface an arbitrary number of times.

\section{Conclusions}
\label{secConclusions}

Siino~\cite{Siino1998b} and Husa and Winicour~\cite{Husa-Winicour:1999}
expected that merging black hole event
horizons should generically have a brief toroidal topology.
While simulations of rotating collapsing matter have shown event horizons
that appear initially with a toroidal topology, the toroidal BBH
event horizon has remained hidden during numerical simulations.
While the $2+1$-dimensional event horizon hypersurface itself does not depend on
the spacetime foliation, the choice of spacetime foliation does
affect the topology of the EH on the slice.
For the case of the inspiral and merger of two equal mass non-spinning black
holes,
we find the event horizon topology transitions directly
from two spheres to one sphere in the \SpEC{} coordinate slicing.
However, we show directly that a toroidal event horizon
is possible through the use of a specially constructed coordinate
transformation.
The topology of the event horizon in the new coordinate system
transitions from two spheres to a
short-lived torus before transitioning finally
to one sphere.
No event horizons of merging black holes prior to this paper
have yielded a toroidal
topology~\cite{Cohen2012, CohenPfeiffer2008, Diener:2003, ponce:11}.

We believe that our reslicing method can be applied to the merger of any
black holes with sufficient asymmetry (\textit{i.e.}, not including a head-on
merger of black holes where the symmetry prevents the possibility of a torus).
Previous work has numerically found a surface of crossover points
during the merger, where generators meet non-neighboring generators as they
join the EH surface.
Because this surface of crossover points is spacelike, we can apply our
coordinate transformation to ``cut a hole'' through the crossover surface,
while keeping the hypersurfaces of constant time spacelike.
We therefore agree with Siino~\cite{Siino1998b}
and Husa and Winicour~\cite{Husa-Winicour:1999} that merging black holes
should, in general, briefly have a toroidal event horizon topology, with the
caveat that the torus may only exist in some foliations of the spacetime.
It is interesting that Siino and Husa and Winicour predict tori generically,
and expect slicings where there is no torus to be an exception to the rule.
It is therefore somewhat surprising that in the time slicing used in \SpEC{}
and all other
numerical codes, it appears that slicings with a toroidal event horizon
are the exception to the rule.

As for topological censorship,
because we are explicitly converting a spherical event horizon into a
toroidal event horizon with our coordinate transformation, we are
satisfying topological censorship by construction.
That is, we can trivially reslice the event horizon back into a spherical
topology, removing the topological torus, implying that the
hole in the event horizon closes faster than the speed of light.
Therefore a photon that appears to probe the topology of the spacetime
by passing through the hole in the EH in one foliation of the spacetime
will simply pass between the event horizons before they merge in another
foliation.
We note that while it is true one can always reslice a
topological-censorship-satisfying torus into a sphere, the reverse
is not always true.

%% file: Tikz/TikzSetup.tex
\definecolor{myteal}{RGB}{27, 158, 119}
\definecolor{myorange}{RGB}{217, 95, 02}
\definecolor{mypurple}{RGB}{117, 112, 179}
\definecolor{mypink}{RGB}{231, 41, 138}
\tikzset{
  griddot/.style={
    draw,
    circle,
    minimum size=12pt,
    color=myteal,
  },
  interpdot/.style={
    draw,
    circle,
    minimum size=7pt,
    color=myorange,
    fill=myorange,
  },
  gridsep/.style={
    draw,
    dashed,
    very thick,
    mypurple
  },
  gridchangedot/.style={
    draw,
    circle,
    fill=black,
    minimum size=5pt,
  },
  plus/.style={cross out, draw,
    minimum size=2*(#1-\pgflinewidth),
    inner sep=0pt,
    outer sep=0pt,
    ultra thick,
    rotate=45},
    plus/.default={5pt},
  cross/.style={path picture={
    \draw[black]
      (path picture bounding box.south east) --
      (path picture bounding box.north west)
      (path picture bounding box.south west) --
      (path picture bounding box.north east);
  }},
}

\pgfmathsetmacro{\xsep}{33}
\pgfmathsetmacro{\ysep}{23}
\pgfmathsetmacro{\regridy}{11}
\pgfmathsetmacro{\ysepone}{0}
\pgfmathsetmacro{\yseptwo}{2}
\pgfmathsetmacro{\ysepthree}{6}
\pgfmathsetmacro{\ysepfour}{12}
\pgfmathsetmacro{\ysepfive}{20}
\pgfmathsetmacro{\timelabelsep}{7}
\pgfmathsetmacro{\gridlabelsep}{3}

%% file: acknowledgments.tex
We thank Aaron Zimmerman for the suggestion of using a negative
amplitude Gaussian to create a baby event horizon, and for helpful
dialogues regarding topologies of event horizon surfaces.
For ongoing feedback and suggestions for finding toroidal event horizons
over the past few years, we thank Jeffrey Winicour.
We also thank Michael Boyle for useful conversations about topologies of
event horizon surfaces and useful paper comments.
We thank Leo C. Stein for verifying the effects of coordinate
transformations and useful paper comments.
For keeping the \SpEC{} code from changing under us while we were locating
event horizons, we thank Daniel A. Hemberger.
We are grateful to Jordan Moxon, Nils Deppe, and Fran\c{c}ois H\'{e}bert
for time slicing conversations and useful comments
during the editing phase of this paper.
We also thank Harald Pfeiffer for providing the BBH simulation with parameters
similar to the system detected by Advanced LIGO.
For helping smooth the visualization of
event horizon surfaces, we thank Curran D. Muhlberger.

We gratefully acknowledge support for this research at Cornell from the
Sherman Fairchild Foundation and NSF grants PHY-1306125 and AST-1333129.
Calculations were performed on the Zwicky cluster at Caltech,
which is supported by the Sherman Fairchild Foundation and by
NSF award PHY-0960291; on the NFS XSEDE network under grant TG-PHY990007N;
at the GPC supercomputer at the SciNet HPC Consortium \cite{Scinet};
SciNet is funded by:
the Canada Foundation for Innovation (CFI) under the
auspices of Compute Canada; the Government of Ontario;
Ontario Research Fund (ORF) – Research Excellence;
and the University of Toronto.
All the surface visualizations were done using Paraview~\cite{paraviewweb}.
The line plots were produced using the
Matplotlib~\cite{Hunter:2007} library with Python.